\documentclass[oneside, 12pt]{article}
\usepackage[spanish]{babel}
\usepackage[mathscr]{euscript}
\usepackage[utf8]{inputenc}
\usepackage[T1]{fontenc}
\usepackage{lmodern}
\usepackage{amssymb}
\usepackage[square,numbers]{natbib}
\usepackage{enumerate} 
\usepackage{makeidx}
\makeindex
\usepackage{graphicx}
\usepackage{float}
\floatstyle{boxed} 
\restylefloat{figure}

\newcommand{\keyword}[1]{\textbf{#1}}

\begin{document}

\pagestyle{empty}
\renewcommand{\indexname}{Índice de nombres}
\setcounter{page}{1}
\pagestyle{plain} 
\date{}
\bibliographystyle{plain}
\begin{titlepage}
\title{INFLUENCIAS ASTRONÓMICAS SOBRE LA EVOLUCIÓN GEOLÓGICA Y BIOLÓGICA DE LA TIERRA (Parte I).}
\author{Carlos A. Olano\\ \small{Facultad de Ciencias Astron\'{o}micas y Geof\'{i}sicas,}\\ \small{Universidad Nacional de La Plata, Argentina}\\ \small{(E-mail: colano@fcaglp.fcaglp.unlp.edu.ar)}}
\maketitle
\end{titlepage} 
\maketitle
\begin{abstract}
 This monograph presents a study of  the nature and origin of meteorites, asteroids and comets;  and of the consequences of  encounters of these cosmic objects with the Earth. The purpose  of this  monograph is mainly of  divulgation for non-specialists,  even though we have not avoided some elementary technical aspects and some own proposals. The section 2 refers briefly to climatic changes occurred  in historical times and their probable causes.  The main theories proposed to explain the formation of the solar system are  exposed  succinctly in the section 3. In the section 4, it is  described characteristics and fall phenomena of meteorites, as well as impact craters of the Earth and of the other planets produced  by  big meteorites and asteroids. The impact structures of Argentine are described with some detail, and it is wonder wether the impacts that formed them contributed to the extinction of the local megafauna. The section 5 shows that the effects of the terrestrial magnetosphere on the trajectories of an interplanetary dust stream that is passing through the Earth  are negligible. The evolution of an expanding bubble originated by the explosion of a cosmic object that releases 50 Mt of energy in  the terrestrial atmosphere, at a height of 5 km  from the ground,  is calculated in the section 6. A short historical review  of the investigations on the famous Tunguska’s explosion of 1908 and an explanation of the white nights observed in Eurasia and of other phenomena related to the Tunguska event are provided in section 7. Besides, by means of a simple model developed by the author in this  article to explain the white nights, it is found that the Earth would have been surrounded by a layer of dust of a few million tons.
Some of the meteor swarms  that originate the  annual meteor showers on the Earth, such as  the Beta Taurids vinculated with the small comet Encke and the fragment of comet that caused the Tunguska explosion, would be the product of the disintegration of a great comet during the last 20000 years. The  impact that the observations of these celestial  phenomena related to the disintegration of the great comet could have in the ancient civilizations is suggested in the last section.
\end{abstract}
\keyword{\textbf{Keywords}: Solar system--Earth--Comets--Asteroids--Meteorites--Impact craters--Meteor showers--The Tunguska explosion of 1908--Megafaunal extinctions.}

\,
\,
\,
\,
\section{Introducción}
 
\begin{quote}
\small \it{Otros sucumbían a los cambios climáticos, otros a la falta de atmósfera. A veces el fin llegaba a causa del choque con densas nubes de polvo o gas, o con enjambres de meteoros gigantes}.\rm

 Olaf Stapledon\index{Stapledon O.} (1886-1950) en ``star maker'' (1937) \footnote{Una edición en español de Hacedor de Estrellas cuenta con un prólogo de Jorge Luis Borges\index{Borges J.L.}, quien concluye  que ``Para los hábitos mentales de nuestro siglo, Hacedor de Estrellas es, además de una prodigiosa novela, un sistema  probable o verosímil de la pluralidad de los mundos y de su dramática historia''. Hasta el presente, los astrónomos llevan descubiertos algunos miles de exoplanetas, es decir sistemas planetarios en otras estrellas, con lo cual la calificación de ``probable o verosímil'' dada por Borges fue sin dudas acertada. De una imaginación casi ilimitada, esta obra de Stapledon inspiró la mayoría de las ideas fundamentales de la ciencia-ficción moderna: las razas simbióticas (ej. E. F.Rusell\index{Rusell F.}), los imperios galácticos (ej. I. Asimov\index{Asimov I.}) y las nebulosas y estrellas inteligentes (ej. F. Hoyle\index{Hoyle F.}).}
\end{quote}  

Carl Sagan\index{Sagan C.},  en su serie televisiva Cosmos, llamaba a la Tierra ``el pequeño planeta azul''. La atmósfera de la Tierra debido a su relativamente alto contenido de oxígeno, característica que la diferencia de los demás planetas del sistema solar\index{sistema solar}, dispersa mayormente la parte azul del espectro lumínico solar. Ese alto porcentaje  de oxígeno no es casual, sino el producto de la evolución de la vida en la Tierra, iniciado en la vida primitiva con el proceso de fotosíntesis. El espesor de la parte más densa  de nuestra  atmósfera (troposfera) es de apenas 10 km, extremadamente delgada  comparada  con el radio de la Tierra ($\approx 6400$ km) a la cual rodea. La vida  depende  críticamente de esta delgada capa de gas. La atmósfera no sólo nos da el aire que respiramos, sino que ésta  es además  en parte el motor térmico que  determina el tiempo y el clima. Y no menos importante, la atmósfera es una coraza que nos protege de los rayos cósmicos y de la radiación ionizante del Sol, como así también de los meteoritos que impactan y se desintegran en ella. 

Sin embargo, nuestra atmósfera es vulnerable a peligros que acechan  tanto  desde el espacio exterior como desde dentro de la misma Tierra. Por ejemplo, si una extensa y densa nube de polvo cósmico penetrara nuestra atmósfera, parte del polvo cósmico podría esparcirse y quedar  suspendido en la atmósfera  oscureciendo completamente al Sol y sumiendo  a la Tierra en una larga noche, que podría durar meses, con consecuencias  catastróficas para la vida. Por la falta de luz solar, la cadena de alimentación se corta al interrumpirse el proceso de fotosíntesis y además la superficie de la Tierra se enfria. El desarrollo de técnicas de detección y protección contra cuerpos cósmicos que puedan amenazar a nuestro planeta constituye una nueva disciplina que abarca diferentes campos científicos \cite{Hazards}.

 En esta monografía\footnote{Intentado como capítulo de un libro con diversos temas astronómicos y físicos, tratados en forma general y no especializada, aunque no se  soslayan algunos aspectos técnicos elementales. Toda crítica, puntualización de errores e imprecisiones  y sugerencias serán muy apreciadas por el autor (emails: a colano@fcaglp.fcaglp.unlp.edu.ar o  carlosolanoA1950@gmail.com)}, estudiaremos el origen y naturaleza de los meteoritos, asteroides y cometas y el fenómeno de sus impactos sobre la Tierra. En particular, describiremos los principales cráteres de impacto descubiertos en la Argentina, y discutiremos brevemente si un gran impacto cósmico pudo ser la causa de la extinción de la megafauna\index{megafauna} local al final del Pleistoceno\index{Pleistoceno}. Por otra parte, trataremos con especial énfasis  la famosa colisión cósmica  que ocurrió al comienzo del siglo XX en la meseta Siberiana cerca del río  Tunguska\index{Tunguska}. Para explicar diferentes aspectos del evento Tunguska\index{Tunguska}, esbozaremos  un modelo nuevo.  En la segunda parte de esta monografía, trataremos el papel que jugaron estos cuerpos menores de sistema solar\index{sistema solar} en la formación de la Tierra y en las grandes extinciones masivas de plantas y animales ocurridas en los últimos 500 millones de años, desde la explosión de vida del Cámbrico\footnote{ver por ejemplo ``Hacia una teoría galáctica del catastrofismo terrestre'' de este autor\cite{Olano1}}.

\section{Cambios climáticos en tiempos históricos}

 La explotación excesiva de los recursos naturales y la contaminación del medio ambiente, resultantes de la actividad industrial del hombre, hacen peligrar el frágil equilibrio que mantiene el clima terrestre y empujan  a la extinción a numerosas especies de animales y plantas. Como es bien sabido, la quema de combustibles fósiles ha provocado un progresivo aumento de los gases de efecto invernadero en la atmósfera. 
 Sin embargo, el cambio climático que puede originar la actividad humana es naturalmente un fenómeno reciente, ocurrido tras la revolución industrial. Se sabe que, tanto en tiempos históricos como en tiempos geológicos, nuestro planeta   sufrió  cambios climáticos naturales.

El Sol es la fuente principal de calor de la Tierra y por lo tanto variaciones de su luminosidad pueden provocar variaciones en las temperaturas medias de la atmósfera. Sin embargo, el Sol es una estrella notablemente estable, que  presenta variaciones de luminosidad  de apenas un $0.1$ por ciento  a lo largo del ciclo solar de 11 años. Una de las manifestaciones del ciclo de 11 años de actividad solar  es el aumento del número de  manchas solares. Aparentemente,  esas pequeñas variaciones  de la luminosidad solar producen cambios en la circulación del calor que no afectan por igual a todo el planeta. Un ejemplo podría ser la pequeña era de hielo que ocurrió entre los años 1650-1720, durante la cual  Europa y América del Norte estuvieron sometidas a  inviernos muy crudos. Notablemente, ese periodo  de 70 años que duró la pequeña edad de hielo coincidió con el Mínimo de Maunder\index{Maunder}, un déficit en la cantidad de manchas solares.
 
Las erupciones volcánicas intensas tienen también efectos transitorios sobre el clima. El polvo volcánico que asciende hasta las capas altas de la  atmósfera se distribuye por todo el planeta y absorbe parte de la luz solar que normalmente llega a la superficie de la Tierra, enfriando la troposfera. Hay muchos ejemplos históricos que dan testimonio de dicho fenómeno. Algunas de las más famosas explosiones volcánicas son la erupción del Tambora\index{Tambora (volcán)} (Indonesia, año 1815), considerada la erupción más importante de los últimos 10.000 años, y  la erupción del Krakatoa\index{Krakatoa  (volcán)} (Indonesia, año 1883). Los volcanes de esa gran magnitud tienen repercusiones climáticas globales que producen, en la troposfera y en la superficie, enfriamientos de hasta $0,6^{\circ}C$ en los años inmediatos a la erupción.

Los científicos coinciden en general que la actividad solar y el vulcanismo pueden sólo inducir cambios climáticos poco significativos  y que por lo tanto no explican el calentamiento global observado en el último siglo. La causa principal del calentamiento  global sería  la emisión de gases de efecto  invernadero. En las secciones siguientes, consideraremos  las grandes catástrofes  climáticas que alteraron (o casi interrumpieron)  el curso evolutivo de la vida en varias ocasiones a lo largo de la historia de la Tierra. Consideraremos también los posibles escenarios astronómicos y geológicos en los que se desarrolló este drama de la vida.

\section{Teorías sobre la formación del Sistema Solar\index{sistema solar}}
La teoría heliocéntrica de Copérnico \index{Copérnico} inició una revolución en la astronomía y la física. Astrónomos, matemáticos  y filósofos  venerables como Kepler,\index{Kepler} Galileo,  \index{Galileo} Descartes, \index{Descartes} Newton\index{Newton} y Kant \index{Kant} trabajaron en el desarrollo de  una nueva física en la cual se unía al cielo con la Tierra. Descartes ideó un sistema de vórtices de éter alrededor de soles y planetas, que explicaría las órbitas casi circulares de los planetas alrededor del Sol. Sin embargo, el éxito de la mecánica Newtoniana opacó la teoría de los vórtices de  Descartes. \index{Descartes} El filósofo alemán, Inmanuel Kant, \index{Kant} en su obra ``Historia general de la naturaleza y teoría del cielo'' publicada en  1755,  presentó su teoría nebular para explicar la formación del sistema solar\index{sistema solar}. Pese a algunos altibajos, la teoría nebular aún se mantiene  en pie como una teoría plausible. Según esta teoría,  el Sol y los planetas se formaron por condensaciones de grumos dentro  una nube interestelar fría, oscura y achatada debido a su rotación alrededor su eje central perpendicular al plano  medio de la nube. El Sol se formó en el núcleo central de la nube. La hipótesis de Kant fue la primera dentro de las variaciones que aparecieron de la hipótesis nebular. El sabio francés Laplace, que no conocía la existencia de la    hipótesis nebular de Kant, publicó en ``Exposición del sistema del mundo'' (año 1796) una idea similar a la de Kant, por ello se suele llamar hipótesis o teoría de Kant-Laplace. La importancia de la teoría nebular radica en que puede explicar el hecho de que los planos orbitales de los planetas son casi coincidentes y además la mayoría de los planetas se mueven en torno al Sol en el mismo sentido y en el mismo sentido que el Sol rota sobre su eje.
 
Las teorías cosmogónicas 
se pueden dividir en dos clases: las {\it evolucionistas} \rm  como la hipótesis nebular  de Kant y Laplace y las {\it catastrofistas}. La primera teoría catastrofista se debe al naturalista francés conde de Buffon (Jorge Luis Leclerc), \index{Buffon conde de} quien sugería que sucesivos choques de cometas contra el Sol desprendieron grandes masas incandescentes del Sol que al enfriarse formaron los planetas. En 1680 el inglés E.  Halley, \index{Halley E.} amigo de I. Newton, \index{Newton} calculó la órbita del cometa que luego llevaría su nombre y encontró que el cometa pasó tan cerca del Sol que casi lo rozó. En ese entonces se creía que los cometas eran cuerpos de gran masa.  En vida del conde de Buffon, en el año 1759,  apareció  nuevamente el cometa Halley en el cielo de la Tierra.  El conde debe haberse  inspirado  en ese acontecimiento para formular su teoría. El mecanismo del conde de Buffon no puede explicar las órbitas cuasi circulares de los planetas, como lo notó Laplace. Las condiciones iniciales de los desprendimientos de  masa solar determinan que estos cuerpos vuelvan a caer sobre el Sol.

Nunca hay que dar por muerta una vieja teoría. El astrónomo Moulton \index{Moulton} y el geólogo Chamberlin,\index{Chamberlin} ambos norteamericanos, rejuvenecieron  en el año 1905 la teoría del conde de Buffón, reemplazando la colisión de cometas por la colisión de una estrella solitaria contra el Sol. Tratando de salvar defectos, se presentaron variaciones de esta teoría catastrofista como la de Sir James Jeans. \index{Jeans J.} Sin embargo, todas ellas chocaron con obstáculos insuperables y fueron abandonadas. La colisión entre dos estrellas de campo no asociadas es un fenómeno poco probable. Se estima que la probabilidad de que esto ocurra es un caso en mil millones y por lo tanto la formación de sistemas planetarios seria poco frecuente (y mucho menos la aparición de la vida en ellos), según estás teorías catastrofistas. Claro que ello de por sí no las invalida.

La hipótesis nebular de Kant y Laplace aún goza de buena salud. Esta hipótesis esta de acuerdo con la observación de discos de  gas y polvo en cuyos respectivos centros yacen estrellas en formación. Todo indicaría que, por alguna  necesidad dinámica,  las estrellas no nacen en general solas. Las estrellas tienden a  formarse  en grupos o cúmulos y se cree que las estrellas solitarias como el Sol se formaron junto con sistemas planetarios.

\section{Cráteres de impacto sobre la Tierra}

Los asteroides,  meteoritos y cometas son el material sobrante de la formación del sistema solar\index{sistema solar}. Estos cuerpos menores fueron los precursores de los planetas (planetesimales), los cuales a través de impactos entre ellos, se juntaron y formaron aglomerados mayores. 
Los asteroides son planetas menores con diámetros desde algunos kilometros, como Icarus \index{Icarus (asteroide)} de solo 2 km, hasta unos pocos cientos de kilometros. Con un diámetro de 1000 km, Ceres\index{Ceres (asteroide)} es el mayor asteroide,  descubierto el 1 de enero de 1801 por Giuseppe Piazzi \index{Piazzi G.} desde el Observatorio de Palermo, Sicilia \index{Sicilia}(Italia).  El naturalista alemán Alexander von Humboldt, \index{von Humboldt A.} en su gran obra \it{Cosmos} \rm del año 1849, no dudaba en llamar a los meteoritos ``los más pequeños de todos los asteroides''.  Tenemos conocimientos de los meteoritos  gracias a que algunos de ellos caen sobre la Tierra de vez en cuando. Exceptuando los tamaños, los asteroides y meteoritos son esencialmente semejantes, incluso con los cometas. El modelo que mejor representa a los cometas es el de bola de nieve sucia. Por lo tanto, resulta atractiva la idea, propuesta primeramente por el astrónomo  Öpik en 1966, \index{Oepik E.J.} de que los cometas son conglomerados de hielo y meteoritos rocosos. Cuando el hielo se evapora, quedan solo los meteoritos.
.

Recién hacia mediados del siglo diecinueve, la idea de que los meteoritos son piedras que cayeron del cielo fue ampliamente aceptada. La caída de un meteorito se observa como una bola de fuego, con una cola o estela luminosa, que surca rápidamente el cielo en su caída. Los efectos lumínicos y sonoros de este fenómeno eran  difíciles  de entender en ese entonces. El desarrollo de la termodinámica permitió comprender que debido al roce con la atmósfera el meteorito pierde energía cinética, con lo cual se calienta la superficie del meteorito y al gas atmosférico circundante a temperaturas de incandescencia. Los estruendos y vibraciones provenientes del bólido, se deben al  paso de la onda de choque que se origina en el frente del meteorito, debido a su interacción con el gas atmosférico. Cuando la velocidad del meteorito es mayor que la velocidad del sonido en la atmósfera, se origina un frente de choque de forma cónica (cono de Mach), \index{Mach} cuyo vértice coincide con el frente del meteorito.
 
  Desde que se reconoció  el origen extraterrestre de los meteoritos, los grandes museos del mundo los coleccionan. La colección  más grande de meteoritos se encuentra en el museo de Viena. \index{Viena} En la Argentina,\index{Argentina} el museo Bernadino Rivadavia \index{Bernadino Rivadavia} y el planetario de la ciudad de Buenos Aires,\index{Buenos Aires} como así también  el museo de ciencias naturales de La Plata \index{La Plata} exhiben  algunos ejemplares. Se han hallado más de 31000  meteoritos bien documentados que van desde pequeños guijarros hasta rocas que pesan decenas de toneladas\footnote{Un cátalogo de meteoricos hallados en territorio Argentino fue  publicado por Rogelio D. Acevedo\index{Acevedo R.D.} y Maximiliano C.L. Roca\index{Roca M.C.L.}\cite{Acevedo}}. El meteorito Hoba\index{Hoba (meteorito)} de 66 ton., que se encuentra en  Namibia, \index{Namibia} es el meteorito de mayor peso en el mundo. Lo siguen los meteoritos Gancedo\index{Gancedo (meteorito)} y El Chaco\index{El Chaco (meteorito)} con $30.8$ y $28.8$ ton., respectivamente, ambos fragmentos de la lluvia de meteoritos del Campo del Cielo \index{Campo del Cielo} ubicado en la Provincia del Chaco\index{Chaco}, Argentina \index{Argentina} (ver sección 4.3). Los meteoritos que han  caído sobre la Tierra eran hasta  hace muy poco  las únicas muestras disponibles del material primigenio del sistema solar\index{sistema solar}  para  efectuar análisis  en el laboratorio. Actualmente, la astronáutica nos da la oportunidad de analizar  asteroides y cometas \it{in situ}. \rm 

Los impactos de grandes meteoritos y asteroides sobre la Tierra primitiva, y la consecuente formación de grandes cráteres sobre ella, jugaron un papel fundamental en el origen y evolución de la Tierra y de la vida misma. Aun la  Tierra ya formada  experimentó  intensos bombardeos de grandes  meteoritos, asteroides y probablemente cometas con efectos catastróficos sobre la vida y el clima del planeta. Basta ver una imagen de la Luna plagada de cráteres, para dimensionar la magnitud de los impactos que recibió la Tierra. La falta de atmósfera de la Luna  favorece la conservación de los cráteres. En cambio en la Tierra, los cráteres muy viejos fueron totalmente  erosionados por los agentes  meteorológicos y disgregados  por los movimientos de la corteza terrestre. Si no fuera por esos procesos que borraron las huellas de la mayor parte de los cráteres de impacto sobre la superficie de la Tierra, habría más cráteres que las 190 estructuras de impacto que han sido documentadas a la fecha. El libro de Paul Hodge\index{Hodge, P.} \cite{Hodge} es un excelente compendio con detalladas descripciones  y fotografías de cráteres de meteoritos y de grandes estructuras de impacto de la Tierra. Quien desee profundizar en el estudio de los meteoritos puede consultar por ejemplo \cite{Heide} y \cite{Norton}. Para lectores en general, el libro de Nigel Calder\index{Calder N.} \cite{Calder} es interesante.

 Hay cráteres que son reconocibles a simple vista, como el hermosamente preservado cráter de Arizona\index{Arizona} \footnote{Este cráter tiene un diámetro de 1300 m y una profundidad de 175 m,  y recibe varios nombres: cráter meteoro, cráter Barringer\index{Barringer (cráter)} y cráter cañón del diablo},  mientras otros son totalmente invisibles, enterrados bajo una gruesa capa de sedimentos como el famoso cráter de Chicxulub\index{Chicxulub (cráter)}                                              
asociado con la extinción en masa de numerosas formas de vida del final del período Cretácico\index{Cretácico}, \index{Cretácico} entre ellos los dinosaurios.
Estos últimos cráteres  de impacto fueron descubiertos en general por algunas anomalías geológicas  del suelo que llamaron la atención y motivaron investigaciones geológicas y geofísicas, y particularmente el mapeado gravimétrico de la región. El mejor criterio para el reconocimiento de un gran cráter de impacto es la existencia, en la zona de la estructura,  de material vidrioso característico de la metamorfosis que sufre la roca alcanzada por la intensa onda de choque del  impacto. El material vítreo  asociado  con los cráteres de impacto consiste generalmente  en pequeños fragmentos de  vidrios de diversos colores que resultan de la cristalización, bajo grandes presiones, de minerales como el cuarzo, \index{cuarzo} el feldespato, \index{feldespato} la monacita \index{monacita} y el circón, \index{circón} entre otros. Dado que esos cristales contienen pequeñas cantidades de uranio \index{uranio} radioactivo que se  transforma en plomo \index{plomo} de manera constante, es posible determinar la edad de los cráteres.

Los meteoritos tienen propiedades químicas y mineralógicas que permiten distinguirlos de las rocas terrestres. A grandes rasgos los meteoritos se dividen  en rocosos y ferrosos. A su vez, estos se subdividen en varias clases. En los meteoritos rocosos se distinguen dos clases principales: condritos \index{condritos} y acondritos.  \index{acondritos} La mayoría de los meteoritos tienen rasgos químicos que los diferencian de las rocas de la corteza terrestre. Ellos son mucho más ricos en elementos  como el iridio (Ir), \index{iridio} el osmio  (Os) \index{osmio} y el 
renio  (Re) \index{renio }. En el impacto contra el suelo, los asteroides y grandes meteoritos se destruyen completamente  fundiéndose con parte del material terrestre desplazado que forma el cráter. Por lo tanto, si analizamos muestras de material de las paredes de un  gran cráter y  ellas arrojan sobreabundancias de elementos como el iridio, entonces podemos asegurar que la  estructura fue formada por un impacto cósmico. De modo que ésta es otra de las pruebas que nos permiten identificar a los cráteres de impacto.

	Diariamente, caen sobre la Tierra numerosos meteoritos muy pequeños (meteoros) que normalmente se queman antes de llegar al suelo por rozamiento con la atmósfera. Durante la noche se  los observa como estelas luminosas, llamadas  ``estrellas fugaces''. En noches despejadas se ven  aproximadamente seis estrellas fugaces por hora. Los meteoros caen  tanto de noche  como de día, pero la probabilidad aumenta hacia el amanecer que es cuando la parte de la superficie terrestre en que nos encontramos  se mueve, debido a la rotación de la Tierra, en el mismo sentido que la Tierra se mueve en su órbita. Las partículas que se desprenden del meteorito al quemarse se precipitan  lentamente hacia el suelo, en parte como esférulas micrométricas, denominadas micro-meteoritos.

La Tierra incrementa su masa  en aproximadamente 100  toneladas al día por este fenómeno.  Hay lluvias de meteoros que ocurren en fechas especificas del año y que están asociadas con material  desprendido por cometas a lo largo de sus órbitas\footnote{La primera asociación entre a cometa y una lluvia de  meteoros fue hecha por el astrónomo italiano  Giovanni Schiaparelli\index{Schiaparelli G.} (1835-1910)}. Los desprendimientos cometarios consisten esencialmente en granos de polvo micrométricos. El astrónomo aficionado Noruego Jon Larsen\index{Larsen J.}, a través de su proyecto ``polvo de estrellas'', incentiva  la búsqueda de  micro-meteoritos \index{micro-meteoritos} que se depositan sobre  los tejados de nuestras casas. 
En la actualidad, se cuenta con una red mundial de observatorios ópticos y de radiofrecuencias  que monitorean el cielo nocturno y diurno  con el fin  detectar  lluvias de  meteoros. Las lluvias de meteoros pueden detectarse aún durante el día mediante técnicas de radar. Uno de dichos observatorios se encuentra en Tierra del Fuego, \index{Tierra del Fuego} al sur de la Argentina. \index{Argentina}

Cuando un meteorito alcanza la superficie de la Tierra, raramente se mantiene intacto. Los  meteoritos  de baja masa, entre $0.1$ y 1 tonelada,  son fuertemente frenados por la atmósfera y el impacto contra el suelo puede no desintegrarlos \footnote{La resistencia a su desintegración depende de la composición  del  meteorito y de las características del terreno. Los meteoritos de hierro son naturalmente los más resistentes} y los cráteres que producen son relativamente pequeños, entre una y unas pocas veces el tamaño del meteorito. Normalmente, el meteorito al entrar a la atmósfera con muy alta velocidad se fragmenta; y si   muchas de esas partes  alcanzan la superficie, sus impactos estarán distribuidos sobre una extensa área, generalmente elíptica. La dirección del eje mayor de la elipse nos indica  la dirección,  casi rasante o de baja inclinación con respecto al suelo, con la cual la corriente de meteoritos chocó contra la superficie. A este fenómeno se lo denomina lluvia de meteoritos. Una de los más grandes lluvias de meteoritos ocurrió  en el año 1969 en el poblado de Allende,\index{Allende} al sur del estado mexicano de Chihuahua\index{Chihuahua}. Miles de piedras cayeron sobre un área elíptica  de 50 km de longitud y 12 km de ancho, en cuyo vértice norte se encontró el meteorito más grande, con un peso  de 110 kg. Si tenemos en cuenta que la mayoría de las piezas superan el kilogramo y caen con velocidades de $\approx 100 $ m s$^{-1}$ (ver próxima sección), no sería nada grato estar bajo semejante lluvia de meteoritos. La probabilidad de que un meteorito golpee a una persona es muy baja. Se estima que esto puede ocurrir una vez cada 1000 años. Sin dudas se cuentan historias de personas muertas por impactos de meteoritos, sin embargo  no se conocen casos  autenticados.

\subsection{Frenado de los meteoritos por el roce atmosférico}
La fuerza aerodinámica de fricción puede expresarse  por medio de la   fórmula clásica
\begin{equation}
 F_{r}= -\epsilon \,\rho(h)\, A \,v^{2},
 \label{Fr}
\end{equation}
donde $v$ es la velocidad del meteorito con respecto a la atmósfera terrestre, a la cual suponemos estática, y $A$ es el área del meteorito que se proyecta en la dirección de $v$. La densidad de la atmósfera $\rho$  depende de la altura $h$ sobre la superficie de la Tierra. $\epsilon$ es el coeficiente de fricción, el cual depende del aspecto del meteorito, pero su valor es usualmente del orden de  2. Observe que $F_{r}$ tiene la dirección de la velocidad pero el sentido es contrario pues se opone al movimiento.

Si llamamos $O$ al punto de entrada del meteorito en la atmósfera, la vertical del lugar es definida por la recta que une el punto $O$ con el centro de la Tierra. La altura de $O$ es la distancia entre la superficie de la Tierra y el punto $O$ medida a lo largo de la vertical. Tomaremos a $O$ como el origen de un sistema Cartesiano de coordenadas, cuyo eje $y$ coincide con la vertical y apunta hacia la superficie de la Tierra.

Las fuerzas que actúan sobre el meteorito en la dirección del eje $y$ son la fuerza de gravedad ($m g$) y la componente $y$ de la fuerza de fricción.
La componente $y$ de la fuerza de fricción se obtiene proyectando $F_{r}$ sobre $y$. Entonces, $F_{y}= F_{r} \, cos\, \phi+ m g$,  donde
$\phi$ es el ángulo entre la vertical y la velocidad $v$, $m$ la masa del meteorito y $g$ la aceleración de la gravedad. Aplicando la segunda ley de Newton, \index{Newton}$F_{y}=m a_{y}= m \frac{dv_{y}}{dt}$ y por lo tanto
\begin{equation}
\frac{dv_{y}}{dt}= -\epsilon \,\rho(h)\,\frac{A}{m} \,v^{2}  cos \,\phi + g, 
\label{Hidrodinamica1}
\end{equation}
donde $v_{y}=v\, cos\, \phi$. Similarmente, se obtiene la desaceleración en la dirección del eje $x$
\begin{equation}
\frac{dv_{x}}{dt}= -\epsilon \,\rho(h)\,\frac{A}{m} \,v^{2}  sin \,\phi, 
\label{Hidrodinamica2}
\end{equation}
Observe que en la dirección $x$ no actúa la gravedad porque $x$ es perpendicular a la vertical, es decir a $y$. Pasando el diferencial de tiempo $dt$ en  (\ref{Hidrodinamica1})  al otro lado del signo igual multiplicando y teniendo en cuenta que $v\, cos\, \phi=v_{y}$, obtenemos para (\ref{Hidrodinamica1}) 
 \begin{equation}
dv_{y}= (-\epsilon \,\rho(h)\,\frac{A}{m} \,v v_{y} + g)\, dt, 
\label{Hidrodinamica3}
\end{equation}
y similarmente para (\ref{Hidrodinamica2}) 
\begin{equation}
dv_{x}= (-\epsilon \,\rho(h)\,\frac{A}{m} \,v v_{x})\, dt, 
\label{Hidrodinamica4}
\end{equation}
Las ecuaciones de movimiento (\ref{Hidrodinamica3}) y (\ref{Hidrodinamica4}) nos permiten calcular la trayectoria de un meteorito en la atmósfera.
Dado que $v=\sqrt{v_{x}^{2}+v_{y}^{2}} $, esas ecuaciones de movimiento están acopladas y deben ser resueltas simultáneamente.
Si el meteorito no se aparta mucho en su caída de la vertical del lugar (eje $y$), podemos suponer que  la densidad atmosférica se distribuye en capas plano-paralelas en las cuales la densidad depende solo de la altura $h$, conforme a una ley exponencial:
\begin{equation}
\rho(h)= \rho(0)\, exp (-\frac{h}{H}),
\label{Atmosfera}
\end{equation}
donde $\rho(0)=1.225$ kg m$^{-3}$ y $H= 8.42$ km.
Comenzaremos a contar el tiempo $t$ a partir del ingreso del meteorito a la atmósfera. Es decir,cuando  $t=t_{0}=0$ el meteorito se encuentra en el punto $O$, cuya altura sobre la superficie de la Tierra la denotamos $h_{0}$. Consideremos además que  $v=v_{0}$ en el punto $O$, velocidad a la cual denominaremos velocidad de ingreso o  cósmica del meteorito, y $\phi=\phi_{0}$, al cual denominaremos ángulo de ingreso. 
En el momento $t=t_{0}=0$, $v_{x} (t_{0})=v_{0}\, sen\, \phi_{0}$ y $v_{y} (t_{0})=v_{0}\, cos\, \phi_{0}$, $h= h(t_{0})=h_{0}$ y  $\rho(h_{0})$ es  dada (\ref{Atmosfera}). Con estas condiciones iniciales y las ecuaciones (\ref{Hidrodinamica3}) y (\ref{Hidrodinamica4}), podemos calcular  la velocidad y posición del meteorito transcurrido un dado diferencial de tiempo $dt$. En efecto, haciendo $v_{x}=v_{x} (t_{0})$, $v_{y}=v_{y} (t_{0})$, $\rho(h)=\rho(h_{0})$ en (\ref{Hidrodinamica3}) y (\ref{Hidrodinamica4}), calculamos la variación de velocidad que sufrió el meteorito  en el intervalo $dt$. Dado que $dv_{y}=v_{y}(t_{1})-v_{y}(t_{0})$ y $dv_{x}=v_{x}(t_{1})-v_{x}(t_{0})$, donde $t_{1}=t_{0}+ dt$, la nuevas velocidades son $v_{y}(t_{1})=v_{y}(t_{0})+dv_{y}$ y $v_{x}(t_{1})=v_{x}(t_{0})+dv_{x}$. En ese intervalo de tiempo el meteorito se desplazó  en la coordenada  $y$ la distancia $dy= (v_{y}(t_{0})+ \frac{dv_{y}}{2})\, dt$ y por lo tanto la altura del meteorito es ahora $h(t_{1})=h(t_{0})-dy$.
En la coordenada $x$, el desplazamiento es $dx= (v_{x}(t_{0})+ \frac{dv_{x}}{2})\, dt$. Reemplazando las  velocidades y posiciones que hemos calculado para el tiempo $t_{1}$ en (\ref{Hidrodinamica3}) y (\ref{Hidrodinamica4}), obtenemos las velocidades y posiciones del meteorito para el tiempo $t_{2}=t_{1}+dt=2 dt$. Repitiendo este procedimiento varias veces podemos determinar la trayectoria y las velocidades del meteorito al atravesar la atmósfera.

\begin{figure}
\includegraphics[scale=1.6]{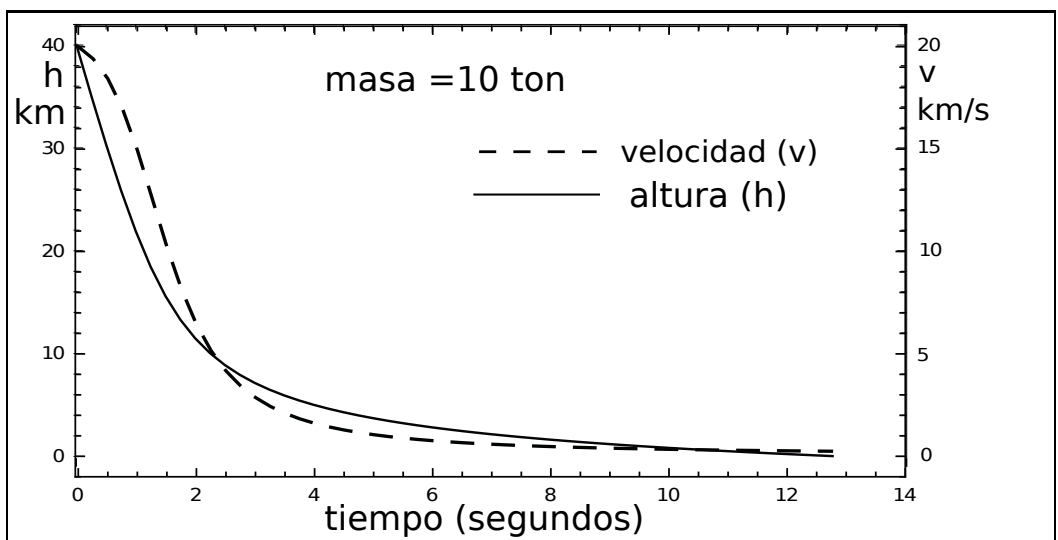} 
\caption{Altura y velocidad de un meteorito rocoso en función del tiempo trascurrido desde el ingreso a la atmósfera. El meteorito tiene una masa de 10 toneladas y ingresa verticalmente a una altura de 40 km con una velocidad de 20 km/s. La línea a trazos representa  la velocidad, la cual se lee en la escala del  eje vertical derecho; y le curva llena da la altura, la cual  se lee sobre el eje vertical izquierdo. El eje horizontal es común a ambas curvas.}
\label{CaidaVHt}
\end{figure}

\begin{figure}
\includegraphics[scale=2.1]{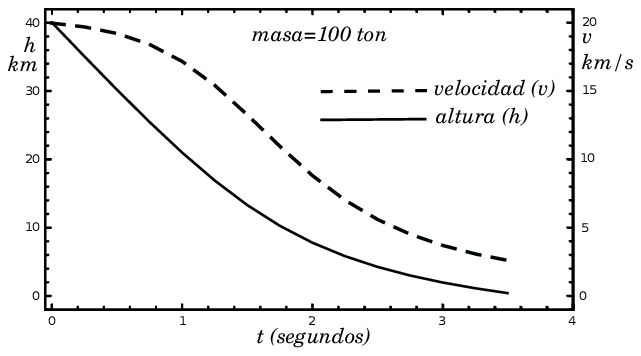} 
\caption{Altura y velocidad de un meteorito rocoso en función del tiempo trascurrido desde el ingreso a la atmósfera. El meteorito tiene una masa de 100 toneladas y ingresa verticalmente a una altura de 40 km con una velocidad de 20 km/s. La línea a trazos representa  la velocidad, la cual se lee en la escala del  eje vertical derecho; y le curva llena da la altura, la cual  se lee sobre el eje vertical izquierdo. El eje horizontal es común a ambas curvas}
\label{CaidaVHt2}
\end{figure}

A fin de calcular la caída de meteoritos de distintos pesos, supondremos por simplicidad que ingresan verticalmente ($\phi_{0}=0$) a la atmósfera a una altura $h_{0}=40$ km con una velocidad cósmica $v_{0}=20$ km s$^{-1}$. Por lo tanto, $x=0$ y $v_{x}=0$ a lo largo de la trayectoria del meteorito y sólo $h$ y $v_{y}=v$ varían con el tiempo. Supondremos además que el meteorito es esférico, con lo cual dados la masa $m$ y la densidad del meteorito obtenemos su radio $R$ y a partir del mismo el área $A (=\pi R^{2} )$ cuyo valor es necesario introducir en las fórmulas (\ref{Hidrodinamica3}) y (\ref{Hidrodinamica4}). Adoptaremos el valor de $4000$ kg m$^{-3}$ para la densidad de los meteoritos, valor que corresponde a la densidad de los meteoritos rocosos que son los más abundantes. Consideraremos además que el meteorito no pierde masa al atravesar la atmósfera. Esto es una simplificación, porque el meteorito en realidad pierde masa por  ablación, sin embargo es válida para lo que queremos demostrar.
La fig. \ref{CaidaVHt} muestra que un meteorito de 10 toneladas llega al suelo $(h=0)$ en $\approx 13$ segundos y que es altamente desacelerado por la atmósfera en los primeros $30$ km chocando contra al suelo con una velocidad muy baja en relación a su velocidad cósmica. En cambio, un meteorito de 100 toneladas impacta contra el  suelo en solo $3.5$  segundos y la pérdida de su velocidad cósmica es mucho menor (ver fig. \ref{CaidaVHt2}).

\begin{figure}
\includegraphics[scale=1.0]{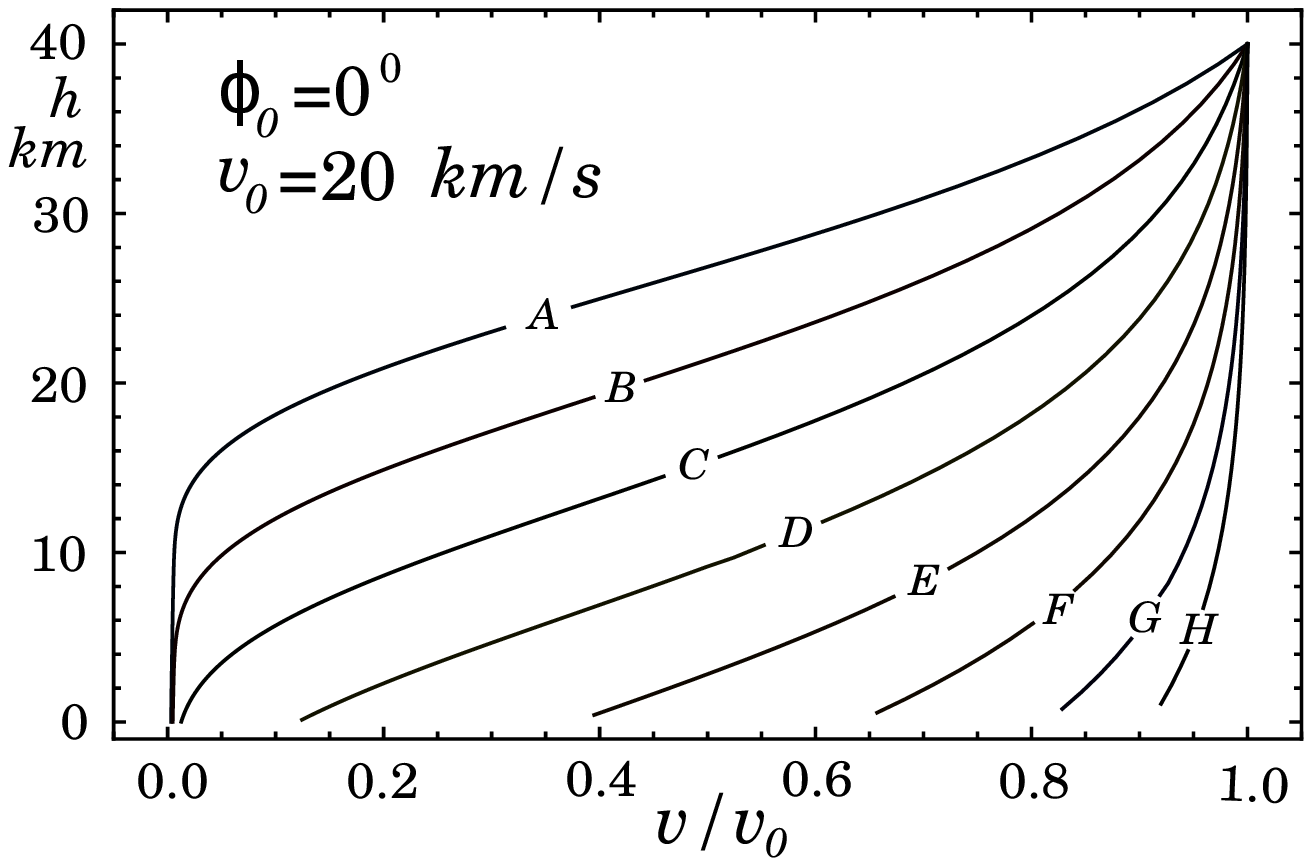} 
\caption{Alturas teóricas en relación con las velocidades  relativas a la velocidad cósmica de meteoritos de diferentes masas.  La letra sobre cada curva refiere a la masa del meteorito: A=0.1, B=1, C=10, D=100, E=1000, F=10000, G=100000 y H=1000000 toneladas}
\label{VrH}
\end{figure}

La fig. \ref{VrH} muestra la variación de la velocidad relativa $v/v_{0}$ con la altura $h$ para meteoritos de distintas masas. En dicha figura, uno ve que el meteorito de $0.1$ ton (curva A) y el de 1 ton (curva B) alcanzan una velocidad constante, llamada  velocidad terminal, a la altura de $\approx 14$ km y de $\approx 6$ km, respectivamente. La velocidad es constante porque a esas alturas la fuerza de roce se iguala a la fuerza de gravedad y por lo tanto la desaceleración es cero (es decir la ecuación (\ref{Hidrodinamica3}) se hace cero). La velocidad terminal es del orden de 100-200 m s$^{-1}$, y por lo tanto la velocidad de impacto de los meteoritos pequeños es relativamente baja. Para estos casos, nuestra atmósfera  es un escudo protector eficaz. El fenómeno lumínico cesa cuando el meteorito alcanza la velocidad terminal y se lo ve caer como un cuerpo oscuro. Solo una delgada capa de la superficie del meteorito es calentada y fundida por las colisiones con las moléculas del aire. Durante el proceso, sin embargo, el resto del meteorito no es significativamente calentado. El proceso es muy rápido, dura solo unos pocos segundos, y no se tiene suficiente tiempo para que el calor penetre  a su interior. Luego, en la etapa final del descenso, la costra del meteorito se enfría. Los efectos del impacto dependen de la dureza del suelo; pudiéndose producir un pozo con el meteorito en su interior o un pequeño cráter  con fragmentos del meteorito. Por ejemplo,  una piedra de 1 ton caída  el 18 de febrero de 1948 en Norton County (Kansas)    penetró el suelo hasta una profundidad de sólo 3 metros \index{Norton County} \index{Kansas}. El proceso de excavación es puramente mecánico, sin que se produzcan ondas de choque de alta presión.

Las velocidades con las que los meteoritos con masas mayores a 10 ton llegan a la superficie de la Tierra superan la velocidad terminal o de caída libre. Los meteoritos con masas mayores a $10^{5}$ ton son levemente frenados, conservando más del 80 por ciento de la velocidad cósmica (ver curvas G y H de la fig. \ref{VrH}). Estos cuerpos, con sus velocidades cósmicas casi intactas y sus altas masas, poseen por lo tanto   energías cinéticas enormes que   deben perderlas instantáneamente  cuando ellos impactan contra la superficie terrestre. A diferencia de los pequeños meteoritos ($ \leq $ 2 o 3 ton) que producen simplemente agujeros en el suelo,  los grandes meteoritos o asteroides  literalmente explotan al chocar contra el suelo, formando cráteres denominados explosivos. La enorme cantidad de energía que se libera en el choque desencadena  procesos físicos  que son cuantitativa y cualitativamente muy  diferentes a aquellos que forman los cráteres no explosivos. Por ello, el mecanismo que forma los cráteres explosivos es tratado aparte en  la sección siguiente.

\subsection{El mecanismo de formación de los grandes cráteres de impacto}
En las últimas tres décadas se ha avanzado mucho en nuestra comprensión de los complejos procesos físicos que  sobrevienen cuando un meteorito gigante o asteroide choca contra una superficie planetaria  a alta velocidad. El astrónomo y astrofísico estonio Öpik \index{Oepik E.J.} \footnote{Fue uno de los primeros que entrevió los efectos catastróficos de una colisión cósmica sobre la Tierra. Por ejemplo en 1970, Öpik  escribió ``parece muy posible que el desarrollo de la vida durante la era Proterozoico\index{Proterozoico} (anterior a 500 millones de años atrás) haya sido afectada, entre otras causas, por colisiones catastróficas''} puntualizó ya en 1936 que cuando  un meteorito con velocidad cósmica impacta la Tierra, los límites elásticos  de ambos, el proyectil (meteorito) y  el blanco (región impactada), son ampliamente excedidos y ambos se comportan como fluidos. Por lo tanto, las ecuaciones hidrodinámicas que describen los movimientos de los fluidos pueden también aplicarse para determinar el comportamiento del proyectil y del blanco. 

En el instante en que un gran objeto cósmico  choca la superficie terrestre, lo hace con  casi su velocidad cósmica original que es mucho mayor que la velocidad del sonido en el suelo. En consecuencia, se genera una  intensa onda de choque  que se irradia desde el punto de impacto hacia  el suelo circundante, y hacia el interior del mismo cuerpo que impacta. Así, la inmensa energía cinética del cuerpo cósmico se convierte en una onda de choque que se genera en una fracción de segundo en la región de contacto entre el borde delantero del cuerpo que impacta  y la superficie del suelo.
La configuración y  evolución del  frente de la onda de choque que penetra el suelo, una envoltura  casi hemisférica,  determinan la morfología final del cráter. Cada elemento del frente de la onda de choque actúa como una topadora empujando el terreno en una dirección perpendicular a sí mismo. La energía de la onda de choque se consume derritiendo y  moviendo el material acumulado en su frente, con lo cual se forma una envoltura de roca fundida.
El material del suelo cercano a la superficie es eyectado fuera del cráter por la onda choque, pues este material es más fácilmente acelerado. Por otro lado,  el cuerpo cósmico  se desintegra totalmente en parte  fundiéndose y en parte vaporizándose por efecto de la onda de choque que lo atraviesa.

En la etapa de excavación, el cráter tiene el aspecto de una superficie parabólica de revolución en torno a la vertical que pasa por el punto de colisión. En otras palabras, el cráter tiene forma de tazón. La mayoría de los grandes cráteres, vistos desde arriba,  son circulares. Ello es así porque independientemente  de la forma del proyectil y de su ángulo de choque  sobre el blanco, toda la energía cinética se deposita casi instantáneamente en el sitio de la colisión  desde el  cual se propaga la onda expansiva de la explosión. En la llamada  etapa de modificación o reacomodamiento del cráter, se produce un rebote de la parte más profunda del cráter, elevando el piso del cráter y  formando en ciertos casos una montañita o  pico en el centro del piso. La formación del cráter se completa en unos pocos minutos, un raro ejemplo de un proceso geológico instantáneo.

\begin{table*}
 \centering
 \begin{minipage}{140mm}
  \caption{Los cráteres de impacto más grandes de la Tierra ($D\geq 100$ km)}
\begin{tabular}{cccccc} \hline \hline
Nombre      & País      & Diámetro (km)  & Edad (Ma)  \\ \hline\hline
Sudbury     & Canadá    & 200           &  1850        \\ \hline
Chicxulub   & México    & 180           &  65         \\ \hline
Acraman     & Australia & 160           & 570         \\ \hline
Vredefort   & SudAfrica & 140           & 1970        \\ \hline
Manicouagan & Canadá    & 100           &  212        \\ \hline
Popigai     & Rusia     & 100           & 35       \\  \hline \hline
\end{tabular}
\label{TablaCrateres}
\index{Sudbury (cráter)}  \index{Chicxulub (cráter)} \index{Acraman (cráter)} \index{Vredefort (cráter)} \index{Manicouagan (cráter)} \index{Popigai (cráter)}
\index{Canadá} \index{México} \index{Australia} \index{SudAfrica} \index{Rusia}

\end{minipage}
\end{table*}

En la Tabla \ref{TablaCrateres}, se listan los cráteres más grandes de la Tierra. En otros cuerpos del sistema solar\index{sistema solar} hay  estructuras de impacto aún mucho mayores. Los cráteres mayores de 300 km de diámetro  son  tradicionalmente llamados cuencas. Las manchas oscuras de la Luna  que otrora se llamaban ``mares'', cuando no se conocía la inexistencia de agua en la Luna \footnote{ Actualmente, se ha detectado la presencia de agua en sitios especiales, pero no está en el estado y  proporciones  para formar mares}, son en realidad enormes cuencas de impacto. Las cuencas se caracterizan por contener en general múltiples anillos topográficos concéntricos que rodean la depresión central \cite{Spudis}. Los anillos, los cuales  pueden describirse como cadenas circulares de montañas, están separados entre si por una depresión anular de la superficie. 
Una conspicua cuenca de impacto  es la cuenca lunar llamada Oriental con seis anillos circulares concéntricos de 320, 480, 620, 930, 1300 y 1900 km de diámetro. La Luna contiene  más de 40 cuencas y a una de las más grandes del sistema solar\index{sistema solar}: la cuenca ``Mar de la Lluvia (Imbrium)'' \index{Imbrium (cuenca de impacto)} cuyo anillo externo tiene 3200 km de diámetro. Las cuencas están también presentes en Mercurio, \index{Mercurio} Marte\index{Marte} y los satélites congelados de Júpiter \index{Júpiter} y Saturno\index{Saturno}. Estudios mediante técnicas de radar revelan que Venus \index{Venus} también posee grandes cuencas de impacto. Así como los otros planetas experimentaron impactos que formaron cuencas, no hay razones para suponer que la Tierra fue una excepción. La mayoría de los cráteres terrestres reconocibles son simples, con forma de tazón,  o bien complejos con picos centrales. Sin embargo, muchos de los cráteres más grandes de la Tierra tienen múltiples anillos; hay 17 cráteres con al menos dos anillos, entre ellos los listados en la Tabla  \ref{TablaCrateres}.

El escenario que se desarrolla con el impacto que forma una cuenca es dantesco y lo que más intriga a los investigadores planetarios es cómo se forman los misteriosos   anillos concéntricos. Si bien se han propuesto algunas hipótesis  para explicar la formación de los múltiples anillos, aún no se ha elaborado un modelo completamente satisfactorio. Probablemente,  la clave para entender este fenómeno de los grandes impactos  esté en que la mayor profundidad excavada por la onda de choque supera el espesor de la corteza del planeta. La corteza es una capa rígida de rocas que  puede  quebrarse bajo la presión de la onda de choque, produciendo  fracturas concéntricas de la corteza en coincidencia con la configuración circular que forma la intersección de la corteza con la onda de choque. El reacomodamiento de esas  placas anulares de la corteza y el manto subyacente generaría los anillos o cadenas montañosas casi circulares de las cuencas de impacto.
 
La relación entre la energía cinética $E_{c}$ del cuerpo  que choca con la Tierra y el diámetro $D$ del cráter que produce es del tipo $E_{c}=k\,\, D^{n}$, donde $k$ y $n$ son constantes. Los valores estimados para $k$ y $n$ yacen dentro de un amplio rango que van de $10^ {22}$ a $10^ {24}$ para $k$ y de $2.6$ a $3.9$ para $n$. Note que $n \approx 3$, indicando que $D^{n}$ es aproximadamente el volumen de la cavidad y $k$ es la energía por unidad de volumen al final del proceso de excavación. Nosotros utilizaremos la siguiente fórmula, debida a Melosh H.J. \index{Melosh H.J.},  
\begin{equation}
E_{c}(erg)=8.45\,\, 10^{22}\, D^{3.89} 
\label{EcDi}
\end{equation}
donde  $D$ es expresado en  km y  la energía $E_{c}$ en ergios. La ecuación (\ref{EcDi}) nos permite calcular $E_{c}$ dado $D$. A su vez $E_{c}=\frac{1}{2} m_{i} v_{i}^2$, donde  $m_{i}$ y $v_{i}$ son la masa y la velocidad del proyectil (asteroide o cometa). Si escribimos $m_{i}$  en términos de la densidad $\rho_{i}$ y el diámetro $d_{i}$ del cuerpo, considerado esférico, $E_{c}=\frac{1}{2} \frac{4}{3} \pi (\frac{d_{i}}{2})^{3 } \rho_{i}  v_{i}^2$. Despejando $d_{i}$ de la última  ecuación, obtenemos $d_{i}= (\frac{12 E_{c}}{\pi \rho_{i} v_{i}^2})^\frac{1}{3}$. Si reemplazamos en esta última ecuación  la expresión de $E_{c}$ dada por (\ref{EcDi}),  resulta
 
\begin{equation}
 d_{i} (km)= 0.32 (\frac{D^{3.89}}{\rho_{i} v_{i}^2})^{\frac{1}{3}},
\label{Dd}
\end{equation} 
donde $D$, $\rho_{i}$ y $v_{i}$ están expresados en km, gr cm$^{-3}$ y  en km s$^{-1}$, respectivamente. Mediante la ecuación (\ref{Dd}), con $v_{i}=20$ km s$^{-1}$ y $\rho_{i}=3.65$ gr cm$^{-3}$ para los asteroides y $\rho_{i}=1.0$ gr cm$^{-3}$ para los cometas, determinamos que los grandes cráteres terrestres, listados en la Tabla \ref{TablaCrateres}, fueron originados por asteroides entre 10 y 30 km de diámetro o por cometas entre 20 y 40 km de diámetro. La formación de cuencas requiere naturalmente el impacto de cuerpos mucho mayores. Según  la relación (\ref{Dd}), para formar una cuenca de 500 km de diámetro se necesita el impacto de  un asteroide de 90 km o de un cometa de 140 km de diámetro.

\subsection{Cráteres de impacto en la Argentina}

\begin{quote}
\it{En la mayoría de los impactos de meteoritos es imposible recuperar las masas intactas de los meteoritos, por lo que no podemos saber la masa total de impacto, ni el ángulo y ni otros detalles. Esto hace que el sitio de Campo del Cielo \index{Campo del Cielo} sea único.} \rm
 
William  Cassidy (1928-2020)\index{Cassidy W. A.} \footnote{Geólogo Norteamericano que organizó, junto a científicos Argentinos,   varias expediciones al Campo del Cielo.}  
\end{quote}

\begin{quote}

\it{ ...se encuentran tierras cocidas y escorias en todos los horizontes desde el de Monte Hermoso hasta los de las capas más recientes de la Formación pampeana...
La presencia de fragmentos de escorias no rodadas es un
hecho muy importante y constituye un argumento muy
serio contra la teoría del origen volcánico de ese material...}\rm

Florentino Ameghino (1854-1911)\index{Ameghino F.}

\end{quote}

Alrededor de  4000 años atrás, un pueblo originario que habitaba una región entre las provincias Argentinas de Chaco \index{Chaco} y Santiago del Estero vio un suceso extraordinario cuyo recuerdo legaron a futuras generaciones con el nombre ``Pinguen Nonraltá'',  que en lenguaje Toba  significa ``Campo del Cielo''.\index{Campo del Cielo}  Este nombre refiere al lugar donde   una gran lluvia de meteoritos cayó sobre  un área de 70 km de largo por 30 km de ancho. Se han encontrado en esa gran extensión de terreno cerca de 30 cráteres, donde el más grande tiene 80 metros de diámetro y 14 metros de profundidad,  y una gran cantidad de meteoritos metálicos, que van desde pequeñas piedras hasta  enormes cuerpos que pesan  decenas de toneladas. 

La historia de los grandes meteoritos metálicos del Campo del Cielo \index{Campo del Cielo} es muy interesante. En el año 1576, los conquistadores españoles enviaron una expedición que encontró un bloque metálico de 15 toneladas al que llamaron ``el Mesón de Fierro'' \index{Mesón de Fierro (meteorito)}. Los Europeos de ese tiempo no creían  que tal enorme cuerpo pudiera  caer del cielo y  solo se interesaron  en él  por la posibilidad de que contuviera plata. El meteorito fue perdido y aún no se lo ha podido localizar. Desde el año 1800 al presente, se descubrieron 11 grandes meteoritos de más de 1 tonelada. En el año 1980, dentro de un cráter de 25 metros de diámetro, se desenterró el meteorito llamado ``El Chaco''\index{El Chaco (meteorito)} de $28.84$ toneladas. El último gran hallazgo  fue realizado  en el año 2016 por la Asociación Chaqueña de Astronomía\index{Asociación Chaqueña de Astronomía} que desenterró un gran meteorito metálico de 30 toneladas, llamado ``Gancedo'',\index{Gancedo}  el cual pasó a ser el segundo meteorito más grande del mundo. 

Gran parte de la región del Chaco\index{Chaco} Argentino, incluyendo el Campo del Cielo, está cubierta por un suelo blando de depósitos sedimentarios  de areniscas y arcillas, llamado loess\index{loess}. Este tipo de suelo fue estudiado por primera vez en el valle del Rin\index{Rhin (rio)}, de allí el término  Alemán ``löss'' que significa suelto o poco compacto. Entonces, el caso del Campo del Cielo se trata en esencia de proyectiles duros (meteoritos metálicos) que impactaron sobre un blanco blando (loess\index{loess}). Ello y el hecho de que el ángulo de impacto con respecto al suelo fue pequeño explica por qué  la mayoría de los cráteres son del tipo cono oblicuo de penetración. También se han encontrado algunos cráteres, con numerosos fragmentos meteoríticos  dispersos alrededor de ellos, que podrían caracterizarse como de explosión y de rebote. La edad aceptada para la caída de los meteoritos del Campo del Cielo es de $3945 \pm 85$ años. Dicha edad se obtuvo mediante el método de carbono 14 aplicado a  restos carbonizados recogidos en uno de los cráteres. 

El ingeniero Mario  Vesconi\index{Vesconi M. A.}, miembro  de la Asociación Chaqueña de Astronomía y estudioso del Campo del Cielo, estima que el cuerpo madre de los meteoritos del Campo del Cielo fue un asteroide del orden de 600 toneladas. Al ingresar a la atmósfera, el asteroide se fracturó  y la mitad de su masa se evaporó. Vesconi concluye que ``de  esas 300 toneladas que impactaron, hemos recuperado un tercio. Hay más material enterrado y muchos elementos dispersos''. 

La mayoría de los cráteres y de los meteoritos más grandes se encuentran sobre una sorprendentemente  estrecha franja de $\approx 20$ km de largo por 2 km de ancho. Tal alargada superficie (de tipo elipsoidal) se orienta en la dirección noreste-suroeste, e indicaría la dirección de la cual provenían los meteoritos. Quizá, los impactos que trazaron tan angosta línea correspondan a los fragmentos del núcleo metálico del asteroide, mientras que las partes menos densas y más fragmentadas del asteroide se dispersaron sobre un área mucho mayor. Si después de la fractura del asteroide, su parte más densa se encontraba dentro de una esfera de diámetro $D$, las trayectorias de  sus fragmentos estarían  en su mayoría inscriptos  dentro de un cilindro de diámetro $D$. Al ángulo $\alpha$ entre el eje del cilindro, o la velocidad media de los meteoritos, y el suelo o horizonte del lugar, se lo llama generalmente  ángulo de impacto. Si el choque de los meteoritos hubiera sido perpendicular al suelo, es decir $\alpha=90^{\circ}$, los cráteres  hubiesen estado dentro de  un círculo de diámetro $D$.  Sin embargo, los impactos del Campo del Cielo se distribuyen sobre una superficie que puede  representarse por una elipse con su eje menor de $\approx 2$ km y su eje mayor de $\approx 20$ km. Ello significa que el ángulo de impacto fue pequeño. En efecto, la intersección de una superficie cilíndrica de diámetro $D$ con un plano inclinado que forma un ángulo $\alpha$ con el eje del cilindro genera una elipse de eje menor igual a $D$, y eje mayor igual a $\frac{D}{sen \alpha}$. En nuestro caso,  $\frac{D}{sen \alpha}=20$ km y $D=2$ km, con lo cual obtenemos que $\alpha \approx 7^{\circ}$. Este resultado es compatible con los ángulos de penetración medidos en algunos cráteres.

Otro notable conjunto de cráteres de impacto se encuentra en la localidad de Río Cuarto, \index{Rio Cuarto} Provincia Argentina de Córdoba. \index{Córdoba} En el año 1990, la vista aérea de una cadena de depresiones alargadas del terreno, extrañamente alineadas,  llamó la atención de  Rubén Lianza,  \index{Lianza R.}  piloto de la fuerza Aérea Argentina y astrónomo aficionado. Dichas  estructuras de impacto resaltan al encontrarse rodeadas por tierras cultivadas en un suelo  plano de sedimentos eólicos (loess\index{loess}) acumulados durante el Pleistoceno\index{Pleistoceno}  y Holoceno\index{Holoceno} hasta profundidades de 25 metros, característico de la Pampa Argentina\index{Pampa Argentina}.

 Posteriores estudios sobre el terreno mostraron la presencia de al menos 10 cráteres elongados de meteoritos alineados en paralelo en la dirección noreste-suroeste. La estructura mayor ($4.5\times 1.1$ km) se ubica en el extremo noreste (cráter A). A 11 km del cráter A hacia el suroeste, se encuentran dos cráteres gemelos y adyacentes entre sí, levemente menores $(3.5\times 0.7)$ km que el cráter A. Los cráteres más pequeños tienen longitudes del eje mayor entre $0.1$ a $0.3$ km.  En general los bordes de los cráteres están entre 3 y 7 metros sobre la planicie. Los pisos de los cráteres mayores yacen a $\approx 10$  metros por debajo de los bordes.  Los 10 cráteres se distribuyen sobre  un angosto corredor de 30 km de longitud.

Los materiales recogidos en los cráteres, tales como  fragmentos de meteorito rocoso (condritos) y de  vidrios de impacto,  indican  el origen meteorítico de los cráteres de Río Cuarto. A partir de la morfología de las estructuras, se piensa que estos cráteres  son el resultado del impacto rasante de un asteroide rocoso de 150-300 metros de diámetro, que fue revotando y fragmentándose  progresivamente desde el primer contacto con el suelo. En base a la relación isotópica  del argón \index{argón}  $^{40} Ar/ ^{39} Ar$ medida en el material vítreo  encontrado en el lugar,  el profesor estadounidense Peter Schultz \index{Schultz P.} y colaboradores  estiman que el evento ocurrió  hace menos de 10000 años.

Otra notable serie de cráteres de posible origen meteorítico se encuentra en un sitio de la Patagonia \index{Patagonia} central argentina, conocido como ``Bajada del Diablo'',\index{Bajada del Diablo}  en el norte de la provincia  de Chubut.\index{Chubut}Dentro de un círculo de 23 km de diámetro, hay más de cien cráteres de entre 60 y 360 metros de diámetro,  y entre 30 y 50 metros de profundidad. El geólogo argentino  Hugo Corbella \index{Corbella H.} fue quien primero notó, en el año 1987,  que tales singulares  estructuras circulares tienen características morfológicas y geológicas propias de estructuras de impacto. El gran área de dispersión de los cráteres se explica por el choque de una extensa nube de cuerpos, resultante de un  agregado cometario poco compacto o un asteroide rocoso que se fragmentó al penetrar la atmósfera. El suelo  sobre el cual se cavaron los cráteres de Bajada del Diablo es muy diferente a aquellos del Campo del Cielo y de Río Cuarto. En efecto, esta parte de la Patagonia es un lugar árido  y rocoso, con mesetas cubiertas por mantos de basalto o por  fragmentos de  roca redondeadas por la erosión. Un cuerpo que choca a gran velocidad contra una superficie dura probablemente se destruya completamente y por lo tanto es difícil encontrar vestigios del mismo en el cráter y su entorno. Con el fin de corroborar más allá de toda duda razonable la hipótesis del impacto y la naturaleza del cuerpo cósmico,   se debe investigar  en los cráteres de Bajada del Diablo la presencia de  residuos del cuerpo cósmico y  material  vítreo formado  en las rocas de los cráteres comprimidas por la onda de choque \cite{Acevedo2}  \footnote{ \cite{Acevedo2} es un interesante artículo de divulgación escrito por científicos  que exploraron el campo de cráteres de Bajada del Diablo}.
  
 La secuencia normal en el descubrimiento de las estructuras de impacto es primero descubrir el cráter y luego identificar los materiales eyectados por el impacto. Raro, pero posible, es el camino inverso. Este fue el caso del famoso cráter de Chicxulub. \index{Chicxulub (cráter)}  Primero se descubrió la capa  de iridio de distribución planetaria en el límite Cretácico-Terciario (K-T)\index{Cretácico-Terciario} y 10 años después se descubrió un cráter (Chicxulub\index{Chicxulub (cráter)}) que por sus dimensiones y edad se correspondía con el impacto de un cuerpo cósmico  capaz de producir semejante eyección de material sobre toda la Tierra. Salvando las distancias, un caso similar se presenta en la capa de \it{loess\index{loess}} \rm que cubre las Pampas Argentinas (Prov. de Buenos Aires),  donde en distintos extractos del suelo se encuentran restos de impactos cósmicos pero aún no se han hallado los correspondientes cráteres.

En los impactos muy energéticos, parte del material afectado se funde y sus salpicaduras  se esparcen en los alrededores y pueden llegar a  sitios muy alejados del cráter de impacto. Dichas gotas de roca fundida se solidifican tomando formas y coloraciones características, por las cuales reciben  nombres tales como  tectitas,  cristitas, impactitas, vidrios de impacto, o genéricamente material vítreo. Un equipo internacional de científicos \footnote{Christian Koeberl \index{Koeberl C.} de la Universidad de Viena, Austria \index{Austria}; Peter Schultz \index{Schultz P.} la Universidad
Brown, Estados Unidos; Marcelo Zárate \index{Zárate M.} del CONICET, Argentina; y otros.}, que inició sus trabajos en el año 1995,  descubrió  materiales vítreos de impacto en los estratos de loess\index{loess} de los acantilados de la costa atlántica bonaerense entre Mar del Plata \index{Mar del Plata} y Necochea, \index{Necochea} y en las cercanías de Bahía Blanca \index{Bahía Blanca}.  Esos materiales originados por impactos yacen  en al menos 7 diferentes capas y sus dataciones por medio de la relación isotópica del argón\index{argón} $^{40} Ar/ ^{39} Ar$ arrojan las siguientes edades:  Holoceno\index{Holoceno} ($6 \pm  2$  Ka), Pleistoceno\index{Pleistoceno} ($114 \pm  26$  ka, $230 \pm 0$  Ka, y  $445 \pm 21$ Ka), Plioceno ($3.27 \pm 0.08$  Ma), and  Mioceno tardío ($ 5.33 \pm  0.05$ y $ 9.23 \pm 0.09$ Ma)\footnote{La abreviatura Ma (del latín: Mega annum) significa un millón de años, y Ka 1000 años}. Entonces, esos materiales provienen de 7 impactos. Sin embargo, en la Prov. de Buenos Aires, solo se han encontrado dos posibles cráteres de impacto: uno de 3 km de diámetro y de 445 Ka de edad ubicado en la localidad de La Dulce\index{La Dulce (cráter)} (partido de Necochea) y otro de 12 km de diámetro y $1.2$ Ma ubicado en las cercanías  de la localidad de  General San Martín. A juzgar por la coincidencia en edad, los materiales vítreos del Pleistoceno\index{Pleistoceno} temprano (445 Ka) desperdigados sobre la costa atlántica habrían sido eyectados por el impacto que originó el cráter de La Dulce\index{La Dulce (cráter)}.

Cabe destacar que el interés en la búsqueda de vestigios de impactos cósmicos en las Pampas Argentinas nació por la existencia de enigmáticas estructuras mineralógicas, en la capa de \it{loess\index{loess}} \rm que cubre la Prov. de Buenos Aires. Esas  enigmáticas estructuras de 1 a 10 cm de tamaño, llamadas localmente ``escorias'' y ``tierras cocidas'', ya eran conocidas  en el siglo XIX y sus posibles orígenes discutidos entre otros por el gran naturalista argentino Florentino Ameghino. \index{Ameghino F.}  Las escorias son fragmentos de aspecto vítreo y las tierras cocidas tienen la apariencia de fragmentos de ladrillo o cerámico, de allí su nombre.  Florentino Ameghino interpretaba, en el marco de su teoría del origen americano del Hombre, que las  escorias y tierras cocidas eran el producto de fogones y de incendios intencionales de pastizales provocados por pobladores prehistóricos. Como hemos visto en el parágrafo anterior, los estudios  petrológicos y geoquímicos modernos  indican que  esos materiales se formaron a partir del \it{loess\index{loess}} \rm sometido  a muy  altas temperaturas y presiones, como consecuencia de impactos cósmicos.

En la capa de loess\index{loess} pampeana se conservan vestigios de notables eventos  ocurridos durante los últimos 10 millones años. Allí se encuentran sepultados no solamente  esquirlas de impactos explosivos, sino también  fósiles de la mega fauna pampeana extinta durante el Pleistoceno\index{Pleistoceno}. Ello podría inducirnos a pensar que uno o más  impactos cósmicos suficientemente energéticos fueron los responsables de la misteriosa extinción de la megafauna\index{megafauna} local. Hemos visto que se registraron varios impactos y algunos de ellos debieron ocurrir  en una región cercana a la de los residuos hallados,  a juzgar por la gran dispersión de los mismos. Sin embargo, a excepción del cráter de  La Dulce\index{La Dulce (cráter)}, aún no se han identificado las fuentes o cráteres que originaron tan profusa dispersión de residuos de impacto. Al no tener identificados los cráteres, no podemos estimar la magnitud del daño que dichas explosiones ocasionaron sobre la flora y fauna de la región pampeana. Por ejemplo, por medio de la fórmula (\ref{EcDi}), obtenemos que la energía $E$ de la explosión que formó La Dulce\index{La Dulce (cráter)}, con un diámetro D=3 km,  es $\approx 6\times 10^{24}$ ergios (150 Megatones). Utilizando la siguiente fórmula 
\begin{equation}
 r_{d}(km)=5 E(Mt)^{1/3}
\label{radioDevastacion}
\end{equation} \footnote{Esta fórmula (\ref{radioDevastacion}) se obtuvo de la fórmula (2) de la referencia bibliográfica \cite{Toon}, con $h=0$ dado que la explosión ocurrió al tocar suelo.}, encontramos   un  radio de devastación  $r_{d} \approx 30$ km, es decir que la alta presión de la onda de choque destruyó la vida   dentro de un radio  de 30 km en torno al punto de la explosión. Ello parece insuficiente para dar cuenta de la extinción de la megafauna\index{megafauna}.

La extinción de la megafauna\index{megafauna} fue  probablemente un proceso multicausal  que se desarrolló a lo largo del Pleistoceno\index{Pleistoceno} y culminó al comienzo del Holoceno\index{Holoceno} ($\approx 9000 $ años atrás). Las dificultades adaptativas de estos animales de gran tamaño y  baja tasa de reproducción, ante  cambios climáticos bruscos, impactos cósmicos catastróficos \footnote{Hay indicios de que una  gran explosión en el aire de un fragmento de cometa ocurrió hace $\approx$ 13000 años sobre Norteamérica, y se especula que este hecho pudo iniciar el enfriamiento climático global de finales del Pleistoceno\index{Pleistoceno}, conocido como ``Dryas Reciente''\index{Dryas Reciente}(ver sección 8).} o nuevas enfermedades, llevaron a una progresiva disminución del número de ejemplares de las especies.   Como el ingreso del Hombre a América ocurrió alrededor de 30000 años atrás, es probable que la megafauna\index{megafauna}  coexistiera con el Hombre por algunos miles de años hasta su extinción total. En efecto, el mismo Florencio Ameghino  \footnote{La ubicación en el mismo substrato geológico de fósiles humanos y de animales extintos llevó a Ameghino a la conclusión de que los ancestros del Hombre  eran muy antiguos y probablemente originarios de América. La teoría de Ameghino \index{Ameghino F.} despertó interés porque se oponía a la teoría de que el Hombre ingresó a América a través del estrecho de Bering en la última glaciación, alrededor de 30000 años atrás, teoría que a la postre se impuso.  Cabe recordar que en la época de Ameghino no se disponía de la datación por el método del carbono 14. Ahora sabemos que los fósiles hallados por Ameghino tienen sólo $\approx$ 9000 años y corresponden al Hombre actual (Homo Sapiens).} y más recientemente el arqueólogo argentino Gustavo  Politis \index{Politis G.} encontraron evidencias de que  especímenes  de la megafauna\index{megafauna} eran cazados por el Hombre para su explotación. Entonces, la aparición del Hombre en la región Pampeana habría significado el golpe de gracia para la megafauna\index{megafauna} local.

\section{El escudo magnético de la Tierra y su interacción con el polvo interplanetario.}

La Tierra es en sí misma un inmenso imán que mantiene un extenso campo magnético que rodea a la Tierra llamado magnetósfera, con sus polos magnéticos situados cerca de los polos geográficos norte y sur. Las fuerzas eléctricas y magnéticas fueron tratadas independientemente hasta que el físico y químico danés Hans Christian Oersted \index{Oersted H. C.} descubrió en 1820  que existe un vínculo entre ambas fuerzas. Oersted encontró que en torno a un cable por el cual circula una corriente eléctrica se establece un campo magnético. El  campo magnético terrestre  es creado por las corrientes eléctricas que se producen en el núcleo de la Tierra que está compuesto de hierro fundido de muy alta conducción eléctrica.

La magnetósfera protege a la Tierra de las cargas eléctricas altamente energéticas que provienen  del Sol con  el viento solar  y que son dañinas para la vida de la Tierra. Sin la protección del campo magnético, el viento solar puede arrastrar gases de la atmósfera superior y llevar a un paulatino adelgazamiento de nuestra atmósfera \footnote{El planeta Marte perdió muy tempranamente su campo magnético y se cree que ello fue una de las causas que originaron un gran cambio climático,  que llevaron de tener agua fluida sobre su superficie al planeta desértico actual.}. Las partículas cargadas son desviadas por la \it{fuerza de Lorentz}, \index{Lorentz} \rm evitando que choquen directamente  contra la atmósfera y la superficie de la Tierra. La fuerza de Lorentz es dada por la igualdad
 \begin{equation}
 \overrightarrow{F}=q\, (\overrightarrow{v} \times \overrightarrow{B}),
 \label{FuerzaLorentz}
 \end{equation}
 donde $q$ es la carga eléctrica de la partícula, $\overrightarrow{v}$ el vector velocidad de la carga y $\overrightarrow{B}$ el vector intensidad del campo magnético. El producto vectorial de dos vectores, indicado por el símbolo $\times$,  es un nuevo vector que es perpendicular a los dos vectores del producto. Por lo tanto, la ecuación (\ref{FuerzaLorentz}) indica que $\overrightarrow{F}$ es perpendicular a ambos $\overrightarrow{v}$ y $\overrightarrow{B}$. Por definición del producto vectorial, el módulo de $\overrightarrow{F}$ es igual al producto del módulo $\overrightarrow{v}$ por el módulo $\overrightarrow{B}$ y por el seno del ángulo $\theta$ entre $\overrightarrow{v}$ y $\overrightarrow{B}$. Es decir, $F= v\, B \,sen\, \theta$. Si descomponemos el vector de la velocidad inicial en una componente paralela  $\overrightarrow{v_{\parallel}}$ y la otra perpendicular $\overrightarrow{v_{\perp}}$ al  campo magnético, $\overrightarrow{v}= \overrightarrow{v_{\parallel}}+\overrightarrow{v_{\perp}}$ y la ecuación (\ref{FuerzaLorentz}) se puede escribir $\overrightarrow{F}=q\, ( \overrightarrow{v_{\parallel}}+\overrightarrow{v_{\perp}}) \times \overrightarrow{B}=q\, ( \overrightarrow{v_{\parallel}} \times \overrightarrow{B} ) + q\, ( \overrightarrow{v_{\perp}} \times \overrightarrow{B} )$. Como el ángulo $\theta$ entre $v_{\parallel}$ y  $\overrightarrow{B}$ es cero,
 $q \,( \overrightarrow{v_{\parallel}} \times \overrightarrow{B})=0$ y por lo tanto 
 \begin{equation}
 \overrightarrow{F}=q\, (\overrightarrow{v_{\perp}} \times \overrightarrow{B}).
 \label{FuerzaLorentz2}
 \end{equation}
 
 En un campo magnético uniforme, la fuerza $F$ desvía constantemente la componente  $\overrightarrow{v_{\perp}}$, sin variar su módulo y el valor de $F$ se determina por la igualdad $F=q\, v_{\perp} B$, dado que $\theta=90^{\circ}$. Por lo tanto, $\overrightarrow{F}$  es una fuerza centrípeta que  produce un movimiento circular uniforme en un plano perpendicular al campo. Este movimiento es descripto por la fórmula 
 \begin{equation}
 m_{p} \frac{v_{\perp}^{2}}{r_{g}}= q\, v_{\perp} B,
 \label{Larmor}
 \end{equation}
 donde $m_{p}$ es la masa de la partícula cargada y $r_{g}$ es el radio de la circunferencia que describe la partícula (también conocido como radio de giro, radio de Larmor\index{Larmor} o radio ciclotrón). Por otro lado,  la velocidad $v_{\parallel}$ no es afectada por $F$. En consecuencia, la partícula se desplaza uniformemente a lo largo de la dirección del campo magnético, mientras que en el plano perpendicular al mismo la partícula rota uniformemente. Combinando ambos movimientos, la partícula cargada describe una línea o órbita helicoidal con eje paralelo a al vector $\overrightarrow{B}$ del campo magnético uniforme.  
 
 Nuestro objetivo final es estudiar la trayectoria de las partículas cargadas que ingresan a la magnetósfera terrestre y como paso previo obtendremos una ecuación equivalente a (\ref{Larmor}), por medio de un método más general. Primero establecemos un sistema de coordenadas cartesianas $(\eta,\xi ,\zeta)$, con origen en el centro de la Tierra y con el eje $\zeta$ apuntando al polo norte geográfico. Consideremos que el vector  
 $\overrightarrow{B}$ es perpendicular al plano $\eta-\xi$, es decir el campo magnético  solo tiene una componente en $\zeta$, que denominaremos $B_{\zeta}$ y por lo tanto $\overrightarrow{B}=(0,0, B_{\zeta})$. Por otro lado, consideramos que el vector velocidad  $\overrightarrow{v}$  de la partícula es perpendicular a $\overrightarrow{B}$, y por lo tanto se encuentra en el plano $\eta-\xi$ y términos de sus componentes $\overrightarrow{v} =\overrightarrow{v_{\perp}} =(v_{\eta},v_{\xi}, 0)$.
  
 Para calcular la fuerza de Lorentz $F$ de acuerdo a (\ref{FuerzaLorentz}) y (\ref{FuerzaLorentz2}), debemos obtener el producto vectorial de  $\overrightarrow{v}$ con $\overrightarrow{B}$, el cual es dado por $\overrightarrow{v} \times \overrightarrow{B}=(v_{\eta},v_{\xi}, 0)\times (0,0, B_{\zeta})= (v_{\xi} B_{\zeta}, -v_{\eta} B_{\zeta})$. Note que el módulo del vector resultante de ese producto vectorial es  $ \mid (v_{\xi} B_{\zeta}, -v_{\eta} B_{\zeta}) \mid = \sqrt{v_{\eta}^{2}+v_{\xi}^{2}}\,\,B_{\zeta}=\mid\overrightarrow{v} \mid \,\, \mid \overrightarrow{B} \mid$, lo cual está de acuerdo con la definición de producto vectorial, si tenemos en cuenta que aquí $sen \, \theta=1$. Entonces, la fuerza de Lorentz puede escribirse para este caso:  $\overrightarrow{F}= (F_{\eta}, F_{\xi})= q\, (v_{\xi} B_{\zeta}, -v_{\eta} B_{\zeta})$. Aplicando la segunda ley de Newton, obtenemos las siguientes ecuaciones de movimiento:
 \begin{eqnarray}
 F_{\eta} & = & m_{p} \frac{d v_{\eta}}{dt} = q\, (v_{\xi} B_{\zeta}) \nonumber\\
  F_{\xi} & = & m_{p} \frac{d v_{\xi}}{dt} = q\, (-v_{\eta} B_{\zeta}) 
  \label{LorentzMov}
 \end{eqnarray}
 
 A partir de las ecuaciones (\ref{LorentzMov}) podemos obtener la posición $(\eta,\xi)$ de la partícula en el tiempo $t$. En otras palabras, $\eta$ e $\xi$ pueden expresarse como funciones de la variable independiente $t$, es decir $\eta(t)$ e $\xi(t)$.  Para ello debemos tener en cuenta que  $v_{\eta}=\frac{d\eta}{dt}$, $v_{\xi}=\frac{d\xi}{dt}$, $\frac{dv_{\eta}}{dt}=\frac{d^{2} \eta}{dt^{2}}$ y $\frac{dv_{\xi}}{dt}=\frac{d^{2} \xi}{dt^{2}}$, con cuyos reemplazos en  (\ref{LorentzMov}) obtenemos 
\begin{eqnarray}
  \frac{d^{2} \eta}{dt^{2}}-Q B_{\zeta} \frac{d\xi}{dt} & = & 0 \nonumber\\
  \frac{d^{2} \xi}{dt^{2}}+Q B_{\zeta} \frac{d\eta}{dt} & = & 0
 \label{LorentzMov2}
\end{eqnarray}
 donde $Q=\frac{q}{m_{p}}$. Este es un sistema de ecuaciones diferenciales \it{acopladas}. \rm Sin embargo por medio de un truco matemático podemos convertir (\ref{LorentzMov2}) en dos ecuaciones  diferenciales independientes  de coeficientes constantes.\rm \,Si integramos con respecto a $t$ los términos de las dos ecuaciones (\ref{LorentzMov2}), convertimos las derivadas segundas en derivadas primeras y las derivadas primeras en solo la variable dependiente. En efecto,
 
 \begin{equation}
   \frac{d\eta}{dt}-Q B_{\zeta} y + A_{1} =  0
    \label{LorentzMov3}
 \end{equation}
 \begin{equation}
 \frac{d\xi}{dt}+Q B_{\zeta} x + A_{2} =  0,
 \label{LorentzMov4}
 \end{equation}
donde $A_{1}$ y $A_{2}$  son constantes de integración. Las condiciones iniciales en $t=0$, $\frac{d\eta}{dt}=v_{\eta}(0)$, $\frac{d\xi}{dt}=v_{\xi}(0)$, $\eta=\eta(0)$ y $\xi=\xi(0)$, deben satisfacer  (\ref{LorentzMov3}) y (\ref{LorentzMov4}), lo cual nos permite obtener el valor de $A_{1}$ y de $A_{2}$. Como nuestro propósito es estudiar trayectorias de  partículas que se acercan a la Tierra, adoptaremos de aquí en adelante  $v_{\xi}(0)=0$. Entonces,
\begin{eqnarray}
A_{1} & = &  -v_{\eta}(0)+ Q B_{\zeta} \xi(0) \nonumber \\
A_{2} & = & -Q B_{\zeta} \eta(0)
\label{A1A2}
\end{eqnarray}

Despejando $\xi$ de  (\ref{LorentzMov3}), obtenemos que $\xi= \frac{1}{Q B_{\zeta}}\frac{d\eta}{dt} + \frac{A_{1}}{Q B_{\zeta}}$, y reemplazando esta expresión para $\xi$ en (\ref{LorentzMov4}) obtenemos
 \begin{equation}
   \frac{d^{2} \eta}{dt^{2}}+ (Q B_{\zeta})^{2}\, \eta + (Q B_{\zeta})\,  A_{2}=0
   \label{EDx}
  \end{equation}
  En forma similar, despejamos $\eta$ de (\ref{LorentzMov4}) y reemplazamos su expresión en (\ref{LorentzMov3}), con lo cual
  
  \begin{equation}
   \frac{d^{2} \xi}{dt^{2}}+ (Q B_{\zeta})^{2}\, \xi - (Q B_{\zeta})\,  A_{1}=0
   \label{EDy}
  \end{equation}
 La ecuación (\ref{EDx})  es una ecuación diferencial ordinaria no-homogénea de coeficientes contantes, cuya solución general $\eta_{g}$ puede escribirse como $\eta_{g}=\eta_{h}+\eta_{p}$, donde   $\eta_{h}$ es la solución de  la ecuación homogénea $\frac{d^{2} \eta}{dt^{2}}+ (Q B_{\zeta})^{2}\, \eta =0$  y $\eta_{p}$ es una solución particular. Dado que la solución general $\eta_{h}+\eta_{p}$ debe cumplir la ecuación (\ref{EDx}), encontramos que  $\eta_{p}= -\frac{A_{2}}{Q B_{\zeta}}$. La ecuación diferencial homogénea está asociada con el movimiento oscilatorio armónico y su solución es $\eta_{h}= C_{1}\, cos (Q B_{\zeta} t) + C_{2}\, sen (Q B_{\zeta} t)$, donde $C_{1}$ y $C_{2}$ son constantes a determinar con las condiciones iniciales. Por lo tanto,
 \begin{equation}
 \eta= C_{1}\, cos (Q B_{\zeta} t) + C_{2}\, sen (Q B_{\zeta} t) - \frac{A_{2}}{Q B_{\zeta}}.
 \label{SolX}
 \end{equation}
En forma similar, obtenemos la solución de (\ref{EDy})
\begin{equation}
 \xi= C_{3}\, cos (Q B_{\zeta} t) + C_{4}\, sen (Q B_{\zeta} t)+  \frac{A_{1}}{Q B_{\zeta}}.
 \label{SolY}
 \end{equation}
donde $C_{3}$ y $C_{4}$ son constantes a determinar con las condiciones iniciales.

En $t=0$, la solución (\ref{SolX}) debe cumplir con las condiciones $\eta=\eta(0)$ y $\frac{d\eta}{dt}=v_{\eta}(0)$, con lo cual obtenemos dos ecuaciones lineales  que nos permiten determinar sus  dos incógnitas $C_{1}$ y $C_{2}$ y si además tenemos en cuenta los valores  $A_{1}$ y $A_{2}$ dados por (\ref{A1A2}), la ecuación (\ref{SolX}) resulta,
\begin{equation}
\eta=\eta(0)+ \frac{v_{\eta}(0)}{Q B_{\zeta}} sen(Q B_{\zeta} t)
\label{SolXf}
\end{equation}
Procedimiento de la misma manera con la ecuación (\ref{SolY}), determinamos $C_{3}$ y $C_{4}$ y con ello obtenemos la siguiente solución para $\xi$:
\begin{equation}
\xi=\xi(0)- \frac{v_{\eta}(0)}{Q B_{\zeta}}+ \frac{v_{\eta}(0)}{Q B_{\zeta}} cos (Q B_{\zeta} t)
\label{SolYf}
\end{equation}

Despejando $r_{g}$ de la ecuación (\ref{Larmor}) y recordando que  $Q=\frac{q}{mp}$ y que en nuestro caso $v_{\perp}=v_{\eta}(0)$ y $B=B_{\zeta}$,
obtenemos $r_{g}=\frac{v_{\eta}(0)}{Q B_{\zeta}}$. Por lo tanto, las ecuaciones (\ref{SolXf}) y (\ref{SolYf}) pueden escribirse $\eta-\eta(0)=r_{g} sen (Q B_{\zeta} t)$ y $\xi-(\xi(0)-r_{g})=r_{g} cos (Q B_{\zeta} t)$. Si elevamos al cuadrado los miembros de ambas ecuaciones, los sumamos miembro a miembro y tenemos en cuenta que $sen^{2}(Q B_{\zeta} t)+cos^{2} (Q B_{\zeta} t)=1$, obtenemos
\begin{equation}
(\eta-\eta(0))^{2}+ (\xi-(\xi(0)-r_{g}))^{2}=r_{g}^{2}
\label{Circ}
\end{equation}
 Esta ecuación (\ref{Circ}) muestra que  la órbita de la partícula es una circunferencia  de centro $(\eta(0),\xi(0)-r_{g})$ y radio $r_{g}$, lo cual naturalmente coincide con el movimiento que describe la fórmula (\ref{Larmor}).
 
 La ecuación (\ref{Circ}) es estrictamente válida solo en un campo magnético uniforme, sin embargo  esta es muy útil para  aprender sobre  el movimiento aproximado  de diferentes partículas cargadas  que ingresan a la magnetosfera terrestre. Primero estudiaremos los efectos del campo magnético de la Tierra sobre  electrones y protones aislados que en general  provienen del Sol y luego sobre granos de polvo sub-micrométricos asociados con desprendimientos de cometas y corrientes de meteoros que chocan contra la Tierra.
 
  El campo magnético terrestre  puede ser aproximado por un dipolo magnético, similar a un imán de barra, cuya intensidad es proporcionada  en coordenadas polares por las siguientes fórmulas:
\begin{eqnarray}
   B_{r} & = & -2 \frac{k_{0}}{r^{3}}\, sin\, \lambda \nonumber \\
   B_{\lambda} & = &  +\, \, \frac{k_{0}}{r^{3}}\, cos \, \lambda, 
   \label{DipoloM}
\end{eqnarray}
donde $k_{0}=8\times 10^{15}$ en unidades de  Tesla por $m^{-3}$, $r$ es la distancia desde el centro de la Tierra y $\lambda$ es la latitud con respecto al ecuador magnético. La diferencia entre la latitud magnética y geográfica es aproximadamente de $11^{\circ}$. La componente $B_{r}$ tiene la dirección del radio vector $r$ y la componente $B _{\lambda}$  es perpendicular a $r$. El módulo de $\overrightarrow{B}=(B_{r}, B _{\lambda})$ es $\mid \overrightarrow{B} \mid= \sqrt{B_{r}^{2}+B _{\lambda}^{2}}=\frac{k_{0}}{r^{3}} \sqrt{1 +\,3 \, sen^{2} \lambda }$, lo cual muestra que la intensidad del campo magnético aumenta hacia los polos.

A fin de aplicar (\ref{Circ}), supondremos que las partículas cargadas se mueven en el plano $\eta-\xi$ que fue definido coincidente  con el plano ecuatorial de la Tierra. Dado que  el plano $\eta-\xi$ no difiere mucho del plano ecuatorial magnético, consideraremos que $\lambda=0$. De las fórmulas (\ref{DipoloM}), tenemos  que $B_{r}=0$ y $B_{\lambda}=\frac{k_{0}}{r^{3}}$.  Por lo tanto, $B=B_{\zeta}=\frac{k_{0}}{r^{3}}$. Como un ejemplo ilustrativo, consideraremos que diferentes partículas  cargadas comienzan sus trayectorias con las mismas condiciones iniciales, a saber: $r=\sqrt{\eta(0)^{2}+\xi(0)^{2}}=3 R_{T}$, donde $R_{T}$ es el radio de la Tierra, y $v=v_{\eta}(0)=-30 $ km s$^{-1}$. El signo negativo de $v$ significa que las partículas se acercan a la Tierra. Además, todas esas partículas están afectadas por el mismo campo magnético, $B_{\zeta}=\frac{k_{0}}{(3 R_{T})^{3}}$, y el único factor que hace diferentes sus trayectorias es $Q$, la relación entre la carga y la masa de la partícula. Para un electrón $Q=-1.76\, 10^{11}$ C kg$^{-1}$, y en consecuencia su radio de giro $r_{g}=\frac{v_{\eta}(0)}{Q B_{\zeta}}=0.15$m y para un protón $Q=9.58\, 10^{7}$ C kg$^{-1}$  y $r_{g}=273$ m. Esto muestra que los electrones y protones permanecen atados al campo magnético, girando en su entorno con relativamente pequeños radios de giro, tal como lo establece la ecuación (\ref{Circ}). Como las componentes de velocidad paralelas  al campo magnético no son afectadas por la fuerza de Lorentz, los electrones y protones se mueven en espiral a largo de las lineas del campo magnético  y pueden penetrar a la atmósfera a través de los polos magnéticos, produciendo las auroras boreales y australes. Sin embargo, las líneas magnéticas convergen hacia los polos aumentando  la intensidad magnética, con lo cual se produce un efecto de espejo que hace retornar a muchas partículas cargadas hacia el ecuador magnético. De modo que la magnetosfera se comporta como una botella magnética que confina las partículas cargadas que se han incorporado desde el exterior. Esas concentraciones  de electrones y protones capturados por el campo magnético que rodea a la Tierra se las conoce como anillos de Van Allen.\index{Van Allen} Además del movimiento en espiral en la dirección de los polos, las partículas cargadas tienen un movimiento de deriva hacia el este o el oeste, dependiendo del signo de la carga eléctrica. 

En general, una partícula de polvo contiene más electrones que protones y por lo tanto la carga eléctrica de un grano de polvo es su exceso de electrones $\mathcal{E} e$, número de electrones menos el nro. de protones, por la carga del electrón. La carga de un grano de polvo depende de muchos factores, por ejemplo  del plasma de partículas eléctricas en el que el grano se encuentra sumergido, de la características  químicas y físicas del grano y de la velocidad relativa  con respecto al medio en que se mueve, entre otras cosas.  Debido a que las cargas del mismo signo se repelen, un exceso de cargas de un signo genera una presión interna en el grano de polvo que puede hacerlo explotar. Por ello, $\mathcal{E} e$ tiene un límite. Nosotros adoptaremos, como caso típico, $\mathcal{E} e$=100 \rm para un grano de polvo con un radio de $0.1 \mu$m. 

Si suponemos que la densidad de las partículas de polvo es de 1 gr cm$^{-3}$, la masa $m_{p}$ de una partícula de  0.1 $\mu$m de  radio es $m_{p}= 4.2\, 10^{-18}$ kg. Por lo tanto, $Q=\frac{\mathcal{E} e \, q_{e}}{m_{p}}=-3.81$ C kg$^{-1}$, donde $q_{e}$ es la carga del electrón. Usando las mismas condiciones iniciales con las cuales calculamos los radios de giro de los electrones y protones, obtenemos que $r_{g}=6.87\,  10^{9}$ km $\approx 1100\, R_{T}$ para los granos de polvo. A través de la ecuación (\ref{Circ}),  podemos   trazar  un tramo de la trayectoria de un grano de polvo en torno a su posición inicial $(\eta(0),\xi(0))$. Dado que  $r_{g}$ es en este caso muy grande, $(\eta-\eta(0))^{2}<< (\xi-(\xi(0)-r_{g}))^{2}$. Entonces, la ecuación (\ref{Circ}) se puede escribir $(\xi-(\xi(0)-r_{g}))= r_{g}$, con lo cual obtenemos que  $\xi=\xi(0)$; ecuación de una recta paralela al eje $\eta$ \footnote{la curvatura de un segmento de una circunferencia de radio $r_{g}$  es $\frac{1}{r_{g}}$. Si $r_{g}$ tiende a infinito, la curvatura tiende a cero, indicando que se trata de una línea recta.}. Es decir, todas las partículas con $\mid \xi(0)\mid\le R_{T}$ se acercan a la Tierra  en línea recta e impactan sobre la atmósfera. En este tratamiento no hemos incluido la fuerza de gravedad de la Tierra. Sin embargo, esas partículas se dirigen directamente al cuerpo de la Tierra  y la gravedad  solo acelera la caída de dichas  partículas, sin alterar grandemente sus trayectorias. Por otra parte, el campo magnético que actuó sobre las partículas fue considerado constante, cuando en realidad éste aumenta a medida que las partículas se acercan a la Tierra. Estas limitaciones serán subsanadas en la siguiente sección.

\begin{figure}
\includegraphics[scale=0.9]{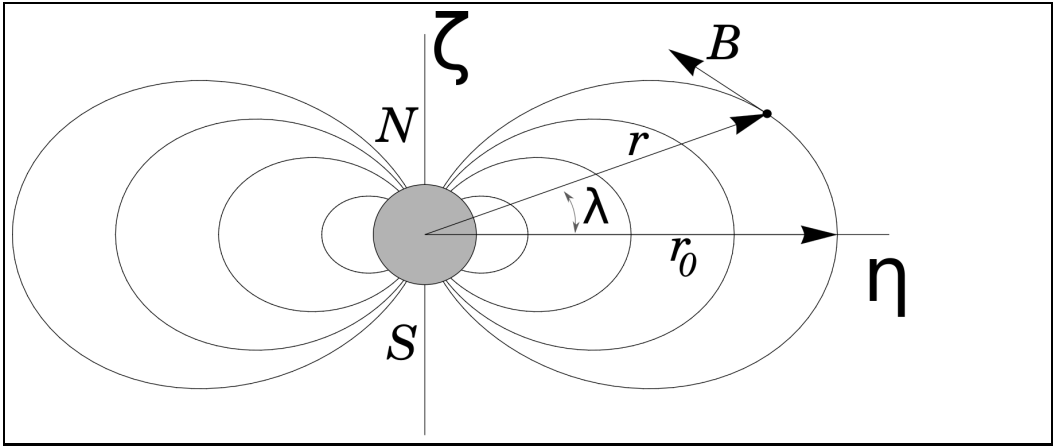} 
\caption{Modelo dipolar de las líneas del campo magnético de la Tierra}
\label{dipolo}
\end{figure}

 Ahora, vamos a estudiar la dinámica de una corriente o flujo de partículas de polvo  que pasa sobre  la Tierra. Ello puede ayudarnos a comprender algunos fenómenos vinculados al encuentro de la Tierra con las corrientes de granos de polvo asociados con desprendimientos cometarios. Calcularemos  las órbitas de las partículas en tres dimensiones y para ello  consideraremos que las partículas están sujetas al campo magnético terrestre dado por las fórmulas (\ref{DipoloM}), a la atracción gravitatoria de la Tierra y a la fuerza fricción atmosférica, cuando las partículas ingresan en ella. Referiremos las órbitas de las partículas a un sistema Cartesiano $(\eta,\xi,\zeta)$ con origen en el centro de la Tierra y el plano $\eta-\xi$ coincidente con el ecuador magnético. Para tal fin, es conveniente expresar el campo magnético  en componentes Cartesianas $B_{\eta}$, $B_{\xi}$ y $B_{\zeta}$, a partir de sus componentes polares  $ B_{r}$ y $B_{\lambda}$ dadas por las relaciones  (\ref{DipoloM}). Si bien el campo magnético es solo determinado por $r$ y $\lambda$, su ubicación  en el espacio es dado por $(r, \lambda, l)$, donde $l$ es un ángulo similar a la longitud geográfica. Las ecuaciones de transformación son
  \begin{eqnarray}
 B_{\eta} & = & B_{r} \, cos\, \lambda\, \, cos\, l - B_{\lambda}\, sen \, \lambda\, cos \, l \nonumber \\
 B_{\xi} & = & B_{r}\, cos \,\lambda \,\,sen\,  l - B_{\lambda} \, sen \, \lambda\, sen \, l  \nonumber \\
 B_{\zeta} & = & B_{r}\, sen \,\lambda+ B_{\lambda}\, cos\, \lambda, 
 \label{Bxyz}
\end{eqnarray}
donde  el ángulo $l$ es medido en sentido antihorario desde el eje $\eta$ positivo. 

En la aproximación dipolar al campo magnético terrestre,  las líneas de fuerza del campo magnético (ver fig. \ref{dipolo}) se obtienen a través de la ecuación $r=r_{0}\, cos\, \lambda$, donde $r_{0}$ es el radio de la línea de campo en el ecuador ($\lambda=0$). El valor de  $r_{0}$ y de la longitud $l$  determinan cada línea de campo. En la fig. \ref{dipolo}, representamos algunas líneas de campo que  se hallan en el plano $\eta-\zeta$, es decir con $l=0^{\circ}$, y el vector $B$ correspondiente a la posición $(r, \lambda, l)$, ó $(\eta,0,\zeta)$. Note que $\overrightarrow{B}$ es tangente a la línea de fuerza en la posición indicada. Utilizando las relaciones (\ref{Bxyz}), obtenemos  $\mid \overrightarrow{B} \mid= \sqrt{B_{\eta}^{2}+B_{\xi}^{2}+B_{\zeta}^{2}}=\sqrt{B_{r}^{2}+B_{\lambda}^{2}}$,  lo cual muestra que el módulo de un vector es independiente del sistema de coordenadas elegido.

A fin de expresar las componentes Cartesianas de $B$ en función de $(\eta,\xi,\zeta)$,
debemos hacer los siguientes reemplazos en (\ref{Bxyz}): 
\begin{eqnarray}
sen\, \lambda & = & \frac{\zeta}{r} \nonumber \\
cos \, \lambda & = & \frac{\sqrt{\eta^{2}+\xi^{2}} }{r} \nonumber \\
cos\, l & = & \frac{\eta}{\sqrt{\eta^{2}+\xi^{2}}} \nonumber \\
sen\, l & = & \frac{\xi}{\sqrt{\eta^{2}+\xi^{2}}},
\label{CoorTrans} 
\end{eqnarray}
donde $r=\sqrt{\eta^{2}+\xi^{2}+\zeta^{2}}$. Consideraremos que sobre las partículas cargadas que atraviesan la magnetosfera actúan la fuerza magnética $\overrightarrow{F_{m}}$, o de Lorentz, la fuerza gravitatoría de la Tierra $\overrightarrow{F_{g}}$ y una fuerza de fricción $\overrightarrow{F_{r}}$ debida al roce con la atmósfera. Dados   $\overrightarrow{B}=(B_{\eta}, B_{\xi}, B_{\zeta})$ y  $\overrightarrow{v}=(v_{\eta},v_{\xi}, v_{\zeta})$,  $\overrightarrow{F_{m}}=q (\overrightarrow{v} \times \overrightarrow{B})$,   con lo cual 
\begin{equation}
\overrightarrow{F_{m}}=q \, ( B_{\zeta} v_{\xi} - B_{\xi} v_{\zeta},\, - B_{\zeta} v_{\eta} + B_{\eta} v_{\zeta}, \, B_{\xi} v_{\eta} - B_{\eta} v_{\xi})
\label{CompM}
\end{equation}

La fuerza de gravedad es una fuerza central que actúa en la dirección del radio vector $r$ que une a la partícula atraída con el centro de la Tierra. El módulo de
$\overrightarrow{F_{g}}$  es $F_{g}=G \frac{m_{p} M_{T}}{r^{2}}$, donde $G$ es la constante gravitatoria, $m_{p}$ la masa de la partícula y  $M_{T}$ la masa de la Tierra. La proyección de $\overrightarrow{F_{g}}$ sobre el eje $x$ proporciona la componente $\eta$ de la fuerza, la cual resulta  igual a $ F_{g} \, cos\, \lambda\, \, cos\, l$ ,  y considerando las relaciones (\ref{CoorTrans}), igual a  $ F_{g} \frac{\eta}{r}$. Procediendo de forma similar, obtenemos que las componentes $\xi$ y $\zeta$ de $\overrightarrow{F_{g}}$ son $F_{g} \frac{\xi}{r}$ y $F_{g} \frac{\zeta}{r}$, respectivamente. Por lo tanto,
\begin{equation}
 \overrightarrow{F_{g}}= G \frac{m_{p} M_{T}}{r^{3}} (\eta,\xi,\zeta), 
 \label{CompG}
\end{equation}
donde $r=\sqrt{\eta^{2}+\xi^{2}+\zeta^{2}}$.

La expresión (\ref{Fr}) de la fuerza de roce con la atmósfera es para el caso particular de un movimiento unidimensional. La fórmula general es  $\overrightarrow{F_{r}}= -\epsilon\, \rho(h)\, A\, v \, \overrightarrow{v} $ y expresado por sus componentes
\begin{equation}
\overrightarrow{F_{r}}= -\epsilon\, \rho(h)\, A\, v\, ( v_{\eta}, v_{\xi}, v_{\zeta})
\label{CompR}
\end{equation} 
Note que $\mid \overrightarrow{F_{r}}\mid$ coincide con la fórmula \ref{Fr}. Emplearemos la ecuación (\ref{Atmosfera}) para calcular la densidad $\rho(h)$ de la atmósfera, pero aquí $h=r-R_{T}$, donde $R_{T}$ es el radio de la Tierra. En este caso, $A$  representa  el área del grano de polvo  expuesto a la fricción con la atmósfera. Adoptaremos $A=\pi a^{2}$, donde $a=0.1 \mu$m  es el radio del grano de polvo. 
 
 De acuerdo con lo supuesto arriba, la fuerza total que actúa sobre una partícula cargada es $\overrightarrow{F}=\overrightarrow{F_{m}}+\overrightarrow{F_{g}}+\overrightarrow{F_{r}}=(F_{\eta}, F_{\xi}, F_{\zeta})$ y usando las relaciones (\ref{CompM}),  (\ref{CompG}) y  (\ref{CompR}), $F_{\eta}=q (B_{\zeta} v_{\xi}-B_{\xi} v_{\zeta} ) -G \frac{m_{p} M_{T}}{r^{2}}\, \eta- \epsilon \rho(h)\, A \,v\, v_{\eta}$, $F_{\xi}=q (-B_{\zeta} v_{\eta}+B_{\eta} v_{\zeta} ) -G \frac{m_{p} M_{T}}{r^{2}}\, \xi - \epsilon \rho(h)\, A \,v\, v_{\xi}$ y  $F_{\zeta}=q (B_{\xi} v_{\eta} - B_{\eta} v_{\xi} ) -G \frac{m_{p} M_{T}}{r^{2}}\, \zeta - \epsilon \rho(h)\, A \,v\, v_{\zeta}$. Por lo tanto, tal como lo prescribe  la segunda ley de Newton,  la aceleración de la partícula es $\overrightarrow{a_{c}}=\frac{\overrightarrow{F}}{m_{p}}=(\frac{F_{\eta}}{m_{p}}, \frac{F_{\xi}}{m_{p}}, \frac{F_{\zeta}}{m_{p}})$.  Dado que $\overrightarrow{a_{c}} = ( \frac{dv_{\eta}}{dt}, \frac{dv_{\xi}}{dt}, \frac{dv_{\zeta}}{dt} )$, y a su vez $v_{\eta}=\frac{d\eta}{dt}$, $v_{\xi}=\frac{d\xi}{dt}$ y $v_{\zeta}=\frac{d\zeta}{dt}$, 
 $\overrightarrow{a_{c}} = (\frac{d^{2}\eta}{dt^{2}}, \frac{d^{2}\xi}{dt^{2}}, \frac{d^{2}\zeta}{dt^{2}})$ y por lo tanto $\frac{d^{2}\eta}{dt^{2}}= \frac{F_{\eta}}{m_{p}}$, $\frac{d^{2}\xi}{dt^{2}}=\frac{F_{\xi}}{m_{p}}$ y $\frac{d^{2}\zeta}{dt^{2}}=\frac{F_{\zeta}}{m_{p}}$. Entonces, 
 \begin{eqnarray}
\frac{d^{2}\eta}{dt^{2}} & =  & \,  Q B_{\zeta} \frac{d\xi}{dt} - Q B_{\xi} \frac{d\zeta}{dt} -G \frac{M_{T}}{r^{2}}\,\eta -\frac{\epsilon \rho(r-R_{T})}{m_{p}} \, A \,v\,\frac{d\eta}{dt} \nonumber\\
\frac{d^{2}\xi}{dt^{2}} & = &  -Q B_{\zeta} \frac{d\eta}{dt}  + Q B_{\eta} \frac{d\zeta}{dt} -G \frac{M_{T}}{r^{2}}\, \xi - \frac{\epsilon \rho(r-R_{T})}{m_{p}}\, A \,v\, \frac{d\xi}{dt} \nonumber\\
\frac{d^{2}\zeta}{dt^{2}} & =  & \, Q B_{\xi} \frac{d\eta}{dt}  - Q B_{\eta} \frac{d\xi}{dt} -G \frac{ M_{T}}{r^{2}}\, \zeta -\frac{\epsilon \rho(r - R_{T})}{m_{p}} \, A \,v\, \frac{d\zeta}{dt},
\label{EDmgr}
 \end{eqnarray}
 donde $Q=q/m_{p}$, $G$, $M_{T}$, $R_{T}$, $\epsilon$ y $A$ son constantes, $r=\sqrt{\eta^{2}+\xi^{2}+\zeta^{2}}$ y $v=\sqrt{\frac{d\eta}{dt}^{2}+\frac{d\xi}{dt}^{2}+ \frac{d\zeta}{dt}^{2}}$. Las componentes del campo magnético, $B_{\eta}$, $B_{\xi}$ y $B_{\zeta}$, se han obtenido usando (\ref{DipoloM}), (\ref{Bxyz}) y (\ref{CoorTrans}) y por lo tanto son también funciones de la posición $(\eta,\xi,\zeta)$ de la partícula. Luego, las ecuaciones (\ref{EDmgr}) son funciones de  $\eta$, $\xi$ y $\zeta$ y de sus derivadas con respecto al tiempo $t$, de modo que forman un sistema acoplado de ecuaciones diferenciales no lineales de segundo orden. La solución de  (\ref{EDmgr}), la cual se obtiene por medio de métodos numéricos (por ejemplo el método de Runge Kutta\index{Runge Kutta}) , nos permite determinar el cambio de posición de una partícula con el transcurso  del tiempo, es decir la trayectoria $(\eta(t), \xi(t), \zeta(t))$ de  una partícula con ciertas  condiciones iniciales ($\eta(0),\xi(0), \zeta(0)$) y ($v_{\eta}(0),v_{\xi}(0), v_{\zeta}(0)$).
  
 Con las herramientas que hemos arriba desarrollado, estudiaremos los movimientos de una corriente o flujo de partículas  de polvo que se acerca a la Tierra. Por simplicidad supondremos que la corriente de partículas es coplanar con el plano $\eta-\xi$  que hemos arriba definido. Sin embargo el tratamiento puede fácilmente generalizarse para  cualquier ángulo de incidencia. Además supondremos, sin pérdida de generalidad, que la  corriente de polvo se acerca a la Tierra en una dirección coincidente con el eje $\eta$. Es decir, cuando la corriente se encuentra lo suficientemente lejos de la influencia  magnética y gravitatoria de la Tierra, el vector velocidad de la corriente en su conjunto y de cada una de sus partículas es $\overrightarrow{v}=(V_{f},0,0)$ donde el valor de $V_{f}$ debe ser negativo porque las partículas se acercan a la Tierra. Por lo tanto,  tomaremos como condición inicial para todas las partículas $v_{\eta}(0)=V_{f}$, $v_{\xi}(0)=0$ y $v_{\zeta}(0)=0$.
 
 Según hemos supuesto, el frente de la corriente de partículas, que consideramos plano,  se mueve  a lo largo del eje $\eta$ en forma perpendicular a dicho eje. Suponemos además que la corriente de partículas  es de forma cilíndrica  con un radio $R_{f}$ y está centrada en el eje $\eta$. Comenzaremos a contar el tiempo $t$ a partir del momento en que el frente plano de partículas  pasa por la posición $\eta_{0}$. En otras palabras, en el instante $t=0$, $\eta(0)=\eta_{0}$ para todas las partículas del frente. Si $b$ es la distancia de una partícula al eje $\eta$ y $\theta$ el ángulo entre el eje $\xi$ y  $b$, $\xi=b, cos\, \theta$ y $\zeta=b \, sen\, \theta$. Por lo tanto, las condiciones iniciales $\xi(0)$ y $\zeta(0)$ de las partículas del frente están dadas por $\xi(0)=b\, cos\, \theta$ y $\zeta(0)=b \, sen\, \theta$, donde $0 \le b \le R_{f}$ y $0 \le \theta \le 360^{\circ}$. La distancia $b$  suele denominarse parámetro de impacto.

\begin{figure}
\includegraphics[scale=1.2]{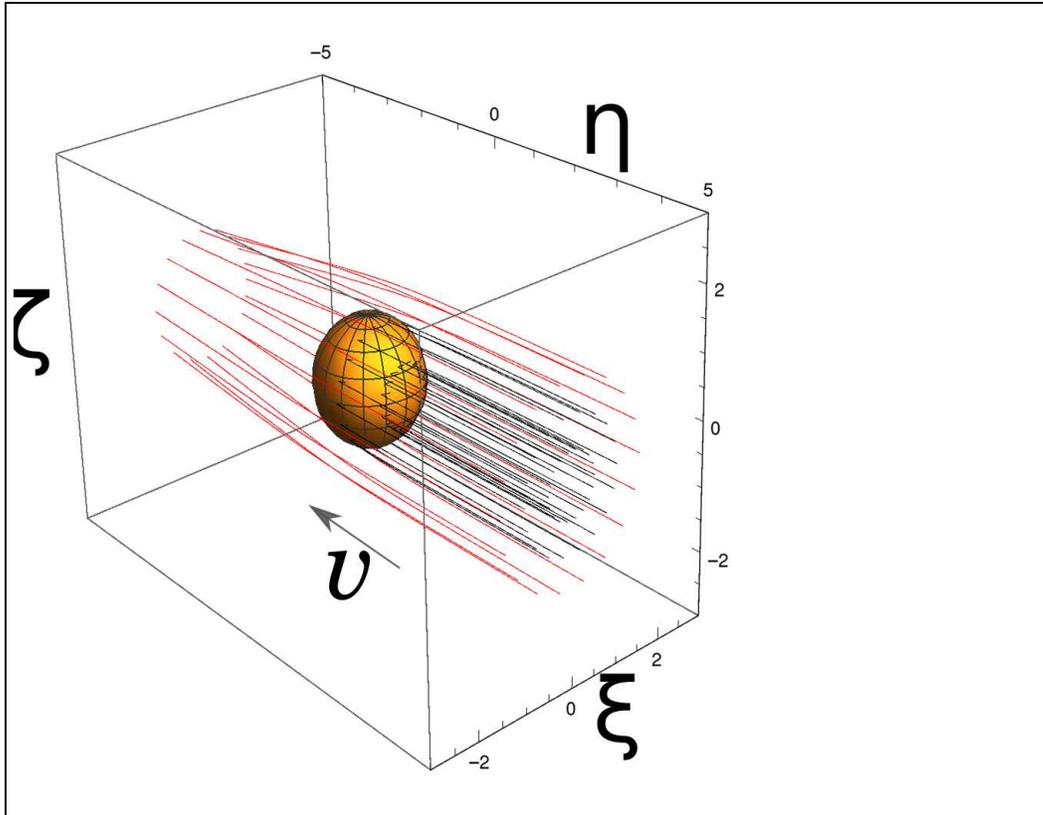} 
\caption{Corriente de partículas de polvo pasando sobre la Tierra con una velocidad $v \,\, (\approx30$\, km $s^{-1})$. Las coordenadas $\eta$, $\xi$ y $\zeta$ están expresadas en unidades del radio de la Tierra $R_{T}$. Las trayectorias de los granos de polvo con el parámetro de impacto $b\le R_{T}$ son representadas en negro, y aquellas con $b> R_{T}$ en rojo.}
\label{CorrienteDust}
\end{figure}   
  
  En resumen, las condiciones iniciales de las partículas que representan el frente de la corriente de polvo son elegidas del siguiente modo: $(V_{f},0,0)$ para la velocidad, igual para todas las partículas, y $(\eta_{0},b\, cos\, \theta, b \, sen\, \theta)$ para la posición, donde $b$ se varía en un paso fijo $\Delta b$ y $\theta$ en $\Delta \theta$. Adoptamos $V_{f}=-30$ km/s (recordemos que ésta es una velocidad relativa a la Tierra), $\eta_{0}=5\, R_{T}$, $R_{f}=1.5 \, R_{T}$, $\Delta b= 0.5\, R_{T}$ y $\Delta \theta=20^{\circ}$. Resolviendo el sistema de ecuaciones (\ref{EDmgr}) para cada una de las partículas adoptadas como representativas del frente de la corriente, con sus correspondientes condiciones iniciales,  obtenemos  las trayectorias de dichas partículas (ver la fig. \ref{CorrienteDust}). La trayectoria o órbita de cada partícula del frente de la corriente puede interpretarse de dos maneras. Primero,  ésta es la línea de puntos imaginarios por los cuales la partícula pasó y dejó atrás. Segundo, ésta es una fila de partículas que se extiende desde el frente a la espalda de la corriente, pues detrás de cada partícula del frente se encuentra otra que tuvo las mismas condiciones iniciales y por lo tanto sigue la misma órbita.  
  
La fig. \ref{CorrienteDust} muestra que las partículas de polvo con $b \le R_{T}$  impactan  la atmósfera, mientras que las trayectorias de aquellas con $b > R_{T}$ se curvan levemente al pasar sobre la Tierra, formando un cono tubular. En conclusión, la  magnetosfera  apenas perturba el movimiento de los granos de polvo  y por lo tanto sus órbitas son casi balísticas. Las partículas que chocan contra la atmósfera son abruptamente frenadas debido al roce con el aire y formarían una capa de polvo  que permanece en suspensión por un cierto tiempo en la alta atmósfera.

\section{Explosiones en la atmósfera: ¿Meteoritos o bombas atómicas?}

En la historia temprana de la Tierra,  los choques con grandes asteroides eran muy frecuentes y jugaron un papel clave en la formación de  la Tierra. En el presente, la probabilidad del encuentro con un gran asteroide es muy baja, pero hay muchos cuerpos relativamente pequeños  que cruzan la órbita terrestre y por lo tanto representan un peligro para nosotros. Uno de los impactos más recientes ocurrió en Rusia\index{Rusia} en la ciudad de Chelyabinsk\index{Chelyabinsk} en 2013 \cite{bolido}. El objeto ingresó a la atmósfera de forma imprevista y explotó a una altura de entre 30 y 50 kilómetros del suelo. La onda expansiva provocó que varias ventanas estallaran en pedazos, que hirieron a más de mil personas. El pequeño asteroide tenía un diámetro de $\approx 20$ metros e ingresó a la atmósfera a una velocidad de entre 15 y 20 kilómetros por segundo. Se estima que la onda expansiva afectó a más de siete mil construcciones de la ciudad.

El 1 de octubre de 1990, sensores satelitales del departamento de defensa de Estados Unidos  \index{Estados Unidos} detectaron sobre el Océano Pacífico \index{Océano Pacífico}  una explosión equivalente a la de un kiloton de TNT \footnote{El Trinitolueno (TNT) \index{Trinitolueno (TNT)} es un compuesto químico  altamente explosivo y para dimensionar la magnitud de una explosión, independientemente del proceso físico que la ocasione, nos referimos a la cantidad de toneladas de TNT que debería detonarse  para producir el mismo efecto. Es decir, nos  referimos a la tonelada  de TNT como una unidad de energía, con cuya explosión se libera una energía equivalente  a  $4.18 \times 10^{16}$  ergios.  La bomba nuclear arrojada en la segunda guerra mundial sobre Hiroshima \index{Hiroshima}  liberó  una energía equivalente a 13.000 toneladas de TNT, es decir 13 kiloton de TNT. Las llamadas bombas nucleares tácticas tienen 1 kiloton de energía
.}. El análisis de la información de esos satélites determinó que probablemente la explosión se debió a la explosión de un meteorito de  100 toneladas  en la atmósfera  a una altura de 30 km. Los efectos de  la explosión de una bomba nuclear son análogos a aquellos provocados por la explosión de un cuerpo cósmico, excepto que este último no está asociado con elementos radioactivos. Ello nos alerta sobre el peligro que implica confundir un evento natural, como el descripto,  con un ataque o  prueba nuclear. De acuerdo con los datos hechos públicos por el centro de estudios de la NASA de objetos cercanos a la Tierra (https://cneos.jpl.nasa.gov/fireballs/), los sensores satelitales del gobierno de Norteamérica detectaron 851 destellos de explosiones en la atmósfera entre los años 1988 y 2020. Es decir, en promedio,  ocurre  un evento de esa naturaleza cada dos semanas en algún lugar de la Tierra.

 El choque de un gran cometa o asteroide contra un planeta es un evento infrecuente. En la Tierra hace 65 millones de años,  uno de tales eventos ocurrió  y provocó  la extinción  de los dinosaurios y de numerosas especies, lo cual significó el final del período Cretácico\index{Cretácico} y el comienzo del Terciario\index{Terciario}. Los cuerpos cósmicos que chocan contra los planetas gaseosos (Júpiter\index{Júpiter} y Saturno\index{Saturno}) siempre explotan en el aire, obviamente. En julio de 1994, 21 fragmentos del cometa Shoemaker-Levy 9\index{Shoemaker-Levy 9 (cometa)} impactaron  sobre Júpiter\index{Júpiter}, tal como se lo había predicho. Astrónomos de todo el mundo  prepararon con antelación  sus telescopios e instrumental  moderno para examinar tan fantástico experimento natural\footnote{El Instituto Argentino de Radioastronomía\index{Inst. Argentino de Radioastronomía} participó en la observación del fenómeno \cite{SL9} \cite{Olano2}}.

Los cuerpos poco compactos, como los cometas o ciertos meteoritos rocosos,  llegan a la superficie de la Tierra solo si ellos son muy grandes. La fuerza de fricción aerodinámica  frena fuertemente y fragmenta  a un cuerpo  relativamente pequeño y de poca densidad. El cuerpo pierde rápidamente la mayor parte de su energía cinética en un tramo relativamente corto de su caída. La energía cinética  perdida por el cuerpo se transfiere  en gran parte al gas atmosférico circundante como  energía térmica (calor). En un corto tiempo, el gas afectado alcanza  una alta temperatura y por lo tanto una alta  presión interna, la cual  al superar grandemente a la presión externa,  produce una explosión. Como se libera una gran cantidad de energía dentro de un relativamente pequeño volumen, la explosión se aproxima a  una explosión puntual.

Si la energía térmica $E_{th}$ generada en el proceso de fragmentación y frenado está dentro de  un volumen $V$ de gas, la presión que este gas caliente ejerce sobre el gas atmosférico que rodea al volumen $V$ es :
\begin{equation}
P=\frac{E}{V},
\label{Explosion1}
\end{equation}
donde hemos supuesto que $E=E_{th}=E_{k}$, a pesar de que $E_{th}$ y la energía cinética del cuerpo  $E_{k}$, antes de ingresar a la atmósfera,  no son estrictamente iguales. También hemos omitido el factor $(\gamma-1)$, donde $\gamma=5/3$ es el cociente de los calores específicos.

Si la presión $P$ es mucho mayor que la del medio circundante,  la capa gas que rodea al volumen $V$, sobre la  cual actúa el empuje de la presión $P$, es obligada a moverse a velocidades supersónicas, con lo cual se genera una onda de choque fuerte. En el caso de una explosión puntual,  el frente de la onda de choque es inicialmente una superficie esférica, centrada en el punto de explosión y que se expande isotrópicamente;  y $V$  es el volumen, naturalmente esférico, que encierra el frente de onda. Por lo tanto $V$ es una función que aumenta con el tiempo y  que denotamos $V(t)$. Por lo tanto, de la fórmula (\ref{Explosion1}), tenemos que la presión también depende del tiempo, $P(t)$.

Utilizaremos el símil de una burbuja para referirnos a la estructura completa del frente de onda y su interior. Suponiendo que la presión interna de la burbuja, $P(t)$, domina cualquier presión externa, la velocidad $v$ de expansión de la burbuja es dada por las relaciones de Rankine-Hugoniot \index{Rankine} \index{Hugoniot} para un choque fuerte, 
\begin{equation}
v=\sqrt{\frac{\gamma+1}{2} \frac{P(t)}{\rho}}.
\label{Explosion2}
\end{equation}
La expresión (\ref{Explosion2}) muestra que si la densidad $\rho$ del medio en que se produce la explosión es uniforme, la burbuja se expande con la misma velocidad en todas direcciones y por lo tanto conserva su forma esférica original. Sin embargo, la densidad atmosférica decrece exponencialmente con la altura. Entonces, las partes más altas de la burbuja de expanden con mayor velocidad y en consecuencia la burbuja se alarga hacia arriba  en la dirección de la vertical local.

Las ecuaciones que gobiernan la propagación de una onda de choque en un medio gaseoso no-uniforme forman  un sistema acoplado de ecuaciones diferenciales parciales que en general solo puede resolverse numéricamente por medio de sofisticadas técnicas  computacionales. Sin embargo, el físico ruso Aleksandr  S. Kompaneets (1914-1974) \index{Kompaneets A.S.} encontró, mediante ciertas simplificaciones, una solución analítica para la propagación de una onda de choque intensa en un medio ambiente exponencialmente estratificado. La motivación original de la investigación de  Kompaneets  fue la descripción de los efectos aerodinámicos de una explosión nuclear de alta energía en la atmósfera superior de la Tierra.

Estudiaremos la evolución de un frente de choque referido a un sistema cilíndrico de coordenadas $(R,z)$, donde el eje $z$ es perpendicular al plano de estratificación de la atmósfera y $R$ es perpendicular a $z$. En el modelo de Kompaneets, el origen de coordenadas está ubicado en la posición de la explosión. Dado que hay simetría en torno al eje z, omitimos el ángulo azimutal $\theta$. La densidad $\rho$  para la fórmula  (\ref{Explosion2}) se obtiene de la expresión\ref{Atmosfera}, haciendo $h=h_{0}+z$, donde $h_{0}$ es la altura, con respecto al suelo, en la que ocurre la explosión. Entonces, 
\begin{equation}
\rho=\rho_{0} \, exp\, (-\frac{z}{H}),
\label{Atmosfera2}
\end{equation}
donde $\rho_{0}=\rho(0)\,  exp\, (-\frac{h_{0}}{H})$ es la densidad atmosférica en el sitio de la explosión.

\begin{figure}
\includegraphics[scale=0.9]{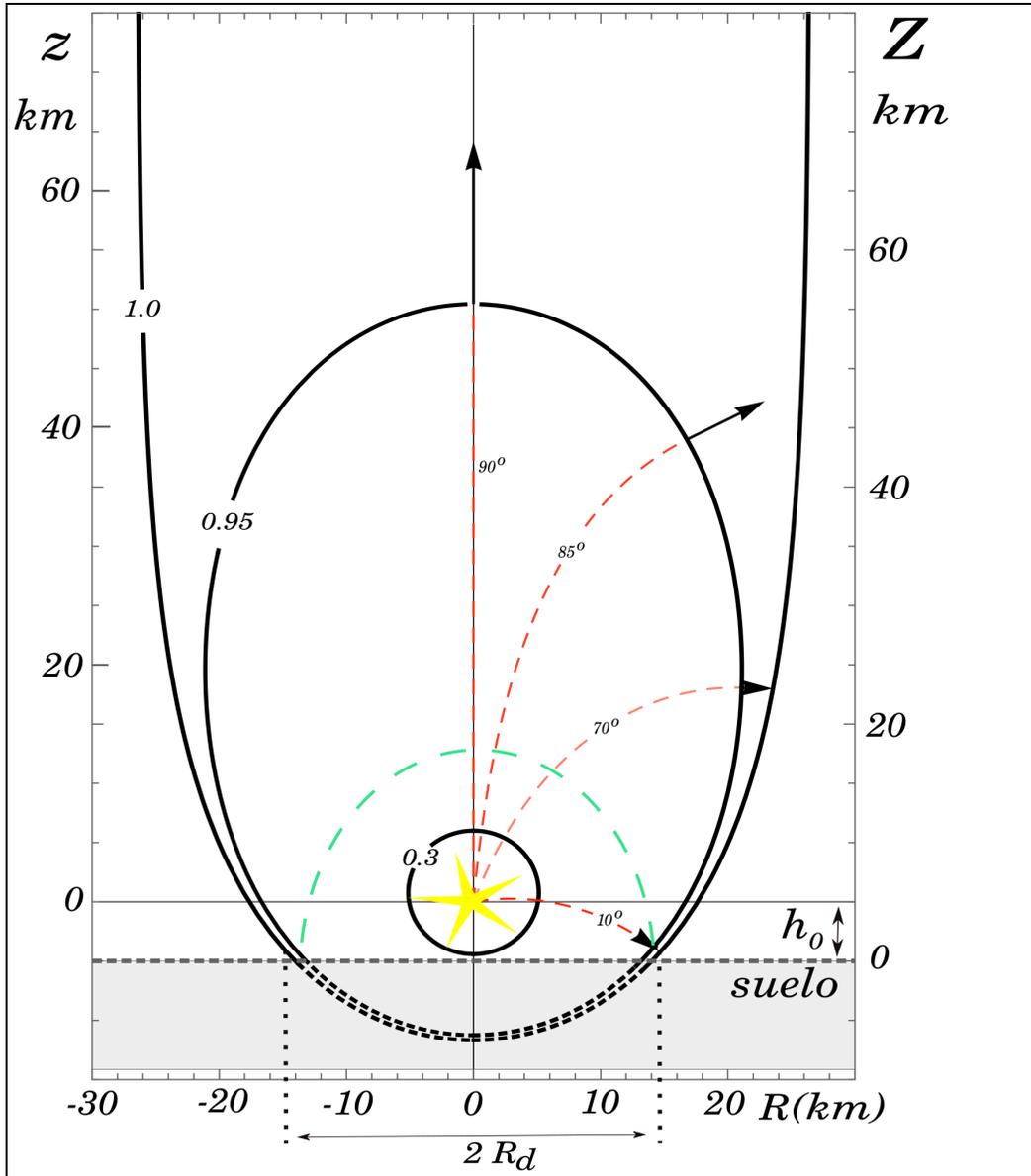} 
\caption{Evolución de la burbuja representada en coordenadas cilíndricas $(R, z)$, centradas en el epicentro de la explosión. Los ejes $R$ y $z$ son paralelos al horizonte y  a la vertical del lugar, respectivamente. El eje $Z$, representado a la derecha, refiere a las posiciones con respecto al suelo y por lo tanto $Z=h_{0}+ z$.}
\label{ExplosionAire}
\end{figure}

\begin{figure}
\includegraphics[scale=1.3]{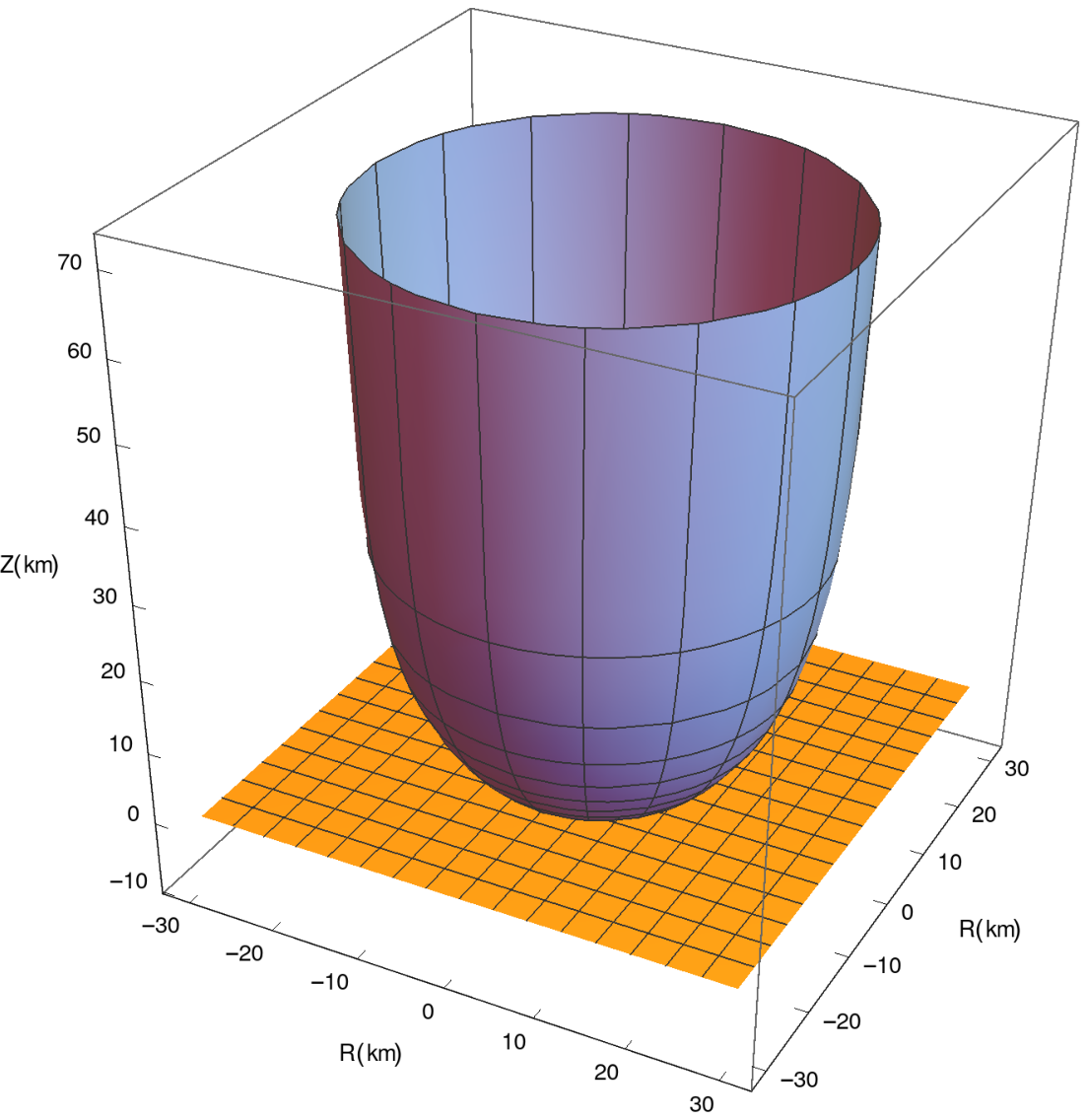} 
\caption{Configuración  tridimensional de la burbuja cuando su parte superior revienta ($t_{\star}=1$). El plano marrón representa el suelo. }
\label{Bubble}
\end{figure}

A partir de las relaciones (\ref{Explosion1}),  (\ref{Explosion2}) y (\ref{Atmosfera2}) y las suposiciones: a) la presión detrás del frente de choque es espacialmente uniforme y depende sólo del tiempo, y b) cada elemento de superficie del frente de choque se mueve perpendicular a si  mismo, Kompaneets dedujo una  ecuación del tipo $R=g(z,t_{\star})$  que describe la forma del frente de choque en función de una variable adimensional  $t_{\star}$, la cual representa el tiempo $t$. En el marco del modelo de Kompaneets, el autor de esta monografía \cite{Olano3} dedujo ecuaciones del tipo $R=f_{1}(\varphi, t_{\star})$ y $z=f_{2}(\varphi, t_{\star})$ que describen en forma completa la evolución de un frente de choque en una atmósfera exponencial. Explícitamente, 
\begin{eqnarray}
R &=& f_{1}(\varphi, t_{\star})= 2\, H\, arctan \,\left(\frac{t_{\star} cos \varphi}{1-t_{\star} sen \varphi}\right)\nonumber \\
z &=& f_{2}(\varphi, t_{\star})=-2\, H \,ln \,\left(\sqrt{1-2 t_{\star} sen \varphi+ t_{\star}^{2}}\right),
\label{SolParametric}
\end{eqnarray}
donde $\varphi$ es el ángulo entre el plano de estratificación, es decir el horizonte del lugar, y la velocidad inicial $\overrightarrow{v_{0}}$ ($t_{\star}=0$)  con la que partió un punto $P$ de la superficie del frente de choque. Por lo tanto, el movimiento de $P$  se efectúa  en el plano determinado por el eje $z$ y el vector $\overrightarrow{v_{0}}$, o en otras palabras el movimiento se realiza en el plano $R-z$ con el ángulo azimutal $\theta=$ constante. Las dos ecuaciones (\ref{SolParametric}) contienen la solución original de Kompaneets, pues si separamos  de ellas las funciones que dependen de  $\varphi$, usando identidades trigonométricas apropiadas, ambas se  funden en la fórmula $R=g(z,t_{\star})$  de Kompaneets. Asimismo, si eliminamos $t_{\star}$ de dichas ecuaciones obtenemos una ecuación para  el órbita de un punto sobre el frente de choque:
\begin{equation}
z=- 2\, H\, ln\, [cos\, \varphi sec \, (\varphi - \frac{R}{2 H})]
\end{equation}
El aspecto de la órbita es determinado por el ángulo inicial de lanzamiento $\varphi$ del punto del frente de choque. Dado el valor de $\varphi=$ const., correspondiente al de  un  punto $P$ del frente de shock, la órbita de $P$ puede también determinarse  mediante  las ecuaciones paramétricas (\ref{SolParametric}) que dan las posiciones $(R(t_{\star}), z(t_{\star}))$ de $P$ en el intervalo de $t_{\star}$ deseado (lineas rojas a trazos en la fig. \ref{ExplosionAire}).
 Fijando  $t_{\star}$ en un valor,   los puntos $(R(\varphi), z(\varphi))$ que se obtienen variando  $\varphi$ entre $0$ y $2 \pi$ en las ecuaciones (\ref{SolParametric}) trazan una curva que representa la intersección de la superficie de la burbuja con el plano $R-z$ (ver la fig. \ref{ExplosionAire}). Rotando dicha curva  en torno al eje $z$, se obtiene  la superficie de la burbuja. La fig. \ref{Bubble} muestra la superficie de la burbuja para $t_{\star}=1$. La burbuja se desarrolla entre  $t_{\star}=0$, momento de la explosión, y  $t_{\star}=1$ que es cuando la parte superior de la burbuja revienta y la presión interna de la burbuja decae abruptamente.
 La altura $z(t)$ del tope de la burbuja, que se calcula usando (\ref{SolParametric}) con $\varphi=\frac{\pi}{2}$, tiende a infinito  cuando $t_{\star} \rightarrow 1$. Por lo tanto, de acuerdo con (\ref{Explosion1}), la presión interna de la burbuja decae abruptamente, pues su volumen aumenta súbitamente.

Nótese que las configuraciones  de la burbuja, calculadas a través de las ecuaciones (\ref{SolParametric}), quedan determinadas por el valor de la escala $H$ de alturas, que en el caso de nuestra atmósfera es $\approx 8.42$ km, y sólo se necesitan conocer la energía $E$ de la explosión y la altura $h_{0}$ en que la misma ocurrió, para determinar la escala temporal. El tiempo está relacionado con $t_{\star}$ a través de 
\begin{equation}
t=\alpha (t_{\star}) \tau,
\label{t-tstar}
\end{equation}
donde $\tau=\sqrt{\frac{8 \pi H^{5} \rho_{0}}{E (\gamma^{2}-1)}}$. Si elegimos $E=50$ Megatones ($10^{6}$ toneladas de TNT) y $h_{0}=5$ km, $\tau=44$ segundos. En la parte final de la evolución de la burbuja, cuando ésta revienta, $t_{\star}=1$ y $\alpha (t_{\star})\approx 2$  (ver el trabajo referido en \cite{Olano3}). Entonces, de acuerdo con (\ref{t-tstar}), todo el proceso dura sólo un minuto y medio. La parte inferior de la burbuja choca contra el suelo, dañando objetos y seres vivos dentro de un radio $R_{d}\approx$ 15 km, centrado en el epicentro de la explosión proyectado sobre  suelo (ver fig. \ref{ExplosionAire}). Un valor similar para $R_{d}$ se obtiene de la ecuación (\ref{radioDevastacion}). Las líneas de trazos debajo de la superficie del suelo (fig. \ref{ExplosionAire}) indican la evolución de la parte inferior de la burbuja si ésta no hubiera chocado con el suelo y hubiera continuado su propagación a través del aire. Sin embargo, esa parte de la onda se reflejó al tocar el suelo. En la fig.  \ref{ExplosionAire},  la onda reflejada es indicada sólo esquemáticamente   por la  línea verde a trazos.  La potencia y altura de la explosión que hemos escogido en nuestro ejemplo se acercan a las estimadas para el caso de la explosión de Tunguska \index{Tunguska} ocurrida en el año 1908 (ver próxima sección).

Las componentes de la velocidad de un punto $P$ (caracterizado por $\varphi$) en el estado evolutivo $t_{\star}$,  se pueden obtener fácilmente derivando las ecuaciones paramétricas (\ref{SolParametric}) del siguiente modo:
\begin{eqnarray}
vR(\varphi, t_{\star})=\frac{\partial r(\varphi, t_{\star})}{\partial t_{\star}} \frac{dt_{\star}}{dt} \nonumber \\
vz(\varphi, t_{\star})=\frac{\partial z(\varphi, t_{\star})}{\partial t_{\star}} \frac{dt_{\star}}{dt},
\label{velocidadBubble}
\end{eqnarray}
donde $\frac{d t_{\star}}{d t}= (\frac{d \alpha(t_{\star})}{d  t_{\star}} \tau )^{-1}$, de acuerdo con \ref{t-tstar}.

 Las ecuaciones  (\ref{SolParametric}) y (\ref{velocidadBubble}) rigen  sólo  entre  $t_{\star}=0$ y  $t_{\star}=1$, y para el estudio de la evolución de la burbuja en su etapa final debemos obtener otras ecuaciones de movimiento. Supondremos que sobre  la cara externa de cada elemento de su superficie actúa  una presión  $P_{atm}$, dada por la presión atmosférica del lugar, y una presión $P_{int}$ sobre su cara interna. Si además consideramos la fuerza de gravedad, la fuerza que actúa sobre cada elemento de la burbuja es 
 \begin{equation}
 \overrightarrow{F}= -\Delta P \overrightarrow{dS}- m g \overrightarrow{z},
 \label{bola1}
 \end{equation} donde  $\Delta P = P_{atm}-P_{int}$ y $\overrightarrow{dS}$ es un vector unidad perpendicular a la superficie del elemento y que apunta hacia afuera de la burbuja, $m$ es la masa del elemento y $\overrightarrow{z}=(0,1)$. Según la segunda ley de Newton, $\overrightarrow{F}= \frac{d\overrightarrow{p}}{dt}$, donde \overrightarrow{p} es el impulso,  definido por $ m \overrightarrow{v}$. Si adoptamos el modelo de bola de nieve, la superficie $\overrightarrow{dS}$ acumula el gas circundante que barre en su recorrido y entonces la masa $m$ del elemento varía con el tiempo $t$. Por lo tanto, 
 
 \begin{equation}
 \overrightarrow{F}= \frac{d\overrightarrow{p}}{dt}=\frac{dm}{dt} \overrightarrow{v} + m \frac{d\overrightarrow{v}}{dt}.
 \label{Impulso}
 \end{equation}

 La masa recolectada por $\overrightarrow{dS}$ en el intervalo de tiempo $dt$ es $dm= \rho (\overrightarrow{dS}.\overrightarrow{v})\, dt$, donde $\rho$ es la densidad atmosférica en la posición del elemento (fórmula (\ref{Atmosfera2})). Con lo cual obtenemos,
 \begin{equation}
 \overrightarrow{F}= \rho (\overrightarrow{dS}.\overrightarrow{v}) \overrightarrow{v} + m \frac{d\overrightarrow{v}}{dt}.
 \label{bola2}
 \end{equation}
 Igualando (\ref{bola1}) y (\ref{bola2}) y despejando  $\frac{d\overrightarrow{v}}{dt}$
 obtenemos
 \begin{equation}
 \frac{d\overrightarrow{v}}{dt} = -\frac{\rho (\overrightarrow{dS}.\overrightarrow{v}) }{m} \overrightarrow{v} -\frac{\Delta P \overrightarrow{dS}}{m}-  g \overrightarrow{z}.
 \label{bola3}
\end{equation}
 El primer término a la derecha del signo igual es equivalente a una fuerza de fricción por unidad de masa, similar a la expresión (\ref{Fr}).  Si el producto escalar $\overrightarrow{dS}.\overrightarrow{v}\le 0$ , $dm=0$ dado que el elemento de superficie  se mueve hacia el interior vacío. Consideraremos que $P_{int}=0$ y que la presión atmosférica es dada por
 
 \begin{equation}
 P_{atm}= P_{0}\, exp (-\frac{Z}{H}),
 \label{PresionAt}
 \end{equation}
donde $P_{0}=101325$ Pascales (1 Pascal= $\frac{N}{m^{2}}$). Por lo tanto, $\Delta P=P_{atm}$.

\begin{figure}
\includegraphics[scale=1.9]{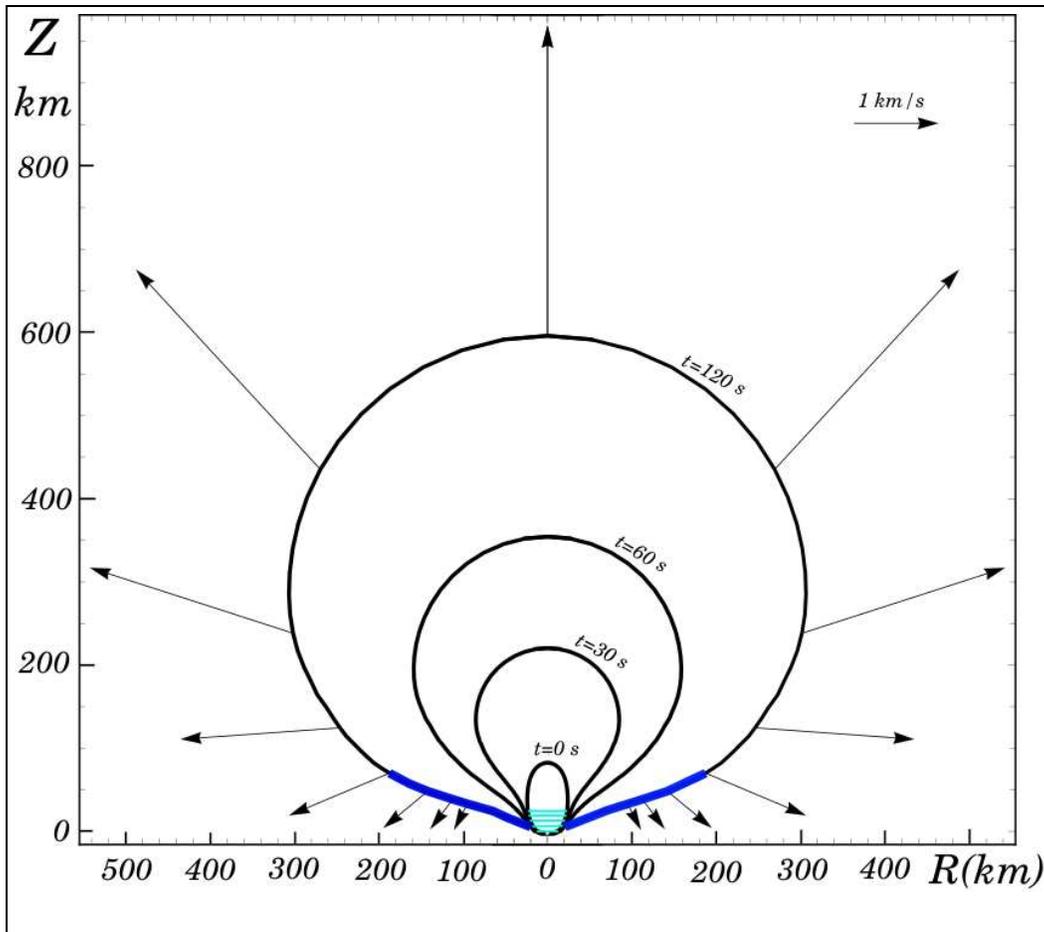} 
\caption{Los últimos estados evolutivos de la burbuja. La línea gruesa azul indica la zona en que la pared de la burbuja interacciona con la troposfera y puede inducir fenómenos meteorológicos. Las líneas celestes representan la parte de la atmósfera que ingresó al interior vacío de la burbuja.}
\label{Huracan}
\end{figure} 

El sistema de ecuaciones diferenciales dado por (\ref{bola3}) nos permite estudiar la evolución de la burbuja a partir del momento en que la presión interna decae abruptamente, es decir en el instante  de singularidad $t_{\star}=1$. A fin de evitar dicha singularidad, tomaremos las velocidades y configuración de la burbuja  en el instante $t_{\star}=0.99$, un infinitesimal antes que $t_{\star}=1$, como las condiciones iniciales para resolver el sistema de ecuaciones (\ref{bola3}). En otras palabras, las posiciones y velocidades dadas por (\ref{SolParametric}) en  $t_{\star}=0.99$ son para el sistema  (\ref{bola3}) las posiciones y velocidades  en el instante $t=0$. Para resolver (\ref{bola3}), debemos recordar que $m$  depende de $t$: $m(t)=\int_{0}^{t} \rho (\overrightarrow{dS}.\overrightarrow{v})dt$.
En la fig. \ref{Huracan}, representamos una secuencia evolutiva de la burbuja, resultante de la solución numérica del sistema de ecuaciones diferenciales   (\ref{bola3}). Para $t=120$ segundos, la burbuja es casi esférica y su parte inferior (resaltada en azul), que es donde se acumuló  mayor masa, empuja la tropósfera desde arriba como si fuera un techo desplomándose  a gran velocidad sobre el suelo. Este  hecho puede desencadenar fenómenos meteorológicas violentos en un radio de al menos 300 km en torno al epicentro de la explosión.  Ese frente de choque, indicado en azul en la fig. \ref{Huracan},  se formó por acumulación de gas de la alta troposfera que tiene muy bajas temperaturas y por lo tanto podría inducir una corriente adiabática descendente que produce vientos huracanados y precipitaciones que inciden sobre el suelo. En efecto, el frente descendente de aire frío acumula vapor de agua que se condensa formando una densa capa de gotas de agua, la cual en su caída empuja el aire subyacente originando intensas ráfagas de viento que chocan contra el suelo.

La región del epicentro indicada por líneas horizontales celestes en la fig. \ref{Huracan} requiere un tratamiento particular. Si la altura $h_{0}$ de la explosión es mayor que $1.39 H\, (= 11.70$ km, ver referencia \cite{Olano3}), la onda de choque no llega al suelo. En el caso representado en la fig. \ref{ExplosionAire} $h_{0}=5$ km, por lo tanto la parte inferior de la onda de choque golpea  el suelo. Esta onda se refleja en la superficie y se eleva creando  un vacío entre la superficie y la onda reflejada que succiona las partículas del polvo desprendidas en el choque, como así también el aire atmosférico que ingresa por los costados (ver fig. \ref{Huracan}). En esta etapa, la burbuja tiene forma de tubo (ver fig. \ref{Bubble})  y el polvo y el aire atmosférico se elevan a través  del interior vacío de la burbuja, tal como el humo se eleva a través de una chimenea. Una pregunta es cuánto tiempo tarda la columna de gas que se eleva en rellenar el espacio vacío creado por la explosión. La fuerza que eleva la columna del gas atmosférico es un caso particular de la fuerza dada por  (\ref{bola1}). En efecto, columna de gas se mueve solo en la dirección $Z$, y de acuerdo con (\ref{PresionAt}),  $\Delta P=P_{0}$, donde $P_{0}$ es la presión atmosférica al nivel del suelo que actúa  sobre la base de la columna de área $S_{0}$ (o área de la base del tubo). De modo que,

\begin{equation}
F_{Z}= P_{0} S_{0}-m g
\label{Relleno1}
\end{equation}

Denotando $Z_{h}(t)$ la altura de la columna de gas en el instante $t$, la masa de la columna de gas es $m= \overline{\rho} \, S_{0}\, Z_{h}(t)$, donde $\overline{\rho}$ es una  densidad media representativa de la baja atmósfera. A fin de simplificar el tratamiento de un fenómeno complejo, trataremos el gas como si fuera un líquido (es decir poco compresible). Por ejemplo, el aumento del volumen del liquido que sube por una  bombilla o sorbete es igual al volumen de liquido que ingresa a la bombilla desde su base. Si la altura $Z_{h}(t)$ de la columna de gas   crece en $dZ_{h}(t)$, la masa de la columna se incrementa en $dm= \overline{\rho}\, S_{0}\, dZ_{h}(t)$. Reemplazando en (\ref{Impulso}) la expresiones halladas para $m$ y $dm$, y teniendo en cuenta que $\overrightarrow{v} =\frac{ dZ_{h}(t)}{dt}$ resulta
\begin{equation}
F_{Z}=\overline{\rho}\, S_{0}\, \left |\frac{dZ_{h}(t)}{dt}  \right | \frac{ dZ_{h}(t)}{dt} + \overline{\rho} \, S_{0}\, Z_{h}(t) \frac{d^{2}Z_{h}(t)}{dt^{2}}  
\label{Relleno2}
\end{equation}

Igualando (\ref{Relleno1}) y (\ref{Relleno2}), obtenemos la siguiente ecuación diferencial de segundo orden 

\begin{equation}
  Z_{h}(t) \frac{d^{2}Z_{h}(t)}{dt^{2}} +  \left |\frac{dZ_{h}(t)}{dt}  \right | \frac{ dZ_{h}(t)}{dt}   +   g Z_{h}(t) -\frac{P_{0}}{\overline{\rho}} + f(t)=0.
 \label{Relleno3}
\end{equation}
 Si $S_{0}$ permanece constante en el tiempo, $S_{0}$ multiplica cada término de la ecuación (\ref{Relleno3}) y por lo tanto esta no depende de $S_{0}$ y $f(t)=0$.
\begin{figure}
\includegraphics[scale=1.7]{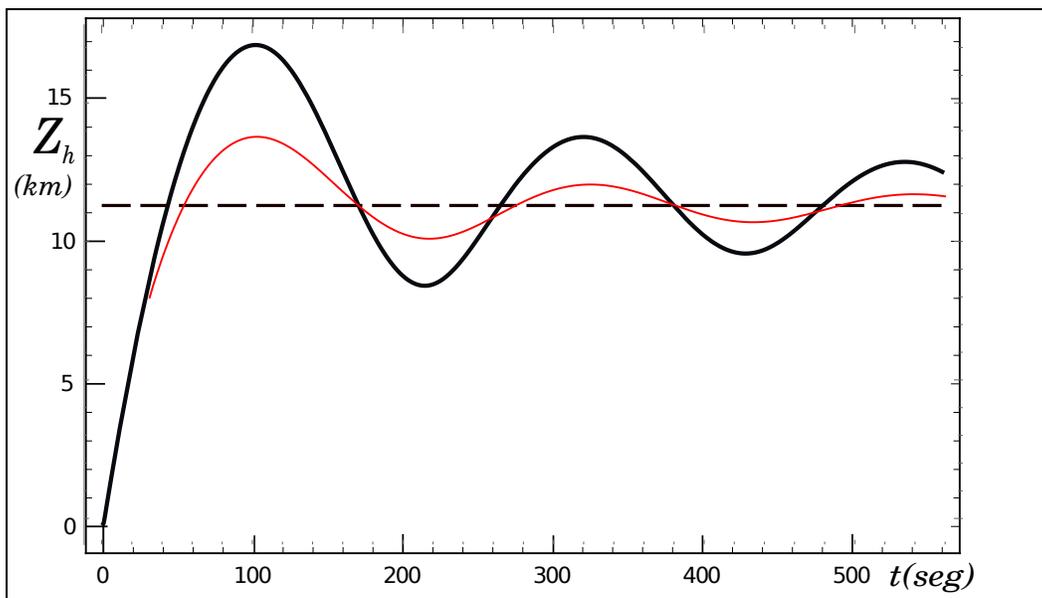} 
\caption{Altura de la columna de gas atmosférico, que ingresa al interior vacío de la burbuja desde su base, en función del tiempo. La línea llena negra representa el caso en que la parte aproximadamente tubular de la burbuja permanece estática,  y la línea roja, el caso en que se expande radialmente. La línea a trazos indica la altura de equilibrio.}
\label{Relleno}
\end{figure} 
Adoptando $\overline{\rho}=0.92$ kg m$^{-3}$, y resolviendo  por métodos numéricos la ecuación (\ref{Relleno3}), obtenemos la solucione representada en la fig. \ref{Relleno} por la línea negra. Si $f(t)=0$, la columna de gas se eleva hasta una altura máxima de 17 km en 100 segundos y luego oscila en forma amortiguada en torno a la altura de equilibrio $Z_{he}$ (indicada por una  recta a trazos en la fig. \ref{Relleno}). La altura de equilibrio es aquella en la que el peso de la columna iguala a la presión que la sostiene, relación dada por los tres últimos términos de la ecuación  (\ref{Relleno3}), con $f(t)=0$. Es  decir, $g Z_{h}(t) -\frac{P_{0}}{\overline{\rho}} =0$, de donde obtenemos que $Z_{h}(t) = Z_{he} =\frac{P_{0}}{g \, \overline{\rho}}\approx 11$ km. 

Si la pared del tubo, que representa la parte inferior de la burbuja, se expande radialmente con una velocidad constante $v_{r}$, el radio del tubo crece según  $r=r_{0}+v_{r} t$, donde $r_{0}$ es el radio inicial, el cual puede vincularse con $S_{0}=\pi r_{0}^{2}$. Por lo tanto el área sobre la cual se apoya la columna de gas, en vez de $S_{0}$, es ahora $S=\pi (r_{0}+ v_{r} t)^{2}$ y $ f(t)=Z_{h}(t) \frac{dZ_{h}(t)}{dt} \frac{1}{S} \frac{dS}{dt}$. Esta expresión para $f(t)$ deriva del hecho de que en este caso la masa depende de dos variables temporales: $m= \overline{\rho} \, S(t)\, Z_{h}(t)$. Adoptando $r_{0}=15$ km y  $v_{r}=0.2$ km s$^{-1}$, obtenemos la solución representada por la línea roja en la fig. \ref{Relleno}.

El hemisferio superior de la burbuja en $t=120$ s se eleva (ver fig. \ref{Huracan}), pero luego debido a la fuerza gravitatoria invierte su movimiento y cae sobre la atmósfera. La pared de la burbuja contiene, además de gas, pequeñas partículas de polvo, las cuales también decantarán sobre la parte superior de la atmósfera. Determinar el área sobre la cual  se esparcen los granos de polvo en la alta atmósfera es  importante  para evaluar las consecuencias de  fenómenos ópticos y meteorológicos. Para tal fin, consideraremos que en  $t=120$ segundos el gas y el polvo de la pared de la burbuja comparten los mismos  movimientos y no interaccionan  entre ellos. Por lo tanto, usaremos las posiciones y velocidades de la pared de la burbuja en $t=120$ s como las condiciones iniciales de las partículas de polvo. Dado que las partículas de polvo están generalmente eléctricamente cargadas, tendremos en cuenta la influencia del campo magnético sobre las trayectorias de los granos de polvo. Por ello, emplearemos el sistema de ecuaciones de movimiento dadas por (\ref{EDmgr}) que refiere las órbitas de las partículas de polvo a un sistema cartesiano con origen en el centro la Tierra, donde el eje $\zeta$ coincide con el eje polar geográfico. Aquí, ubicamos el eje $\eta$ en el plano meridional del sitio de la explosión.
\begin{figure}
\includegraphics[scale=1.1]{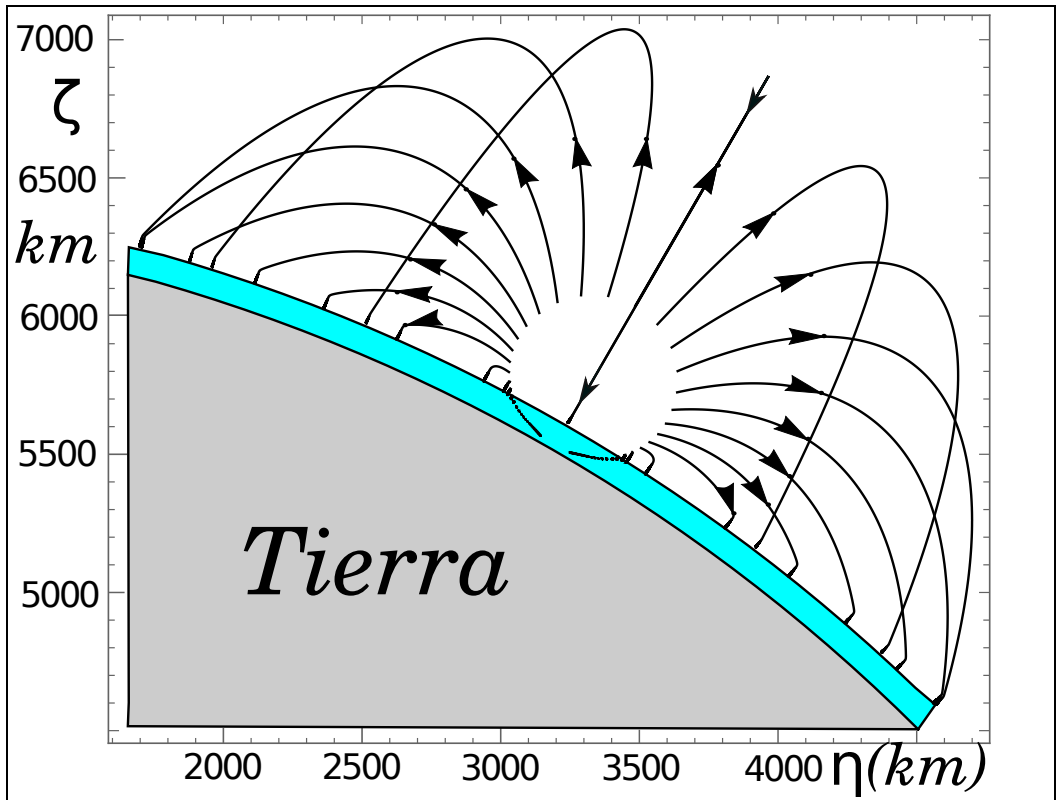} 
\caption{Trayectorias de las partículas de polvo eyectadas por la explosión. La flecha sobre cada órbita  indica el sentido del movimiento. Las flechas más pequeñas sobre la trayectoria vertical a la superficie de la Tierra muestran la inversión del movimiento.  La parte superior de la franja celeste corresponde a la  mesosfera donde las partículas de polvo son frenadas.}
\label{SurgentePolvo}
\end{figure} 
\begin{figure}
\includegraphics[scale=0.7]{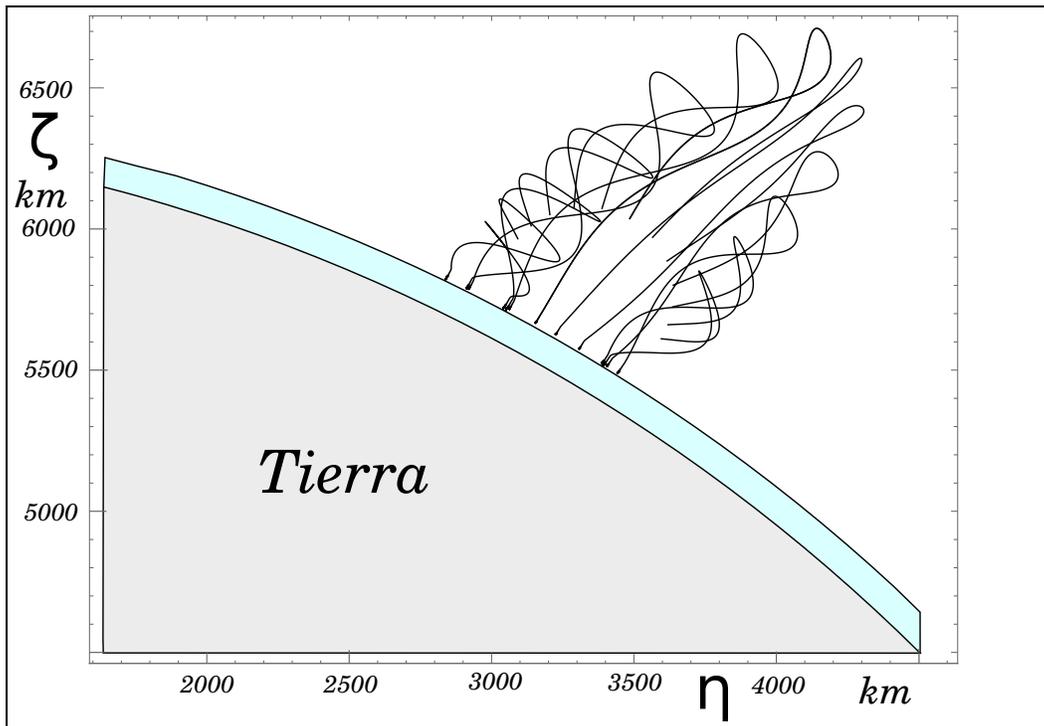} 
\caption{Lo mismo que en la fig. \ref{SurgentePolvo}, pero para partículas de polvo con cargas eléctricas mayores ($\mathcal{E} e =10^{4}$). En este caso, el campo magnético  local conduce a las partículas de polvo a moverse en torno a las líneas de fuerza magnéticas.}
\label{SurgentePolvo1}
\end{figure} 
Las fórmulas de conversión de  una posición  $(R,Z)$ de la burbuja en t=120 (ver fig. \ref{Huracan}), considerada aquí una posición inicial, en la correspondiente al sistema $(\eta, \xi, \zeta)$, son:
\begin{eqnarray}
\eta  &=& r\, cos \lambda \nonumber \\
\zeta &=& r\, sin \lambda \nonumber \\
\xi &=& 0,
\end{eqnarray}
donde $r=\sqrt{R_{T}^{2}+ R^{2}+ Z^{2}- 2 R_{T} \sqrt{R^{2}+ Z^{2}} cos (\pi-\arctan (\frac{R}{Z}))}$, \,  $\lambda=\lambda_{0}-\arctan (\frac{R}{R+Z})$ y $\lambda_{0}$ es la latitud geográfica del sitio de la explosión. Consideraremos que $\lambda_{0}=60^{\circ}$, similar la latitud del sitio donde ocurrió la explosión de Tunguska\index{Tunguska} (ver sección siguiente), y  que el valor del factor $Q$ de las partículas de polvo es igual a $-3.81$ C kg$^{-1}$, como en la sección anterior. Es decir, dado que $Q=\frac{\mathcal{E} e \, q_{e}}{m_{p}}$, relación entre la carga eléctrica y la masa de la partícula, hemos adoptado $\mathcal{E} e =10^{2}$ y  la masa del grano de polvo $m_{p}= 4.2\, 10^{-18}$ kg y $q_{e}$ es la carga del electrón. Con dichos  parámetros y las condiciones iniciales de una dada  partícula obtenemos su trayectoria mediante la solución del sistema de ecuaciones  (\ref{EDmgr}). En la fig. \ref{SurgentePolvo},  se muestran  trayectorias de varias partículas de polvo, representativas del conjunto. Las partículas eyectadas por la explosión siguen trayectorias casi balísticas (parabólicas). Con el valor adoptado para $Q$, la influencia del campo magnético sobre las trayectorias de las partículas es despreciable. Las partículas lanzadas verticalmente llegan a una altura máxima y luego retornan al sitio de explosión. Aquellas lanzadas con  ángulos menores a 90 grados con respecto al suelo caen sobre la mesósfera, donde son frenadas, a distancias del epicentro que no superan los 1200 km. En aproximadamente 30 minutos,  todas las partículas de polvo se depositan sobre la mesósfera  dentro de  un circulo de 1200 km de radio 
en torno al  epicentro.

Como consecuencia de las altas temperaturas desarrolladas en el proceso explosivo, las uniones  moleculares  y atómicas del gas del  interior de la burbuja pueden romperse y desprender  partículas eléctricamente cargadas,  formando lo que en física se llama ``plasma''. Acoplados al plasma se encuentran también partículas  micrométricas de polvo. Las partículas de polvo que forman parte de este plasma sucio, pueden tener valores de ``Q'' más altos que el que hemos arriba empleado. Por lo tanto, aquí supondremos que $\mathcal{E} e =10^{4}$, en lugar del valor de $10^{2}$ empleado antes, y conservamos el mismo valor para la masa $m_{p}$. Usando las mismas condiciones iniciales con las que obtuvimos las trayectorias representadas en la fig. \ref{SurgentePolvo}, mediante las ecuaciones (\ref{EDmgr}), calculamos  las trayectorias de las partículas de polvo con  una mayor carga eléctrica (ver fig. \ref{SurgentePolvo1}). Esta figura muestra que los granos de polvo se elevan moviéndose en espiral en torno a las líneas del campo magnético local. Como las componentes de velocidad vertical de los granos son desaceleradas por la atracción gravitatoria, las partículas invierten sus movimientos, retornando al sitio de la explosión en aproximadamente 30 minutos. Vemos que el radio de dispersión de estas partículas sobre el suelo es mucho menor que en el caso de la fig. \ref{SurgentePolvo}.
\section{El evento Tunguska} \index{Tunguska}

Temprano en la mañana del 30 de junio de 1908, en una región boscosa, despoblada e inhóspita de Siberia\index{Siberia}  cerca del rio Tunguska, \index{Tunguska} ocurrió un hecho extraordinario cuya reverberación fue creciendo con el paso del tiempo en los círculos científicos. Sin embargo, transcurrieron muchos años antes de que el mundo se enterara de lo que realmente había sucedido. A los 2 días del episodio, un periódico de Irkutsk, \index{Irkuts} una ciudad importante de Siberia que está al borde del lago Baikal y es una de las paradas del tren transiberiano, informó  que  labriegos  que vivían a 300 km del  río Tunguska\index{Tunguska}  habían visto una  inmensa bola de fuego más brillante que el Sol acercándose al suelo, seguido por una enorme nube de humo e intensas detonaciones. El maquinista del  expreso transiberiano que transitaba en ese momento a 600 km de la región de Tunguska\index{Tunguska} frenó el tren, creyendo que   el estruendo y la sacudida de los vagones  se debían a una explosión ocurrida en el mismo tren. En la noche del 30 de junio al 1 de julio, el cielo sobre gran parte de Eurasia fue extrañamente luminoso. Por ejemplo, en el Reino Unido, a más de 5000 km de Tunguska\index{Tunguska}, se podía jugar al cricket y leer un periódico a medianoche por el resplandor del cielo nocturno. Además, en ese día, muchos observatorios del mundo registraron ondas sísmicas y de presión, que solo mucho tiempo después pudieron asociarse como diferentes aspectos del mismo evento. En particular, el sismógrafo del observatorio de Irkutsk\index{Irkuts}, ciudad alejada en cerca de 1000 km del río Tunguska\index{Tunguska},  registró un intenso movimiento sísmico, lo cual evidentemente estaba relacionado con el fenómeno informado  al observatorio por numerosos testigos que vieron una enorme bola de fuego cruzar el cielo de sur a norte.

El valle del río Tunguska\index{Tunguska} es una vasta región de pantanos, ríos y bosques donde solo incursionaban cazadores nómades de un pequeño grupo étnico, llamado Tungús \index{Tungús}(de allí el nombre del lugar). El difícil acceso al lugar y las  dificultades de la Primera Guerra Mundial y la Revolución Rusa impidieron llegar pronto al sitio de  los hechos. Sin embargo,  la sospecha de que el peculiar evento de Tunguska\index{Tunguska} podría tratarse de la caída de un gran meteorito permaneció  firme entre los inquietos especialistas que conocían la noticia. El investigador ruso de meteoritos, Leonard Kulik, \index{Kulik Leonard} fue el primero que exploró la región de Tunguska\index{Tunguska},  y logró avances  en el esclarecimiento de lo que ha pasado a la fama como “el evento de Tunguska\index{Tunguska}''.  Kulik dirigió dos expediciones, una en el año 1921, y otra en el año 1927 en la cual logra penetrar al sitio de la explosión del año 1908. En el primer viaje,  Kulik se dirigió en el tren transiberiano desde San Petersburgo\index{San Petersburgo} a Kansk\index{Kansk}, ciudad donde aprendió que el impacto habría tenido lugar a cientos de kilómetros hacia el norte, cerca del río Tunguska\index{Tunguska}.

Cuando Kulik\index{Kulik Leonard}  llegó al lugar de los hechos, vio un panorama devastador. Dentro un radio de $\approx  25$ km entorno al epicentro de la explosión, los árboles gigantes de la densa  foresta fueron partidos y derribados como si fueran cañas por la   onda expansiva (ver figs. 4 y 5 de \cite{Vasilyev}). Los tallos derrumbados están orientados en direcciones radiales desde el epicentro, como si fueran flechas que indican el flujo que siguió la onda  de choque. En la zona central del bosque aplastado, los arboles estaban también carbonizados. El hecho singular es que no hallaron un gran cráter, sino sólo un gran llano fangoso, cierto número de agujeros pequeños y  algunas partículas meteoríticas de una décima de milímetro de grosor. La explicación más plausible es que se trata de una explosión de impacto, y a juzgar por la falta de un gran cráter, esta explosión habría ocurrido en la atmósfera a una cierta altura sobre el sitio. 

Así como el descubrimiento de la tumba del faraón Tutankamón \index{Tutankamón} alentó las investigaciones   de muchos   arqueólogos y egiptólogos,  la evidencia  de estar frente a las consecuencias de  la explosión  más grande de un cuerpo cósmico que se haya registrado en tiempos históricos suscitó la atención de geólogos, astrónomos e investigadores de variadas disciplinas. Se han realizado muchos estudios \it{in situ} \rm empleando técnicas modernas de exploración. El geoquímico y astrónomo ruso Kiril Florenski\index{Florenski K.} dirigió una expedición en 1961-62 al sitio de la explosión de Tunguska\index{Tunguska}, en la cual se recolectaron  esferas micrométricas de magnetitas y silicatos enriquecidas en iridio y otros metales que son abundantes en meteoritos y cometas pero raros en la Tierra. Este hallazgo demostró  el origen extraterrestre del cuerpo que explotó en Tunguska\index{Tunguska}. Desde entonces, varios equipos internacionales realizaron expediciones  a la región de la explosión de Tunguska\index{Tunguska}, entre ellos un grupo de científicos italianos, coordinados por el físico Giuseppe Longo\index{Longo Giuseppe} de la Universidad de Bolonia\index{Bolonia}, que llevó a cabo dos expediciones, una en  1991 y la otra en 1999. Sin embargo,  nadie ha logrado hasta la fecha  recuperar un posible fragmento del objeto cósmico, lo cual no les ha permitido saber a los especialistas si el cuerpo que explotó fue un gran meteorito o un fragmento de cometa. A partir de la extensión  de la región boscosa aplastada y de las intensidades de las ondas atmosféricas y sísmicas registradas fue posible deducir que el objeto explotó en el aire a una altura de 5-10 km y que la energía liberada fue equivalente a 10-15 Megatones,\footnote{1 Megaton (Mt), o Megatonelada de T.N.T,  es equivalente a $4.2 \,\,10^{22}$ ergs} mil veces superior a la energía de la bomba de Hiroshima\index{Hiroshima}. Algunos autores obtienen  valores mayores  para la energía de la explosión de Tunguska\index{Tunguska}; a saber, 50-100 Megatones. Si esto hubiese ocurrido sobre una zona densamente poblada, hubiese sido una gran tragedia de consecuencias históricas. Renos y otros animales salvajes fueron muertos por la explosión, pero no se conoce con certeza que haya habido alguna víctima humana. Giuseppe  Longo\index{Longo Giuseppe} en el libro ``Comet/Asteroid Impacts and Human Society'' comenta que si el cuerpo cósmico hubiera alcanzado la Tierra cuatro horas después, hubiera explotado sobre la Capital Rusa de San Petersburgo\index{San Petersburgo}, que se encuentra en la misma latitud geográfica que Tunguska\index{Tunguska}. En esta situación contrafáctica, la participación rusa en la primera guerra mundial y la revolución rusa no hubiesen sido probablemente posibles. Este ejemplo hipotético muestra cómo un suceso azaroso puede alterar el devenir de la historia del mundo.

Cuando se trata de un suceso imprevisto, como es  la caída de un meteorito, los datos que puedan proveer los observadores circunstanciales del episodio son de gran ayuda para los investigadores. La más completa colección de testimonios reúne  386 observaciones directas de la explosión de Tunguska\index{Tunguska} desde diferentes posiciones geográficas. Considerando las noticias publicadas en periódicos locales, informes y comunicaciones oficiales, el número de testimonios presenciales se eleva  a un total de 708. Por ejemplo, uno de los testimonios muy citados es   el de  un granjero que se encontraba a 200 km del epicentro y recordaba,  casi 20 años después, que : ``cuando me senté a desayunar al lado de mi arado, sentí repentinamente explosiones, como si fueran disparos de un arma de fuego. Mi caballo cayó sobre sus rodillas. Desde el lado norte sobre los árboles, una llamarada surgió. Luego vi que los arboles se inclinaron por el viento y pensé en un huracán. Me aferré a mi arado con ambas manos para no ser arrastrado... y luego el huracán levantó una pared de agua del río Angará\index{Angará (rio)}''. Es interesante la referencia que hace el testigo sobre  la cortina de agua que acompañaba al huracán y que él atribuyó a agua arrancada del río cercano.  En la sección anterior,   hemos visto que  una fuerte explosión en el aire  origina una extensa burbuja de gas  en expansión (fig. \ref{Huracan}). La parte inferior de la burbuja, que se mueve a gran velocidad, empuja al gas atmosférico circundante desde lo alto de la atmósfera  hacia el suelo. Nosotros especulamos que dicho proceso podría inducir  una  corriente convectiva descendente que  provoque  una tormenta con precipitaciones y vientos muy fuertes, a más de 200 km h$^{-1}$,  incidiendo sobre el suelo. Este fenómeno podría explicar el gran alcance de los efectos de la explosión y los hechos descriptos por el mencionado testigo, que se encontraba considerablemente alejado del epicentro.

Los testimonios presenciales permiten estimar la dirección desde la cual provino el cuerpo cósmico y así calcular los parámetros orbitales del cuerpo. En base al estudio de  varios testimonios del evento de Tunguska\index{Tunguska}, el astrónomo Checoslovaco Lubor Kresák\index{Kresák L.} encontró  que el punto del cielo   desde donde apareció  el proyectil de Tunguska\index{Tunguska}  coincidía con el punto radiante de la lluvia de meteoros de Beta Tauridas\index{Tauridas Beta (corriente de meteoros)}. Se sabe que los meteoritos de Beta Tauridas  \index{Tauridas Beta (corriente de meteoros)} son desprendimientos del cometa Encke\index{Encke (cometa)}, los cuales forman una gruesa corriente que cruza la órbita terrestre y producen una lluvia anual de meteoros\footnote{Esta corriente no es visible a ojo desnudo, pues choca contra el hemisferio de la Tierra en que es de día. La corriente fue descubierta mediante técnicas de radar usando el radiotelecopio de Jodrell Bank \index{Jodrell Bank} en la década de 1950}. Por lo tanto, tal como concluyó  Kresák, el proyectil de Tunguska\index{Tunguska} podría ser un fragmento del cometa Encke\index{Encke (cometa)}. El hecho de que la densidad de los cometas es baja comparada con la  de los asteroides puede explicar mejor  porqué la explosión  desintegró totalmente el proyectil de Tunguska\index{Tunguska}. Si esta asociación es correcta, la velocidad del fragmento de cometa que produjo la explosión de Tunguska\index{Tunguska} debió ser igual a la velocidad con que ingresa a la atmósfera la lluvia de meteoros, es decir a  una velocidad  $v\approx$ 33 km/s. Si adoptamos un valor intermedio  de $E=50$ Megatones ($6.3 \,\, 10^{23}$ ergios) para la energía de la explosión, con la fórmula de la energía cinética,  calculamos la masa cometaria  $m=\frac{2 E}{v^{2}}$, la cual  resulta del orden de 400000 toneladas. Dado que la densidad de los cometas es  $\approx 1 $ gr cm$^{3}$, el tamaño del fragmento de cometa  que explotó en Tunguska\index{Tunguska} fue del orden de 100 metros, es decir como un estadio de fútbol. Parece un cuerpo demasiado chico en relación con el daño que produjo. Sin embargo recordemos que lo que mata no es  solo el plomo de la bala,  sino el plomo a una alta velocidad.

Nos resta analizar el fenómeno de las noches blancas \it{anómalas} \rm \footnote{aquí no nos referimos a las noches blancas normales que se observan en verano en el Norte a latitudes geográficas altas.} que se observaron en Euroasia y que evidentemente están relacionadas con el evento de Tunguska\index{Tunguska}. Una  explicación es que  partículas de hielo y polvo, ubicadas fuera del cono de sombra de la Tierra, reflejaban difusamente la luz solar hacia la superficie oscura de la Tierra. El problema es que no conocemos bien los procesos físicos y dinámicos que expliquen la presencia de las partículas de hielo y polvo a tan alta altura. Una posibilidad es que como consecuencia de la explosión se eyectó hacia la alta atmósfera una gran cantidad de polvo y vapor de agua. A las bajas temperaturas de la alta atmósfera, el vapor de agua se condensa en torno a las partículas de polvo, formando cristales de hielo. En la sección anterior,  simulamos  el proceso de eyección de granos de polvo por una explosión de 50 Mt y encontramos que   las partículas se precipitan sobre la zona de la explosión  dentro de un área de a lo sumo $\approx 1000$ km de radio (ver fig. \ref{SurgentePolvo}). Este tipo de explosiones tiene solo efectos locales, y no planetarios. Además, resulta difícil entender que en  24 hs 
parte del material eyectado  haya recorrido 5000 km hacia el oeste para iluminar Londres\index{Londres},  a modo de muchas Lunas llenas. Cabe destacar que la luminiscencia nocturna también se vio al otro lado del Atlántico en Norteamérica\index{Norteamérica}. A fin de salvar esa objeción, el científico ruso Zotkin\index{Zotkin I.T.} propuso en 1961 que el cuerpo que  impactó en Tunguska\index{Tunguska}  fue un fragmento de cometa  que poseía una extensa cola de partículas de hielo y polvo. Una parte de la cola interaccionó con la atmósfera durante al menos un día. La hipótesis de Zotkin, planteada originalmente en 1961,  se consolidó aún más con el vínculo establecido por Kresák\index{Kresák L.} en 1978, entre el cuerpo que explotó en Tunguska\index{Tunguska},  la corriente de meteoros de Beta Tauridas\index{Tauridas Beta (corriente de meteoros)}  y el cometa Encke\index{Encke (cometa)}.

Como la corriente de partículas cometarias  provenía de posiciones del cielo cercanas a la del Sol, el acercamiento y las desintegraciones de esas pequeñas partículas en la atmósfera  ocurrían a pleno día y por lo tanto no eran visibles a simple vista. El flujo restante de partículas de la cola cometaria que no fue interceptado por la Tierra siguió su camino con una velocidad de $\approx$ 33 km/s. Otro hecho interesante que nos ayuda a inferir la naturaleza del fenómeno es que varios observatorios del mundo notaron una significativa  pérdida de transparencia de la atmósfera, y otras anomalías ópticas, desde al menos el día previo al evento de Tunguska\index{Tunguska}\footnote{En el artículo  de N.V. Vasilyev \index{Vasilyev N.V.} \cite{Vasilyev}, pueden encontrarse las referencias originales sobre este hecho y otros relacionados}. También se observaron inusualmente intensos y prolongados halos solares. Por lo tanto, el cuerpo que  chocó en Tunguska\index{Tunguska} fue la parte más densa  de una extensa nube o coma. Como consecuencia de la interacción de las partículas de polvo y hielo de la coma y cola cometaria con la alta atmósfera (mesósfera), una gran cantidad de polvo y vapor de agua debe haber quedado en suspensión a grandes alturas. El vapor de agua vuelve a condensarse sobre los granos de polvo, formando cristales de hielo. 

  En el año 1953,  Edward George  Bowen\index{Bowen E.G.}, físico y radioastrónomo  de origen Galés\index{Gales}, encontró que  las fechas de los máximos  de lluvias del hemisferio Norte se correspondían con las fechas de los máximos registrados en el hemisferio Sur, fenómeno difícil de explicar por razones climatológicas. Él encuentra que este efecto está estrechamente relacionado con las lluvias más intensas de meteoros. La explicación sería la siguiente. El fino polvo cósmico frenado y temporalmente retenido 
en la alta atmósfera se precipita  al cabo de un cierto tiempo sobre la baja atmósfera con efectos meteorológicos. Esas partículas de polvo al penetrar las nubes de la troposfera actúan como núcleos de condensación del vapor, estimulando a la nubes a producir lluvias. Las moléculas del vapor de agua que se adhieren a la superficie de un grano de polvo pueden unirse con mayor facilidad para formar gotas de agua. Este fenómeno también se manifestó después del evento de Tunguska\index{Tunguska} con intensas precipitaciones, registradas en toda Europa entre dos y tres semanas después del evento. Es impresionante que partículas interplanetarias o rayos cósmicos que llegan a la Tierra  de regiones espaciales lejanas influyan sobre procesos vitales como las lluvias.

 El fenómeno lumínico anómalo debe también haberse observado en la región polar sur, la cual  en esa época del año estaba en plena noche polar de seis meses. En efecto, en las anotaciones de la expedición Antártica de Shackleton\index{Shackleton}, se menciona con especial énfasis la ocurrencia de auroras notablemente intensas en la fecha del 30 de junio de 1908. Esto refuerza la idea de que el evento de Tunguska\index{Tunguska} formó parte de un fenómeno global que involucró a toda la atmósfera y comenzó cierto tiempo antes de la explosión de Tunguska\index{Tunguska}, aunque con afectos no catastróficos exceptuando  a aquellos que afectaron a la región de Tunguska\index{Tunguska}.
 
  En una carta publicada el 2 de junio por el London \it{Times},  \rm  una lectora que vio a la medianoche del 1 de junio el sorprendente brillo del cielo reclamó que alguien diera una explicación de tan extraordinario fenómeno. El requerimiento de la lectora del London \it{Times} \rm aún sigue vigente. \index{London Times}  Esta apreciación coincide con el comentario  que hace el astrónomo australiano Duncan Steel\index{Steel D.} \cite{Steel} sobre que  no se ha explicado propiamente si el fenómeno de las noches blancas fue debido a la reflexión difusa de la luz solar por granos de polvo a  grandes alturas, alguna forma de luminiscencia atmosférica (airglow en inglés) inducida 
\footnote{La luz solar y partículas muy energéticas como los rayos cósmicos excitan e ionizan las moléculas de la alta atmósfera contra  las cuales chocan. También, las lluvias de meteoros en interacción con la alta atmósfera dejan detrás trazos  de gas ionizado. Las moléculas y átomos afectados se  desexcitan  y recombinan rápidamente   emitiendo  luz de diferentes colores. Las auroras polares son un ejemplo de este fenómeno de luminiscencia nocturna, por el cual el cielo nunca está completamente oscuro.} o alguna causa física no conocida. Einstein\index{Einstein A.} dijo que los científicos son  oportunistas (claro que en sentido positivo). A más de 110 años del caso Tungusta, aún éste guarda misterios y la oportunidad de revelarlos. 
A través de un camino inesperado, Michael Kelley\index{Kelley M.} junto a otros científicos de la Universidad de Cornell encuentran un vínculo con el evento de Tunguska\index{Tunguska}.  Ellos descubren en el año 2009 que la eyección de vapor de agua en la termósfera por el transbordador espacial, una acción similar a la de un cometa, produce partículas de hielo que se trasladan miles de kilómetros hacia regiones del Ártico\index{Ártico} y de la Antártida\index{Antártida}, formando nubes luminiscentes. Ello constituye otra evidencia de que probablemente, el  30 de junio de 1908,  parte de  un cometa chocó contra la Tierra, originando entre otros efectos la explosión  de Tunguska\index{Tunguska}.

Si bien no hay evidencias físicas de que hayan ocurrido otros sucesos tipo Tunguska\index{Tunguska} en tiempos históricos,  hay relatos que son muy sugestivos. Para  las civilizaciones antiguas, el cielo es el lugar de las deidades. De modo que acontecimientos como eclipses o lluvias de meteoritos, impresionaban como señales divinas y por lo tanto eran descriptos en muchos casos  desde una perspectiva  moralizante y alegórica. Un caso interesante es el relato bíblico de la súbita destrucción de las ciudades de  Sodoma y Gomorra\index{Sodoma y Gomorra}: ``Lo mismo sucedió en tiempos de Lot: comían y bebían, compraban y vendían, sembraban y edificaban. Pero, el día en que Lot salió de Sodoma, llovió del cielo fuego y azufre y acabó con todos.
(Lucas 17:28-29)''. Esas ciudades se encontraban al borde del Mar Muerto. Excavaciones arqueológicas de la región han revelado restos de ciudades y evidencias de grandes incendios entre los escombros que coinciden con el relato bíblico. La pregunta que se han hecho algunos investigadores es si estamos frente a un caso similar al de Tunguska\index{Tunguska}. 

\ \

\ \

\ \

\ \

\ \

\ \

\ \

\ \

\subsection{Las noches luminiscentes relacionadas con el evento Tunguska: una explicación posible} \index{Tunguska}

\begin{quote}

\small \it{Quizás nunca hubo puesta del sol más hermosa, ni se vio más azulado cielo. El astro radiante pareció amortajarse en un lecho de   púrpura y oro. Un disco rojo se ocultó en el horizonte; pero las estrellas no se mostraron. ¡No llegó la noche! Al día solar sucedió otro cometario y lunar, iluminado por intensa luz análoga a la de las auroras boreales, pero más viva,... Este foco cometario salió por la parte de Oriente casi al mismo tiempo que la Luna llena, que pareció ascender con él en los cielos,... La noche siguió alumbrada por el extraño resplandor cometario que se cernía en los cielos, por la lluvia meteoros que aun duraba y por los numerosos incendios.}\rm 

Camile Flammarion \index{Flammarion C.} (1842-1925) en su novela ``El fin del mundo'' publicada en el año 1894.
\footnote{Flammarion fue un famoso astrónomo francés, prolífico divulgador científico y escritor de ciencia ficción. La trama de la novela de marras transcurre en el siglo 25, donde el encuentro  con un gran cometa amenaza destruir al mundo.}
\end{quote}

En la sección anterior, hemos mencionado que un extraño resplandor del cielo nocturno fue observado en gran parte de Europa en la fecha del evento de Tunguska\index{Tunguska}, y en unas pocas noches siguientes, y que dicho fenómeno hablaba en favor de que el cuerpo que impactó en Tunguska\index{Tunguska} fue un pequeño núcleo cometario, acompañado por una extensa y poco densa nube de polvo que rodeó la Tierra. En esta sección intentaremos ahondar un poco más en ese fenómeno óptico,  también referido como ``noches blancas anómalas''. 

La mayor parte de la luz que llega a nuestros ojos no lo hace directamente de las fuentes, sino en forma indirecta. Cuando miramos un árbol, vemos luz solar reflejada difusamente. Cuando miramos una nube en el cielo, vemos luz solar dispersada. La dispersión de la luz va acompañada generalmente por absorción. Una hoja de árbol se ve verde porque la hoja dispersa preferentemente la luz verde y absorbe la luz de otros colores. 

Nosotros pensamos que los granos micrométricos de polvo de la nube cometaria suspendidos en la alta atmósfera  dispersaron la luz solar, ocasionando las noches blancas. Cuando una onda electromagnética de longitud $\lambda$ se encuentra con una pequeña partícula, con tamaño del orden de $\lambda$, la onda tiende a bordear la partícula desviándose  de su trayectoria original,  fenómeno conocido  como difracción, lo cual explica la dispersión de la luz por pequeñas partículas sólidas. Gustav Mie \index{Mie G.} en 1908 resolvió en forma exacta este problema aplicando las ecuaciones electromagnéticas de Maxwell con condiciones de contorno a esferas homogéneas.

\begin{figure}
\includegraphics[scale=0.65]{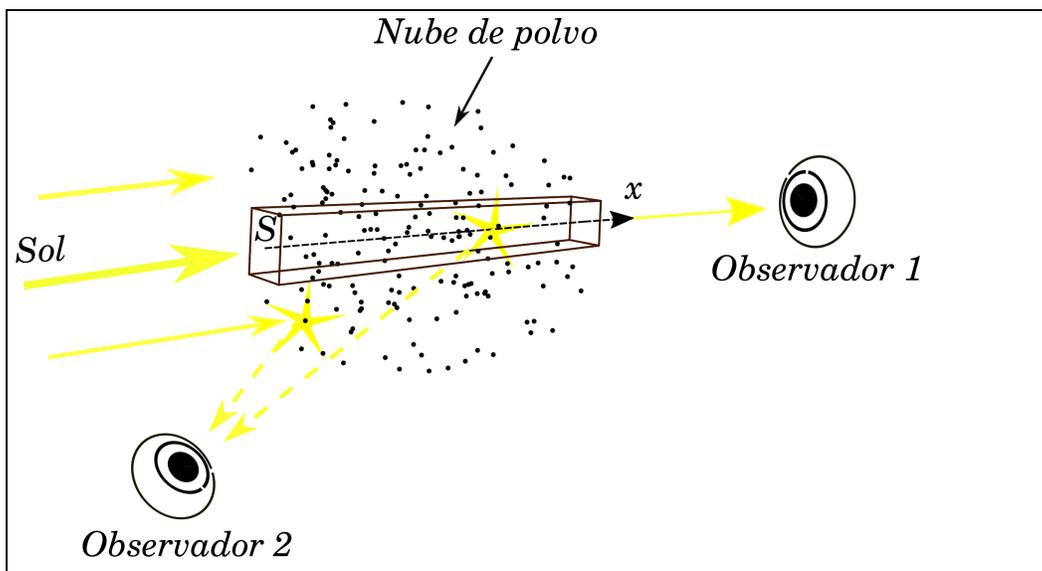} 
\caption{Nube de polvo iluminada por el Sol. El observador 2, que tiene al Sol detrás,  ve brillar la nube por la luz solar reflejada por la nube. Para el observador 1, la nube se interpone entre él y el Sol. El observador 1 ve disminuido el brillo del Sol pues una parte de los rayos de luz dirigidos hacia él son absorbidos y otra parte desviados por los granos de polvo.}
\label{Extincion}
\end{figure}

Ahora estudiaremos como  dos observadores verían,   desde posiciones distintas,   una nube de polvo iluminada por el Sol (ver fig. \ref{Extincion}). Consideremos primero el caso del observador 1, al cual llegan los rayos de luz solar en forma directa. Cada  grano de polvo de radio  $a$  interpone una superficie perpendicular a los rayos de luz   dada por $\pi a^{2}$,  la cual es llamada sección geométrica. Cuando  $a$ es del orden de la longitud de onda $\lambda$, los efectos ondulatorios de la luz son importantes y la energía de los rayos de luz interceptados por el grano de polvo es proporcional $\pi a^{2} Q_{ex}$, donde $Q_{ex}$ es un factor efectivo de extinción que depende de  $\lambda$. A su vez, $Q_{ex}=Q_{s}+ Q_{a}$, donde $Q_{s}$ representa la parte de la radiación  dispersada en todas direcciones por la partícula de polvo   y $Q_{a}$ representa la parte absorbida por la partícula.

Los rayos de luz solar, que forman un frente plano,  inciden perpendicularmente a la superficie $S$ del paralelepípedo rectangular y en su trayecto son en parte absorbidos y dispersados por las partículas de polvo contenidas en el paralelepípedo (ver fig. \ref{Extincion}). El eje $x$ del  paralelepípedo es paralelo a los rayos solares incidentes. El origen de $x$ $(x=0)$  se ubica sobre la cara $S$ y la cara opuesta a $S$ yace en $x=L$, donde $L$ es el diámetro de la nube. El eje $x$ apunta al observador 1. Dividiremos el paralelepípedo en $n$ rebanadas perpendiculares al eje $x$ y de espesor $dx$, es decir $dx=\frac{L}{n}$. Por lo tanto cada rebanada tiene un volumen $dV=S\, dx$. 

Suponemos que la nube de polvo está en las cercanías de la Tierra, y por lo tanto el brillo del Sol $B_{\odot}$ (energía por segundo y por unidad de superficie) que ilumina la superficie $S$ del paralelepípedo es $B_{\odot} =\frac{L_{\odot}}{4 \pi d_{ST}^{2}}$, donde $L_{\odot}$ es la luminosidad del Sol y $d_{ST}$ la distancia Sol-Tierra. Por lo tanto,  el flujo de energía solar  que ingresa a la primera rebanada del paralelepípedo (en $x=0$) es $F_{0}=B_{\odot}\, S$. La energía radiante que los granos de polvo de la rebanada 1 absorben y dispersan es $\frac{F_{0}}{S}  n_{p}\pi a^{2} Q_{ex}$, donde $n_{p}$ $(= \rho_{p} dV=\rho_{p} \, S\, dx)$ es el número de granos de polvo contenido  en la rebanada y $\rho_{p}$ es la densidad numérica de granos de  la nube de polvo. Por lo tanto, el flujo restante de rayos que  no fueron absorbidos ni desviados es $F_{0}-F_{0}  \pi a^{2} Q_{ex} \rho_{p} dx$. Entonces, el flujo de energía solar que entra en la rebanada 2 es
\begin{equation}
F_{1}=F_{0} (1-\pi a^{2} Q_{ex} \rho_{p} dx).
\label{extincion1}
\end{equation}
El mismo procedimiento se aplica a la  rebanada 2, pero ahora el flujo que ingresa es $F_{1}$ y el flujo restante que sale e ingresa en la rebanada 3 es
$F_{2}=F_{1} (1-\pi a^{2} Q_{ex} \rho_{p} dx)$. Si reemplazamos en esta última ecuación la expresión para $F_{1}$ dada por la ecuación (\ref{extincion1}), obtenemos
\begin{equation}
F_{2}=F_{0} (1-\pi a^{2} Q_{ex} \rho_{p}  dx)^{2}.
\label{extincion2}
\end{equation}
De las ecuaciones (\ref{extincion1}) y (\ref{extincion2}), inferimos que 
\begin{equation}
F_{n}=F_{0} (1-\pi a^{2} Q_{ex} \rho_{p}  dx)^{n}.
\label{extincion3}
\end{equation}
$F_{n}$ es el flujo que emerge de la última rebanada y llega al observador 1. Dado que $dx=\frac{L}{n}$, la ecuación  (\ref{extincion3}) 
puede escribirse  $F_{n}=F_{0} (1-\frac{\tau}{n})^{n}$, siendo $ \tau=\pi a^{2} Q_{ex} \rho_{p} L$, número a-dimensional llamado profundidad óptica que indica el grado de  transparencia u opacidad de la nube.
Se puede demostrar que 
\begin{equation}
\lim_{n \rightarrow \infty} (1-  \frac{\tau }{n})^{n}=e^{-\tau}. \nonumber\\
\label{tau}
\end{equation}

Entonces, el brillo que recibe el observador 1 por la luz del Sol que ha dejado pasar la nube de polvo puede escribirse:

\begin{equation}
B_{n}=\frac{F_{n}}{S}=B_{\odot} e^{-\tau}.
\end{equation}
Si $\tau>>1$, el Sol es totalmente oscurecido.

La nube de polvo que habría rodeado al cuerpo del evento Tunguska\index{Tunguska} se acerco a la Tierra de día. Por lo tanto,  la nube de polvo se interpuso entre el Sol y la Tierra. Esta situación se corresponde con la del observador 1 representado en la fig. \ref{Extincion} \footnote{Note que en la cita de la novela de Flammarion, \index{Flammarion C.} el gran cometa se acerca desde una dirección opuesta a la de nuestro caso. Allí la Tierra se encuentra entre el Sol y la nube cometaria, que es el caso del observador 2 de la fig. \ref{Extincion} }. Si bien la transparencia atmosférica habría disminuido previamente al evento, el Sol no fue ocultado ni mucho menos. Esto indica que la nube de polvo era ópticamente  delgada ($\tau<<1$).

Ahora calcularemos el brillo que recibe el observador 2 de la nube de polvo iluminada por el Sol. La energía que entra al paralelepípedo  es $F_{0}$ y la que sale del paralelepípedo  en la dirección $x$ es  $F_{n}=F_{0} e^{-\tau}$. Por lo tanto, la energía que se absorbió y dispersó en el paralelepípedo es $\Delta F= F_{0}-F_{0} e^{-\tau}=F_{0} (1- e^{-\tau})$. Si $\tau<<1$, $e^{-\tau}\approx 1-\tau$ y en consecuencia $\Delta F =F_{0} \tau= B_{\odot} S \tau$. Si ajustamos el volumen de la nube de polvo con  $m$ paralelepípedos orientados en  la misma forma que el paralelepípedo representado en la fig. \ref{Extincion}, y denotamos la contribución de cada paralelepípedo $i$ por $\Delta F_{i} = B_{\odot} S_{i} \tau_{i}$, la energía total que la nube extrae del flujo solar es $F_{t}=\sum_{1}^{m} \Delta F_{i}=B_{\odot} \sum_{1}^{m}  S_{i} \tau_{i}$. La profundidad óptica del polvo contenido en el paralelepípedo $i$ es  $\tau_{i}=\pi a^{2} Q_{ex} \rho_{p} L_{i}$, donde el valor de  $L_{i}$ es menor o a lo sumo igual el diámetro de la nube. Así, $F_{t}=B_{\odot} \pi a^{2} Q_{ex} \rho_{p} \sum_{1}^{m}  S_{i} L_{i}$, pero $\sum_{1}^{m}  S_{i} L_{i}$ es igual al volumen de la nube,  que denotaremos $V_{c}$. Si además recordamos que $Q_{ex}=Q_{s}+ Q_{a}$, podemos escribir $ F_{t}=F_{a}+F_{s}$ donde

\begin{equation}
 F_{a}=B_{\odot} \pi a^{2} Q_{a} \rho_{p} V_{c} 
 \label{Flujo1}
\end{equation}

\begin{equation}
 F_{s}=B_{\odot} \pi a^{2} Q_{s} \rho_{p} V_{c} 
 \label{Flujo2}
\end{equation}

La fórmula (\ref{Flujo1}) corresponde a la energía de la radiación absorbida por los granos de polvo. Esa energía  es reirradiada a longitudes de ondas largas que son   invisibles al ojo. De modo que la única radiación que ve el observador 2 (el cual se encuentra de espaldas al Sol) es aquella que es desviada o dispersada en su dirección por los granos de polvo. $Q_{s}$ es una función compleja que depende de la relación $\frac{a}{\lambda}$ y del ángulo de dispersión
 Para nuestros propósitos es suficiente suponer que la dispersión es isotrópica. En la deducción de las fórmulas (\ref{Flujo1}) y (\ref{Flujo2}) se haya implícita la aproximación de que   una vez dispersado un rayo éste no vuelve a  sufrir otro evento de dispersión o absorción. Si la distancia entre el observador 2 y el centro de la nube es $d_{c}$ y $\frac{L}{d_{c}}<<1$, el brillo de la nube $B_{c}$ para el observador 2 es $B_{c}=\frac{F_{s}}{4 \pi d_{c}^{2}}$ y usando (\ref{Flujo2}) 
\begin{equation}
B_{c}=\frac{B_{\odot} \pi a^{2} Q_{s} \rho_{p} V_{c}}{4 \pi d_{c}^{2}}
\label{Bcloud}
\end{equation}

\begin{figure}
\includegraphics[scale=0.65]{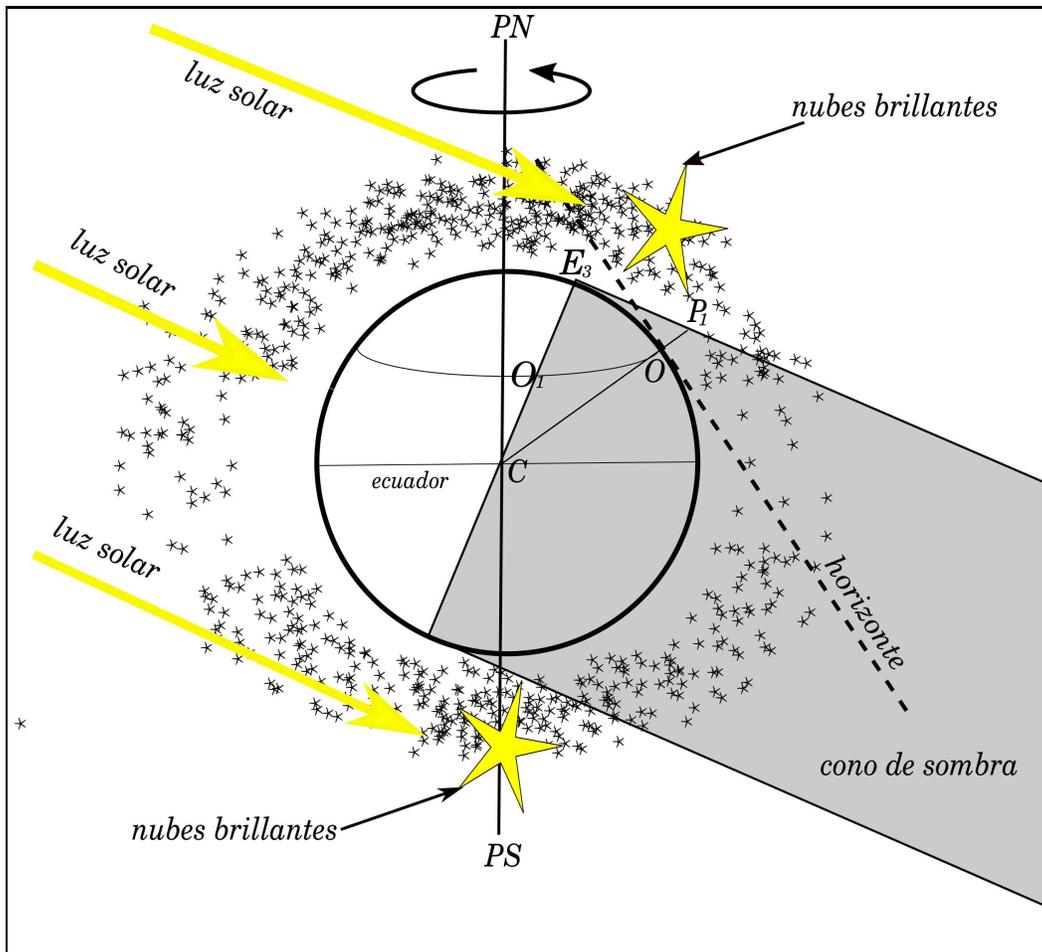} 
\caption{Modelo para explicar las noches blancas anómalas observadas después del evento de Tunguska\index{Tunguska}.}
\label{NochesBlancas}
\end{figure}

\begin{figure}
\includegraphics[scale=0.65]{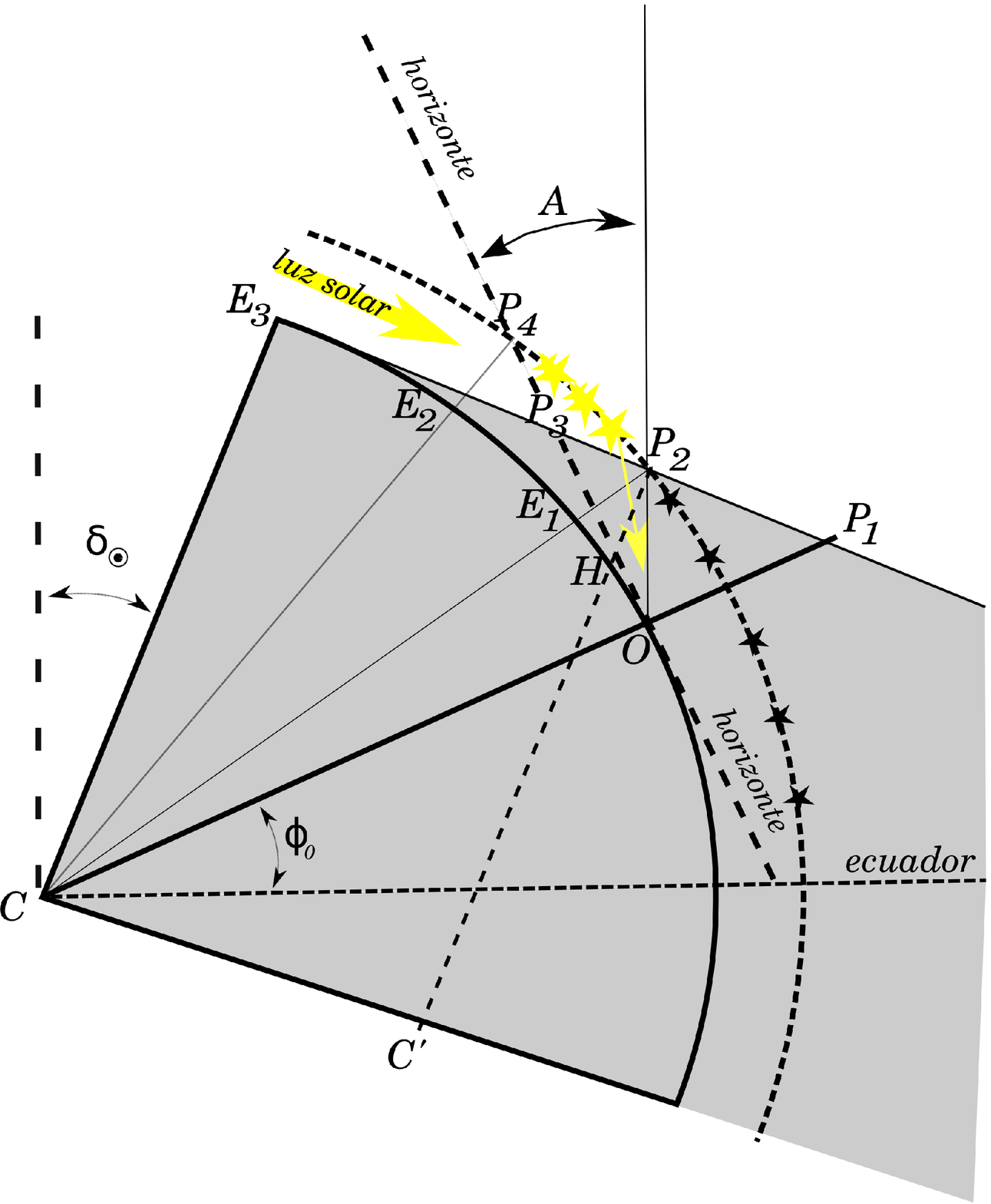} 
\caption{Modelo para explicar las noches blancas anómalas observadas después del evento de Tunguska\index{Tunguska}.}
\label{NochesBlancas2}
\end{figure}

\subsubsection{Modelo sobre la envoltura de polvo que cubrió la Tierra.} 

Los granos de polvo no son en general eléctricamente neutros y por lo tanto al ingresar a la magnetosfera terrestre interaccionan con ella. Sin embargo, hemos calculado en la sección 5 que las trayectorias de los granos de polvo  no son esencialmente alteradas por el campo magnético terrestre y que una corriente de pequeñas partículas de polvo que se dirige hacia el cuerpo de la Tierra choca directamente contra la atmósfera (ver fig.\ref{CorrienteDust}). Si suponemos que el fragmento de cometa que chocó en Tunguska\index{Tunguska} fue acompañado por una corriente de granos de polvo (con probablemente algunas concentraciones discretas) y que  la Tierra  quedó sumergida en dicha corriente por al menos un día, entonces las finas partículas de polvo que chocaron la atmósfera se frenaron y formaron una capa de polvo en la alta atmósfera que rodeó a la Tierra. En un día, la Tierra completa su rotación y toda la atmósfera fue expuesta al flujo de partículas de polvo.
 
La fig. \ref{NochesBlancas} representa esquemáticamente nuestro modelo para explicar el fenómeno de las noches blancas. En la figura de marras, suponemos que el observador $O$  está en la medianoche y por lo tanto    el Sol se encuentra sobre el mismo plano meridiano del observador  pero iluminando el otro hemisferio de la Tierra, naturalmente. El observador $O_{1}$, que se encuentra sobre la línea que separa el hemisferio iluminado del hemisferio que está dentro del cono de sombra, está entrando en la noche. Ahora, nuestro propósito  es calcular a qué altura sobre $O$ se encuentra el borde del cono de sombra; para ello debemos obtener la distancia $OP_{1}$, a la que llamaremos $h_{z}$.  El punto $C$, que indica el centro de la Tierra, con los puntos $E_{3}$ y $P_{1}$ forman un triángulo rectángulo, donde el ángulo con vértice $E_{3}$ es recto. El ángulo entre el lado $CO$ y el plano ecuatorial de la Tierra es la latitud geográfica $\phi_{O}$ del observador $O$. El ángulo con que inciden los rayos solares con respecto al plano ecuatorial es igual a la declinación $\delta_{\odot}$ del Sol. El ángulo entre el lado $CE_{3}$ y  el eje polar de rotación es igual a $\delta_{\odot}$. Al ángulo suma de $\delta_{\odot}$ y $\phi_{O}$ le  llamaremos $i$, de modo que de aquí en adelante consideraremos que 
\begin{equation}
i=\phi_{O}+\delta_{\odot}
\label{inclinacion}
\end{equation}
Ahora estamos en condiciones de determinar la altura $h_{z}(=OP_{1})$, pues tenemos que el ángulo entre el lado $CE_{3}$ y el lado $CP_{1}$ es igual a $90-i$ y $CE_{3}=R_{T}$, donde  
$R_{T}$ es el radio de la Tierra. Por lo tanto, resolviendo el triángulo $CE_{3}P_{1}$, 
$CP_{1} = \frac{CE_{3}}{cos(90^{\circ}-i)}$ y dado que $CO=CE_{3}=R_{T}$, la altura  es 
\begin{equation}
h_{z}=OP_{1}= CP_{1}-CO=(\frac{1}{cos(90^{\circ}-i)}-1) R_{T}.
\end{equation}
Como el fenómeno ocurrió en el comienzo del verano en el hemisferio norte, $\delta_{\odot} \approx 23^{\circ}$ (la latitud del Trópico de Cáncer), y adoptando  $\phi_{O}=50^{\circ}$ que es una latitud representativa para Europa, $i=73^{\circ}$ y  $h_{z}=290$ km. En consecuencia, si la altura de la capa de polvo y hielo sobre la superficie de la Tierra fue menor que 290 km, como seguramente fue el caso, solo el sector de la capa recostado hacia el norte podía recibir la luz solar y reflejar difusamente parte de esa luz  hacia el observador, produciendo el resplandor de las noches blancas. En cambio el sector ubicado entre el Zenit del observador y el sur estaba dentro del cono de sombra. En la fig. \ref{NochesBlancas}, hemos representado por claridad en el dibujo a la capa de polvo y hielo por encima del punto $P_{1}$. 
 
 Con el fin de calcular la altura real de las nubes brillantes sobre la superficie terrestre, hemos representado en la fig. \ref{NochesBlancas2} un sector de la fig. \ref{NochesBlancas} con mayor detalle. En la nueva figura, la altura sobre la superficie terrestre del punto  $P_{2}$,  intersección del borde del cono de sombra (en el plano meridiano) con el borde superior de la capa de polvo y hielo,  nos da  la altura superior de dicha capa, a la que llamaremos $h_{c}$. Llamaremos altura angular $A(P_{2})$ al ángulo   entre el horizonte y la dirección a las nubes brillantes que se vieron a mayor altura en el cielo en el plano meridiano, es decir hacia el punto $P_{2}$ (ver fig. \ref{NochesBlancas2}). Para calcular el valor de $h_{c}$, debemos resolver primero el triángulo $OP_{1}P_{2}$, cuyo  ángulo con vértice en $P_{1}$ es igual a $i$. En efecto, ese ángulo es parte del triángulo rectángulo $CP_{1}E_{3}$, cuyo ángulo con vértice en $C$ es 
$90^{\circ}- i$. El ángulo con vértice $O$ es igual a $(90^{\circ}-A(P_{2}))$. Por lo tanto, llamando  $\alpha$ al ángulo restante, el cual  tiene vértice en $P_{2}$,  resulta que  $\alpha = 90 + A(P_{2})-i$. Además, sabemos que  el lado $OP_{1}=h_{z}=290$ km. Aplicando el teorema del seno, obtenemos que 
\begin{equation}
OP_{2}=\frac{sen(i)}{sen \, \alpha} h_{z}.
\label{OP2}
\end{equation}
El mayor brillo del cielo nocturno se vio en Alemania\index{Alemania}. Las partes más intensas del brillo se ubicaron sobre el horizonte norte hasta alturas angulares entre $20^{\circ}$ y $40^{\circ}$. Por lo tanto, adoptando $A(P_{2})=40^{\circ}$ y usando la ecuación que hemos obtenido para  $OP_{2}$, obtenemos $OP_{2}=301.3$ km.

Ahora estamos en condiciones de calcular el valor de $h_{c}$. Para ello debemos resolver el triángulo $COP_{2}$, del cual conocemos el lado $CO=R_{T}$ y el lado $OP_{2}$ dado por la ecuación (\ref{OP2}) y su ángulo interno con vértice $O$ que es igual a $90^{\circ}+A(P_{2})$. Aplicando el teorema del coseno, $CP_{2}^{2}=OP_{2}^{2}+CO^{2}-2\, OP_{2}\times CO \times cos(90^{\circ}-A(P_{2}))$ y teniendo en cuenta que  $CP_{2}=CE_{1} + h_{c} = R_{T} + h_{c}$, obtenemos  
$h_{c}=E_{1} P_{2}=\sqrt{OP_{2}^{2} +  R_{T}^{2}+ 2 R_{T} \times OP_{2} \times cos(90^{\circ}+A(P_{2}))}- R_{T}$. Por lo tanto, $h_{c}=217$ km.

\begin{figure}
\includegraphics[scale=0.65]{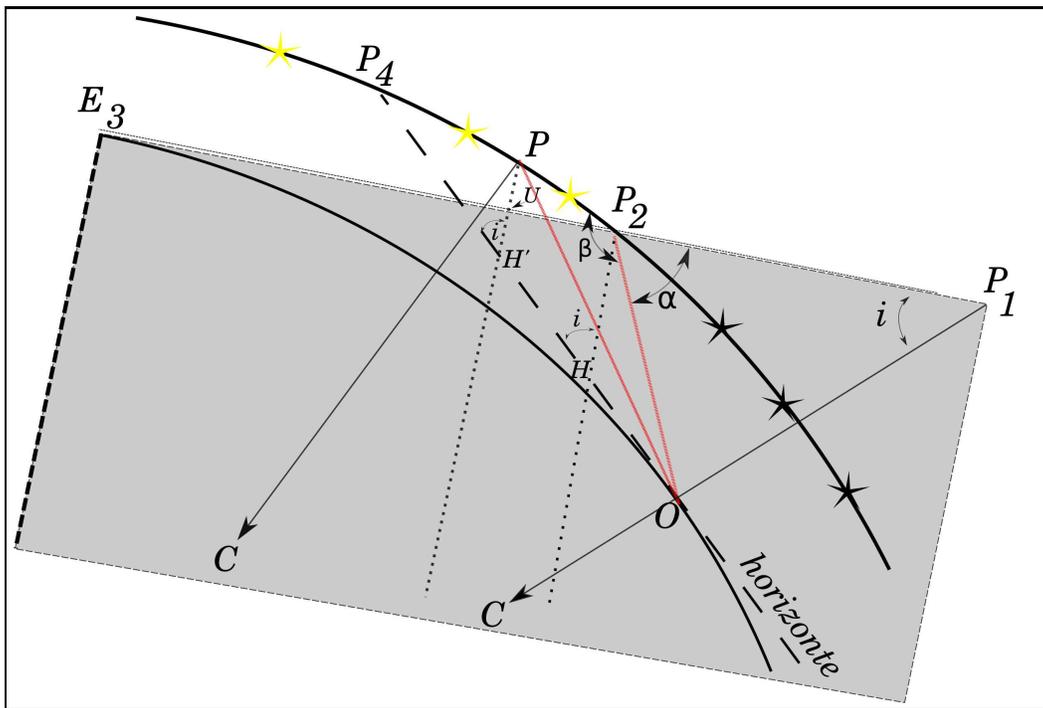} 
\caption{Sector de la fig. \ref{NochesBlancas2} aumentado a fin de facilitar la visualización de las relaciones  geométricas.}
\label{NochesBlancas3}
\end{figure} 

Haciendo $A=0$ en las ecuaciones de arriba, obtenemos la altura del  punto $P_{3}$ (ver fig. \ref{NochesBlancas2}), que es la altura mínima que pudo tener el polvo iluminado por el Sol. A la latitud de $50^{\circ}$, esa altura mínima es de $\approx 70$ km y en consecuencia la capa de polvo yacía entre 70 y 217 km. Ello  es consistente con que la capa de polvo estaba ubicada en la mesósfera,  donde los pequeños meteoritos se desintegran produciendo las estrellas fugaces. Para latitudes menores que $35^{\circ}$, la altura del punto $P_{3}$ supera la altura de 217 km;  es decir, la capa de polvo iluminada por el Sol queda  debajo del horizonte del observador. Por dicha razón, el fenómeno de una medianoche luminosa no pudo observarse  a latitudes intermedias. Sin embargo, si la capa de polvo envolvió uniformemente la Tierra, debieron observarse en todo el Planeta crepúsculos más brillantes de lo normal. 

En base a la fig. \ref{NochesBlancas2}, hemos obtenido una sección de la capa de polvo visible al observador que resulta de la intersección del plano meridional del observador (plano del dibujo) y la capa de polvo. A fin de determinar la porción total de la capa de polvo que el observador ve en el cielo, haremos cortes transversales al plano de la fig. \ref{NochesBlancas2}, mediante planos perpendiculares al eje del cono de sombra. La recta que pasa por los puntos $P_{2}$ y $H$ (línea de puntos en la fig. \ref{NochesBlancas3}) es la intersección de uno de esos planos, los cuales son  perpendiculares al eje del cono sombra  y perpendiculares al plano de la fig. \ref{NochesBlancas3}). El horizonte del observador es tangente a la esfera terrestre en la posición del observador y  perpendicular al plano del dibujo, la intersección de ambos planos es dada por la línea de trazos de la fig. \ref{NochesBlancas3}. El cruce de  esa línea de trazos con la línea de puntos  es representado  por el punto $H$. 

La resolución del triángulo $OP_{2}H$ nos permite determinar  el lado $HP_{2}$, al cual denominaremos $e$. El ángulo $OP_{2}H$ es complementario del ángulo $\alpha$ (ver fig.  \ref{NochesBlancas3}) y por lo tanto es igual a $(90^{\circ}-\alpha)$.  Dado que también conocemos el ángulo $HOP_{2}$ ($=A(P_{2})$),  el ángulo restante del triángulo es igual a $180^{\circ}-i$. Aplicando el teorema del seno, obtenemos que $e=HP_{2}=\frac{OP_{2} sen\, A(P_{2})}{sen\, i}$ y teniendo en cuenta que $OP_{2}=301.3$ km
\begin{equation}
e=HP_{2}=301.3 \, \frac{sen\, A(P_{2})}{sen \, i}.
\end{equation}
 Supongamos que el observador, ubicado en $O$, mira hacia el punto $P$  de la capa de polvo iluminada que tiene la altura angular $A(P)$ con respecto al horizonte, entonces  la distancia entre $O$ y $P$ puede obtenerse resolviendo el triángulo $OPC$, donde $CO=R_{t}$ y $CP=R_{t}+h_{c}$ y el ángulo con vértice en $O$ es $90^{\circ}+A(P)$. Aplicando el teorema del coseno, tenemos $(R_{T}+h_{c})^{2}=R_{T}^{2} + OP^{2} - 2 R_{T}\, OP\, cos (90^{\circ}+A(P))$, donde  nuestra incógnita $OP$ forma una ecuación cuadrática cuya solución es 
 \begin{equation}
 OP= R_{T}cos (90^{\circ}+A(P)) \pm \frac{1}{2} \sqrt{(2 R_{T}\, cos (90^{\circ}+A(P)))^{2}+ 4(2 R_{T}\, h_{c}+ h_{c}^{2})}.
 \end{equation}
  
 Ahora nuestro objetivo es usar el triángulo $OPH'$ para calcular la longitud del lado $H'P$, al cual llamaremos $e'$. Note que el ángulo con vértice $H'$ es $180^{\circ}-i$ y el ángulo con vértice en $O$ es $A(P)$. Aplicando el teorema del seno, 
 \begin{equation}
 e'=H'P=\frac{OP sen\, A(P)}{sen\, i}.
 \end{equation}
 Similarmente, obtenemos el valor del lado $OH'$ al que llamaremos $y_{0}$:
 \begin{equation}
 y_{0}=O H'=\frac{sen\, (i-A(P))}{ sen\, A(P)} e'.
 \label{yOrigen}
 \end{equation}
 
 Otro dato que necesitamos conocer es la longitud del lado $PP_{2}$ del triángulo $POP_{2}$, para lo cual aplicamos a dicho triángulo el teorema del teorema del coseno, obteniendo
 
 \begin{equation}
 PP_{2}=\sqrt{OP^{2} + OP_{2}^{2} - 2\,\, OP \times OP_{2}\times cos (dA)},
 \end{equation}
 donde $dA=A(P_{2})-A(P)$. Aplicando el teorema del seno al mismo triángulo, obtenemos el valor del ángulo $\beta$:
 
 \begin{equation}
 \beta=180^{\circ} - arcsen (\frac{OP \,\, sen(dA)}{PP_{2}}).
 \end{equation}
  Note que el lado $PH'$ tiene una parte iluminada que denominaremos $e'_{i}=PU$ y una parte que yace en el cono de sombra que denominaremos $e'_{s}=UH'$. Por lo tanto, $e'=e'_{i}+e'_{s}$. Con el fin de obtener el valor de $e'_{i}$, resolvemos el triángulo rectángulo $P P_{2} U$, cuyo ángulo con vértice en $P_{2}$ es igual a $\alpha +\beta - 180^{\circ}$ y consecuentemente 
\begin{equation}
   e'_{i}= P P_{2}\, sen (\alpha +\beta - 180^{\circ}).
\end{equation}
 Similarmente, el lado $U P_{2}$, al que llamaremos $l'$, es dado por 
 \begin{equation}
  l'= U P_{2}= P P_{2}\, cos (\alpha +\beta - 180^{\circ}).
  \label{lP}
 \end{equation}
 
 \begin{figure}
\includegraphics[scale=0.65]{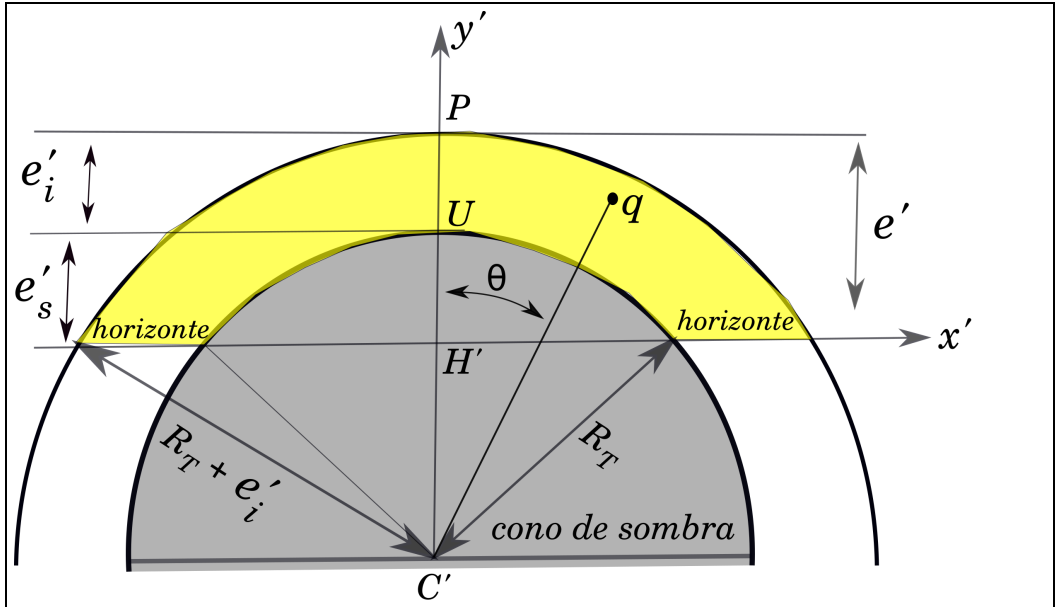} 
\caption{Corte perpendicular al plano del dibujo de la fig. \ref{NochesBlancas3} que genera la recta $PH'$ }
\label{NochesBlancas4}
\end{figure} 
 
 La fig. \ref{NochesBlancas4} muestra el corte perpendicular al plano del dibujo de la fig. \ref{NochesBlancas3} que genera la recta $PH'$. El plano del corte es a su vez perpendicular al eje del cono de sombra, donde la intersección de ambos es el punto $C'$.  La sección del corte representada en amarillo en la fig. \ref{NochesBlancas4} es la parte de la nube de polvo iluminada por el Sol que ve el observador $O$. Tomemos un punto $q$ de dicha sección y expresemos su posición en coordenadas Cartesianas $(x', y')$ con origen en $H'$. El eje $x'$ coincide con la recta que resulta de la intersección del plano de corte con el horizonte del observador y el eje  $y'$ coincide con la recta $PH'$. Por lo tanto,
\begin{eqnarray}
x' &=& \rho \, sen \,  \theta \nonumber\\
y' &= & \rho \, cos \, \theta - (R_{T}-e'_{s}), 
\label{xyprima}
\end{eqnarray}
donde $\rho$ es la distancia entre $C'$ y $q$, y $\theta$ es el ángulo entre la recta $C'q$ y la recta $C'P$. Las coordenadas polares $\rho, \theta $ de los puntos que pertenecen al sección amarillo (o sección  de la nube brillante) yacen en los intervalos 
\begin{eqnarray}
 R_{T} &  \leq &   \rho \leq R_{T} + e'_{i} \nonumber\\
  -\theta_{max} &  \leq &  \theta \leq \theta_{max},
  \label{dlP}
\end{eqnarray}
donde $\theta_{max}=arcos \frac{ R_{T} - e'_{s}}{ R_{T} + e'_{i}}$. Variando $\rho$ y $\theta $ dentro de dichos intervalos, las ecuaciones (\ref{xyprima}) dan las coordenadas $(x', y')$ de todos los puntos de la sección de la nube noctilucente. Dado que, para el observador terrestre, el plano del horizonte es el principal plano de referencia, transformaremos las coordenadas  $(x', y')$  en las coordenadas cartesianas $(x,y,z)$ de un sistema de referencia apoyado sobre el plano del horizonte del observador. Los ejes $x$ e $y$ yacen sobre el plano del horizonte y están centrados en $O$, el observador. El eje $x$ es paralelo al eje $x'$ y tiene el mismo sentido que este último. El eje $y$ coincide con la recta $O H'$ y apunta hacia el norte. El eje $z$ es perpendicular el plano del horizonte y apunta hacia el Zenit. Por lo tanto, las ecuaciones de transformación de coordenadas son
 \begin{eqnarray}
x &= & x'\nonumber\\
y &=&  y_{0} + y'\,cos\, i \nonumber\\
z &=& y'\, sen \,i,
\label{xyzH}
\end{eqnarray}
 donde  $(x', y')$, $y_{0}$ y  $i$ son dados por (\ref{xyprima}),  (\ref{yOrigen}) y (\ref{inclinacion}), respectivamente.
  
 A partir del hecho de que se podía leer un periódico a medianoche sin luz artificial, estimaremos el número de partículas de polvo y hielo que dispersaron  la luz solar y  produjeron la iluminación nocturna que el mencionado hecho requería. Si utilizamos como referencia la iluminación de la luna llena a medianoche (cuando la Luna cruza el meridiano del lugar), es  claro que  se necesitaría un cierto número $N_{L}>1$ de lunas llenas brillando simultáneamente en el cielo para que la iluminación de ellas permita leer un periódico. Mantendremos a $N_{L}$ como una incógnita y al final de esta sección 
 estimaremos su valor. 
 La energía de la luz solar interceptada por el disco lunar es $B_{\odot} \pi r_{L}^{2}$, donde $r_{L}=1738 $ km es el radio de la Luna.  La Luna refleja difusamente, hacia afuera del hemisferio lunar iluminado, solo una fracción de la energía solar recibida. Esa fracción, que denotaremos  $\gamma_{L}$,  es el albedo lunar ($\gamma_{L}\approx 0.1$). En Luna llena, toda la energía reflejada ($B_{\odot} \pi r_{L}^{2} \gamma_{L}$) se dispersa hacia el hemisferio que apunta al observador en la Tierra. De modo que, en la posición del observador, esa energía se distribuye sobre la superficie del hemisferio $2 \pi d_{TL}^{2}$, donde  $d_{TL}= 384400$ km es la distancia Tierra-Luna. Por lo tanto, el brillo de la Luna llena resulta
 
\begin{equation}
 B_{L}=\frac{B_{\odot} \pi r_{L}^{2} \gamma_{L}}{2 \pi d_{TL}^{2}}\approx 10^{-6}  B_{\odot}. 
 \label{Luna}
\end{equation}  
 Tal como hemos dicho, vamos a suponer que el brillo de la nube de polvo que iluminó las noches blancas tuvo 
 $N_{L}$ veces el brillo de la Luna llena.  Es decir, $B_{c}=N_{L} B_{L}=N_{L} \,10^{-6}  B_{\odot}$, donde el brillo de la nube de polvo $B_{c}$ se puede obtener a partir de la ecuación (\ref{Bcloud}), del siguiente modo:
 \begin{equation}
 B_{c}=\frac{B_{\odot} \pi a^{2} Q_{s} \rho_{p}}{4 \pi} \sum  \frac{dV_{c}}{ d_{c}^{2} } cos\, \zeta,
 \label{Bc}
 \end{equation}
donde $dV_{c}$ es un elemento de volumen de la nube,  $d_{c}$ la distancia de dicho volumen al observador y  $\zeta$ el ángulo entre  la dirección a $dV_{c}$ y la vertical del lugar del observador. El $ cos\, \zeta$ tiene en cuenta el hecho de que el haz de luz se reparte sobre  una superficie mayor cuando su  ángulo de incidencia  $\zeta$ con respecto a la vertical se acerca a $90^{\circ}$. La suma de todos los elementos de volumen es  $\sum dV_{c}=V_{c}$, donde $V_{c}$ es el volumen total de la nube visible al observador. Para una dada posición $(\rho, \theta)$ dentro de los  intervalos especificados en (\ref{dlP}),  el valor de  $dV_{c}$  se calcula mediante la ecuación $dV_{c} =\rho\,  d\theta \, d\rho \, dl'$, donde $dl'$ es la distancia entre los dos  planos de corte que delimitan al elemento de volumen y se calcula a través de la fórmula (\ref{lP}). La posición $(x,y,z)$ de $dV_{c}$ se obtiene mediante las expresiones (\ref{xyzH}) y su distancia 
 al observador $O$ es $d_{c}=\sqrt{x^{2}+ y^{2} + z^{2}}$. El valor de $cos\, \zeta$ podemos obtenerlo usando la propiedad del producto escalar entre dos vectores, en nuestro caso  $\overrightarrow{d_{c}}=(x,y,z)$ y $\overrightarrow{z}=(0,0,z)$, con lo cual $ cos\, \zeta=\frac{\overrightarrow{d_{c}}. \overrightarrow{z}}{d_{c}\, z}$.
 
\begin{figure}
\includegraphics[scale=0.65]{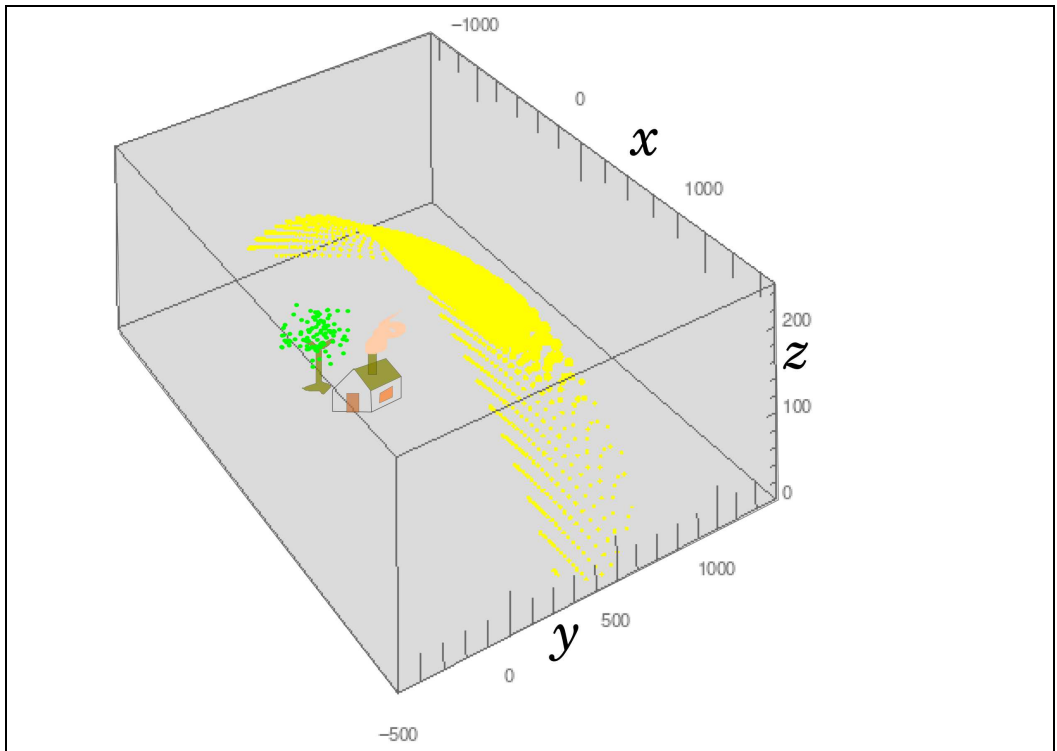} 
\caption{Representación de la capa  de polvo (puntos amarillos) que reflejó la luz solar e iluminó el cielo de medianoche visto desde la latitud de $ 50^{\circ}$. Las coordenadas $x$, $y$ y $z$ están expresadas en km. El observador se encuentra en el origen de coordenadas y los objetos  que lo simbolizan, ubicados en sus cercanías sobre el horizonte, no se encuentran naturalmente dibujados a escala.} 
\label{NocheBlanca3D}
\end{figure}

 La posición y brillo de cada  elemento de la capa de polvo son representados, en la fig. \ref{NocheBlanca3D}, por la coordenada $(x,y,z)$ del punto medio del elemento de la capa, y  por el ancho del punto, que es proporcional al brillo, respectivamente. Según nuestro modelo, un observador ubicado a una latitud de $\approx 50^{\circ}$ vio una semi-bóveda brillante recostada hacia el norte que iluminó el cielo a medianoche.
 
 Igualando $B_{c}$ a  $N_{L} B_{L}$, dados  por (\ref{Bc}) y (\ref{Luna}) respectivamente, y despejando la densidad numérica $\rho_{p}$,  obtenemos que 
 \begin{equation}
 \rho_{p} = 4\, 10^{-6} \, N_{L}
  (a^{2}\, Q_{s}\, \sum  \frac{dV_{c}}{ d_{c}^{2}} cos\, \zeta)^{-1}.
 \label{densidadNube}
 \end{equation}
El tamaño del polvo cometario es submicrométrico. Por ello, adoptaremos $a=0.1 \mu$m. La longitud de onda $\lambda$ de la luz amarilla es del orden de $ 0.6 \mu$m y por lo tanto $\frac{2 \pi a}{\lambda} \approx 1$ y $Q_{s} \approx 0.1$,  de acuerdo con la teoría de Mie. Aplicando (\ref{densidadNube}) con los valores arriba adoptados, $\rho_{p}=  0.13 \, N_{L}$ granos de polvo por $cm^{3}$. También, hemos calculado que el espesor medio de la capa de polvo $E_{e}$ y su volumen $V_{c}$ son $\approx 70$ km y  $1.15\, 10^{23}$ cm$^{3}$, respectivamente. Con ello, 
el número total de granos de polvo de la nube que iluminó la noche al observador $O$ es igual a $\rho_{p} V_{c}=  1.5\, 10^{22}\, N_{L}$. Si suponemos que la masa $m_{p}$ de un grano de polvo es de $4.2\, 10^{-15}$ gramos, la masa total de la nube es $\rho_{p} V_{c} m_{p}= 63 \, N_{L}$  toneladas y  la masa total de la envoltura de polvo $M_{e}$ que rodeó a la Tierra es 
\begin{equation}
M_{e}=\rho_{p} (4 \pi R_{T}^{2} E_{e} ) m_{p}\approx 20000 N_{L} \,\,\rm ton
\label{MasaEnv}
\end{equation}

La máxima iluminación de la luna llena es del orden de 1 lux y una muy buena iluminación para leer requiere entre 400 y 500 lux. Por lo tanto,  $N_{L}=200$ es un valor razonable. Reemplazando ese valor de $N_{L}$ en (\ref{MasaEnv}), obtenemos que $M_{e}=$ 4 millones de toneladas. Aquí nos referimos al polvo que no se desintegró totalmente al ingresar a la atmósfera y pudo reflejar la luz solar. En consecuencia, la cantidad total de polvo  podría ser mayor a $M_{e}$. Según resulta de nuestro modelo, varios millones de toneladas de polvo cubrieron la Tierra, resultado que coincide con la estimación que hizo el Dr. Ramachandran Ganapathy\index{Ganapathy R.} \cite{Ganapathy}
a partir de su descubrimiento de que a una profundidad de $\approx 10.5$ metros en el hielo Antártico, resultante de nieve que cayó en 1908 y 1909, hay un gran exceso de Iridio, un elemento raro el la superficie de la Tierra pero abundante en cometas y asteroides.
Si bien  Ganapathy \index{Ganapathy R.} atribuye  la capa de polvo descubierta en la Antártida al polvo del cuerpo cósmico eyectado por su propia explosión en  Tunguska\index{Tunguska}, el resultado de Ganapathy es independiente de su interpretación. El hecho observacional  descubrimiento por  Ganapathy nos indica que el evento de Tunguska\index{Tunguska} fue parte de un fenómeno que involucró a toda la Tierra.

Probablemente la suposición  de que el polvo que rodeó a la Tierra estuvo uniformemente  distribuido lleve a una sobrestimación de $M_{e}$, ya que esta resulta de una extrapolación de  regiones en las que el fenómeno se manifestó notablemente debido a que sobre ellas la acumulación de polvo fue mayor. Este parece también ser  el caso de la Antártida donde  se observaron anormalmente intentas auroras, algunas horas  antes  de la explosión de Tunguska\index{Tunguska}. Cabe recordar que la región del polo Sur se encontraba en la noche de seis meses.

\subsubsection{Modelo sobre la corriente de polvo que atravesó la Tierra y su interacción con la atmósfera: ¿más de un Tunguska\index{Tunguska}?}

\begin{quote}
\small \it{...The next point is that the Earth and the atmosphere are spinning round and that the Cloud will be hitting the atmosphere from one side only.
 ``from what side ?'', asked Parkinson.
``The earth's position in its orbit will be such that the Cloud will come at us from the approximate direction of the Sun'' ,  explained Yvette Hedelfort.

``Although the Sun itself won't  be visible'',  added Marlowe. 
``So the Cloud will be hitting the atmosphere during what would normally be the daytime?'' .
``That's right. And it will not be hitting the atmosphere during the night'' .}\rm

Fred Hoyle \footnote{Astrofísico británico, uno de los  creadores de la teoría cosmológica según la cual el universo se encuentra en un estado estacionario. Fred Hoyle fue quién acuñó el nombre de ``
Big Bang'' para llamar a la teoría rival a la suya.} \index{Hoyle F.} (1915-2001) en su novela de ciencia ficción ``La Nube Negra'' (The Black Cloud)  publicada  en el año 1957.
\end{quote}

Nos interesa estimar la densidad media $\rho_{f}$  de partículas de polvo de la sección de la corriente de las Beta Tauridas\index{Tauridas Beta (corriente de meteoros)} que cruzó la Tierra en ocasión del evento Tunguska\index{Tunguska}. Como hemos visto, el evento Tunguska\index{Tunguska} se asocia con el paso de la Tierra a través de la corriente de las Beta Tauridas\index{Tauridas Beta (corriente de meteoros)}. Se estima que la Tierra puede tardar un día en cruzar la corriente de las Beta Tauridas\index{Tauridas Beta (corriente de meteoros)}. Por lo tanto, el ancho de esta corriente $D_{f}$ puede calcularse en forma  aproximada  por medio de 
\begin{equation}
D_{f}= v_{T} t,
\end{equation}
donde $v_{T}=30 $ km s$^{-1}$  es la velocidad orbital de la Tierra y $t=24$ horas. Como la velocidad $v_{f}$ del flujo con respecto a la Tierra  fue del orden de  de  33  km s$^{-1}$, casi el mismo valor que $v_{T}$,  la longitud de la corriente $L_{f}$ que pasó sobre la Tierra fue del mismo orden que $D_{f}$, es decir $ D_{f} \approx L_{f}  \approx 3\,\, 10^{6}$ km (10 veces la distancia Tierra-Luna).

Ahora nos preguntamos cuál fue la luminosidad del cielo nocturno debido al reflejo de la luz solar por las partículas de polvo cuando la Tierra se encontraba en medio de la corriente de polvo de las Beta Tauridas\index{Tauridas Beta (corriente de meteoros)}. Dividamos el espacio en torno a la Tierra en cáscaras concéntricas, como las capas de una cebolla, cada una con  un espesor $\Delta d_{c}$. Consideremos una capa de radio $d_{c}$, con centro en la Tierra, un diferencial de volumen de esta capa expresado en coordenadas esféricas es $dV=d_{c}^{2}\, sen \theta \,d\theta \,d\phi\, \Delta d_{c}$, donde $\theta$ es el ángulo entre la vertical del lugar del observador y la dirección a $dV$. La sección de volumen entre 
$\theta$ y $\theta + d\theta$, a la que llamaremos aquí $dV_{c}$, se obtiene integrando $dV$ con respecto a $\phi$ entre 0 y $2 \pi$, con lo cual   $dV_{c}=2 \pi d_{c}^{2}\, sen \theta\, d\theta\,  \Delta d_{c}$. Aplicando (\ref{Bcloud}), obtenemos el brillo de esta sección de volumen
\begin{equation}
dB_{c}=\frac{B_{\odot} \pi a^{2} Q_{s} \rho_{f} dV_{c}}{4 \pi d_{c}^{2}} cos \theta
\label{Bcloud2}
\end{equation}
El $cos\, \theta$ en (\ref{Bcloud2}) tiene en cuenta el hecho de la energía lumínica se reparte en una superficie mayor cuanto aumenta la inclinación con que incide la luz.
Si reemplazamos la expresión de $dV_{c}$ en (\ref{Bcloud2}) e integramos con respecto a $\theta$ entre 0 y $\pi/2$, dado que el observador ve solo un hemisferio, obtenemos el brillo completo de la semicáscara que se encuentra entre $d_{c}$ y $d_{c} + \Delta d_{c}$; el cual resulta $dB_{c}=\frac{B_{\odot} \pi a^{2} Q_{s} \rho_{f}}{4 } \Delta d_{c}$. Entonces,  el brillo de fondo del  cielo nocturno  se obtiene sumando el brillo de todas las cáscaras: $B_{c}=\frac{B_{\odot} \pi a^{2} Q_{s} \rho_{f}}{4 } \sum_{n=1}^{N} \Delta d_{c}$, donde $N$ es el número de cáscaras que contribuyeron al brillo nocturno. Teniendo en cuenta $\sum_{n=1}^{N} \Delta d_{c}=N \Delta d_{c}= \frac{D_{f}}{2} = \frac{v_{T} t}{2}$, 
\begin{equation}
B_{c}=\frac{1}{8 }B_{\odot} \pi a^{2} Q_{s}  \rho_{f} D_{f}
\label{BrilloFlujo}
\end{equation}
\begin{figure}
\includegraphics[scale=0.65]{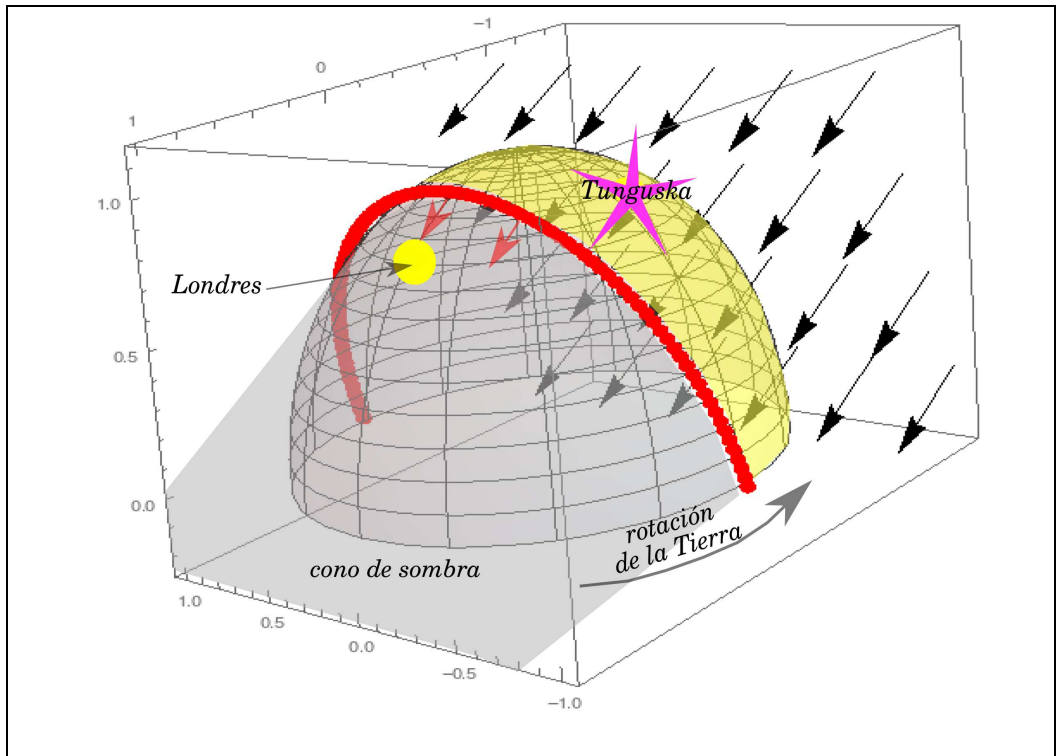} 
\caption{Esquema de la posición del anillo de gas y polvo (línea gruesa roja) formado en la alta atmósfera  como consecuencia del encuentro rasante con  granos de polvo de la nube que acompañó al fragmento de cometa que explotó  en Tunguska\index{Tunguska}. Las flechas negras indican la dirección de las partículas de polvo, que  consideramos coincidente con la de los rayos solares, y por lo tanto el anillo indicado en rojo separa el hemisferio diurno del nocturno. Las posiciones de Londres\index{Londres} (en amarillo) y de Tunguska\index{Tunguska} (estrella violeta) en el momento de la explosión son indicadas. Las flechas rojas muestran el desplazamiento del propuesto anillo de polvo y gas hacia Londres\index{Londres} y Europa en general. }
\label{anillo}
\end{figure}
El brillo de fondo del cielo nocturno visto de diferentes latitudes tiene que haber sido menor que un cuarto  luna  para pasar desapercibido. Haciendo $B_{c}= \frac{1}{2} B_{L}$ en (\ref{BrilloFlujo}) y despejando $\rho_{f}$ encontramos que $\rho_{f}=0.42$ m$^{-3}$. Por medio de la expresión $\rho_{f} \pi R_{T}^{2} D_{f} m_{p}$ encontramos que la masa de esta componente de la envoltura de polvo que rodeó a la Tierra fue de solo $\approx 3000$ toneladas. Recordemos que la masa total de la envoltura de polvo necesaria para explicar las noches blancas es de $\approx 4$ millones de toneladas (ver sección anterior). En consecuencia, la mayor parte de la masa de la envoltura de polvo se debió al encuentro de la Tierra con al menos una nube discreta de polvo de la corriente de las Beta Tauridas. Esta nube de polvo fue probablemente la coma o envoltura del fragmento de cometa que explotó en Tunguska\index{Tunguska}.
 
 La explosión de Tunguska\index{Tunguska} ocurrió el 30 de junio a las 7 horas y 17 minutos de la mañana, hora local,  y con respecto al meridiano de  Greenwich\index{Greenwich} a la 0 hora y 14 minutos. Es decir,  en Londres\index{Londres} era medianoche y por lo tanto el hemisferio de la Tierra  iluminado por el Sol estaba aproximadamente centrado sobre el meridiano de $180^{\circ}$ de longitud geográfica (el antimeridiano de Greenwich). Como la corriente de las Beta Tauridas\index{Tauridas Beta (corriente de meteoros)} ingresa a la atmósfera desde el lado diurno de la Tierra, la alta atmósfera sobre Asia, Oceanía, parte del Océano Pacífico y parte oeste de América del Norte pudo ser afectada por la interacción con la nube de polvo.

 \begin{figure}
\includegraphics[scale=0.65]{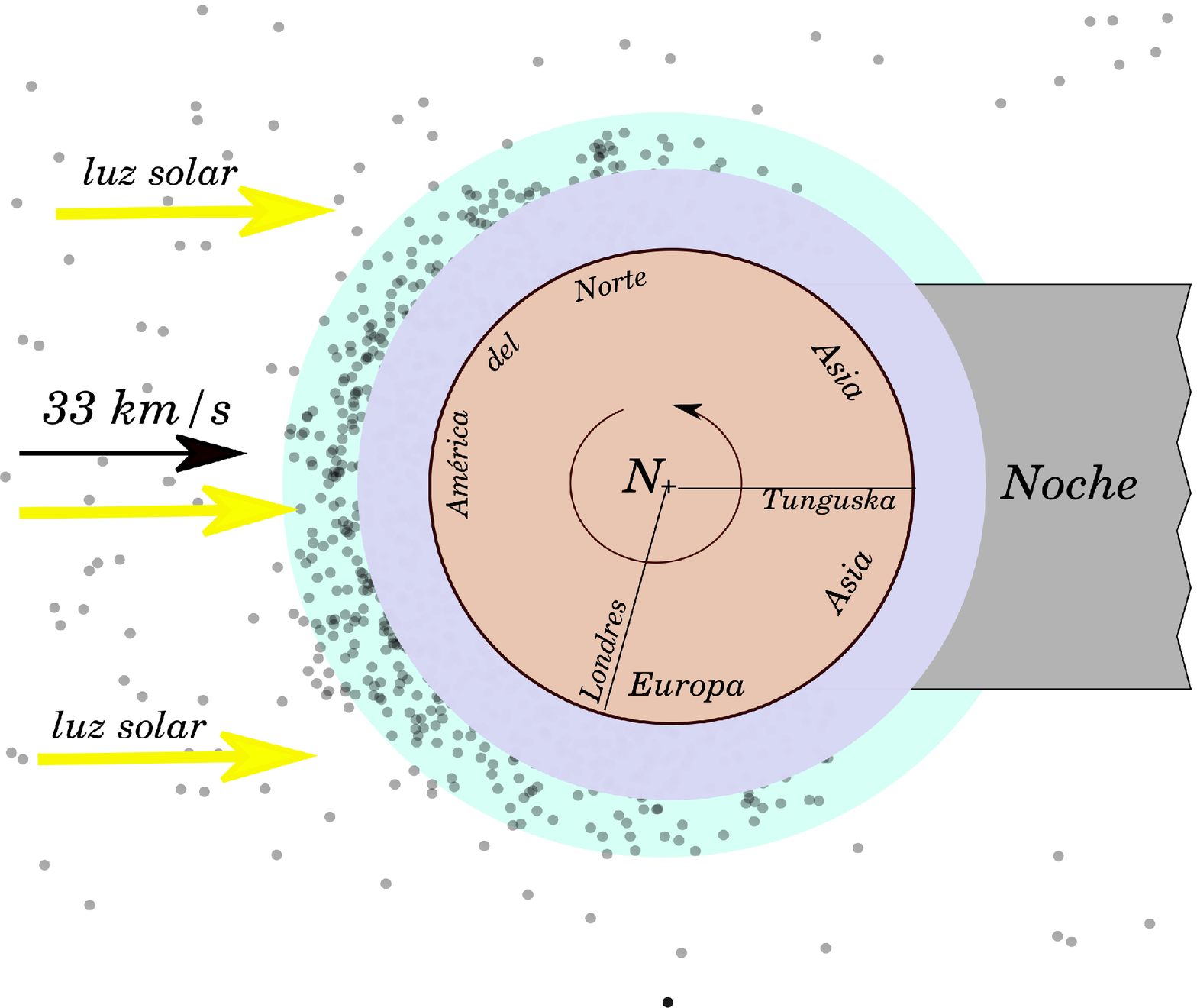} 
\caption{Representación del hipotético choque de una nube de polvo contra la Tierra, ocurrido 7 horas antes de la explosión de Tunguska\index{Tunguska}. Los granos de polvo de la nube (puntos negros) inciden sobre el hemisferio diurno de la Tierra. El dibujo muestra en forma esquemática una sección perpendicular al eje polar, correspondiente  a una latitud cercana a la de Tunguska\index{Tunguska} y a la de Londres\index{Londres}.  El sector que yace en el cono de sombra es menor que el iluminado por el Sol, reflejando el hecho de que la noche es más corta que el día,  puesto que el fenómeno ocurrió al comienzo del verano en el Norte. La zona celeste representa la mesósfera, donde la mayoría de los granos de polvo se frena, y la zona azul la parte densa de la  atmósfera terrestre.  El espesor de la atmósfera es exagerado con fines de visualización.  Los meridianos de Tunguska\index{Tunguska} y Londres\index{Londres} son indicados. La flecha curva, en torno al polo norte (N), muestra el sentido de la rotación terrestre}
\label{DosNubes1}
\end{figure}

 Europa, y naturalmente Londres\index{Londres}, se encontraba en el lado oscuro de la Tierra y en consecuencia quedó fuera de la zona sobre la cual  el polvo cósmico se precipitó directamente en la atmósfera. Esto plantea la incógnita  de  como llegó al lugar la capa de polvo que produjo  las notables noches blancas observadas en Londres\index{Londres} y en el resto de Europa. Hemos demostrado que  el polvo esparcido por la explosión del fragmento de cometa en Tunguska\index{Tunguska}   abarcó  un área relativamente pequeña en torno al epicentro y por lo tanto debemos descartar este mecanismo para el trasporte del polvo cósmico. En base a la forma de la devastación de los árboles que ocasionó el impacto del cuerpo cósmico, se estima que la trayectoria del cuerpo fue  en dirección noroeste y su ángulo de inclinación con respecto al horizonte fue relativamente pequeño ($\approx 30^{\circ})$. De modo que el cometa y el sector de  la nube de polvo cercano a él chocó en forma casi rasante con la atmósfera.  
Si el tamaño de la nube de polvo tuvo como máximo el tamaño de la Tierra, la duración del choque fue de 6 minutos a lo sumo.
Como resultado de la interacción de parte de la nube de polvo con la atmósfera superior probablemente se formó una corriente de gas y polvo  moviéndose a gran altura hacia el noroeste con respecto al suelo.

 \begin{figure}
\includegraphics[scale=0.65]{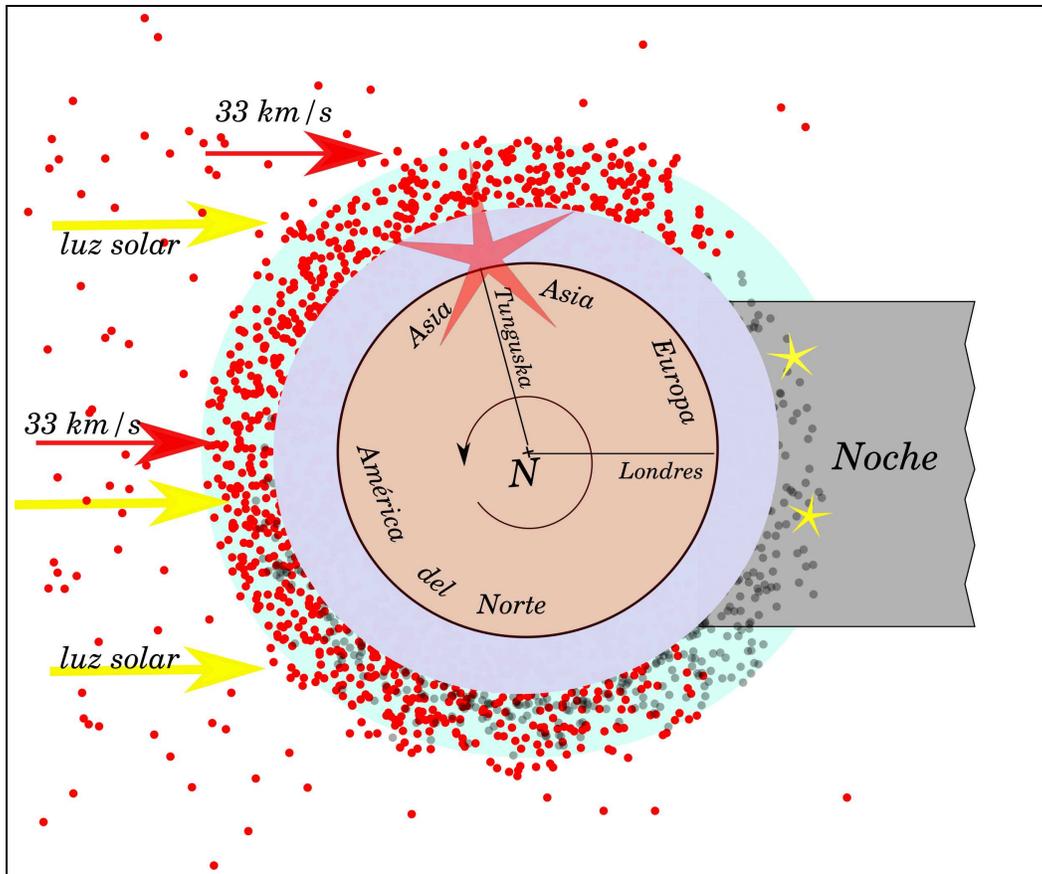} 
\caption{Choque del fragmento de cometa que explotó en Tunguska\index{Tunguska} y de la nube de polvo que acompañaba al cuerpo. Los granos de polvo son representados por los puntos rojos, para diferenciarlos de los granos del evento anterior (puntos negros).}
\label{DosNubes2}
\end{figure} 

 Si  la coma del fragmento de cometa, debido a  su gran tamaño,  incidió sobre todo o gran parte del hemisferio diurno,  
 la parte superior de la atmósfera sobre la cual los granos de polvo  incidieron rasante se la puede representar como  un anillo de gas que  divide a ambos hemisferios.  El impulso de las  partículas  de polvo que  el anillo frena e incorpora a su cuerpo  se transfiere al anillo y lo obliga a moverse hacia el hemisferio que no estuvo  expuesto a la corriente de polvo. Es decir, el anillo de gas y polvo  se mueve horizontalmente a gran altura como un frente de viento con  una velocidad media $v_{w}$ relativa a la superficie de la Tierra. La sección del frente de viento  que se mueve sobre el meridiano de Greenwich es la que recorre la menor distancia para llegar a Londres\index{Londres} ($\approx$ 2300 km). El frente de viento cubre el cielo de Londres\index{Londres}, originando su gran noche blanca, aproximadamente 24 horas después de producido el evento de Tunguska\index{Tunguska} y el anillo de gas y polvo.  Por lo tanto, $v_{w}=\frac{2300 \rm km}{24 \rm horas} \approx 100$ \rm km h$^{-1}$. En la fig. \ref{anillo} representamos la situación en el momento de la explosión de Tunguska\index{Tunguska}. En 24 horas la Tierra gira  $360^{\circ}$ y la posición de Londres\index{Londres} con respecto al cono de sombra es la misma que la representada en la fig. \ref{anillo}, pero ahora parte del anillo de polvo se extiende sobre el cielo de Londres\index{Londres} y produce la singular noche blanca observada en la noche del 30 al 31 de junio.
 
 Si $M_{p}$ es la masa total de las partículas de polvo contenidas en la sección del anillo de gas y polvo que se extendió sobre el cielo de Londres\index{Londres}  y $v_{p}=33$ km\,$s^{-1}$ es la velocidad de las partículas de polvo con la que chocaron sobre la atmósfera, podemos hacer la siguiente aplicación aproximada  de la ley de conservación del impulso: 
\begin{equation}
 M_{p} v_{p}= (M_{p} + M_{g}) v_{w},
 \label{ImpulsoRing}
\end{equation}
 donde $M_{g}$ es la masa total de la componente gaseosa de la sección del anillo. 
 En la sección anterior, estimamos  que $M_{p}=63 N_{L}=12600$ toneladas para la capa de polvo que produjo la noche blanca en Londres\index{Londres}. De (\ref{ImpulsoRing}) obtenemos que $M_{g}=(\frac{v_{p}}{v_{w}}-1) M_{p}=15 \times 10^{6}$ toneladas. La densidad gaseosa de la capa de polvo es $\rho= \frac{M_{g}}{V_{c}}$, donde $V_{c}= 1.15\times 10^{23}$ cm$^{3}$ según lo obtenido en la sección anterior. Por lo tanto, $\rho=1.3 \times 10^{-7}$ kg m$^{-3}$, densidad similar a la densidad atmosférica a la altura $h=$135 km, de acuerdo con la fórmula (\ref{Atmosfera}). Estos resultados  son congruentes con nuestra hipótesis.

 Otra posibilidad para explicar la presencia de polvo cósmico sobre Europa es que, durante el 30 de junio, pero horas antes de la explosión de Tunguska\index{Tunguska} mientras era de día en Europa,  otra nube de polvo chocara contra la Tierra envolviendo en forma directa la región  con una capa de polvo. Esta posibilidad y la anteriormente expuesta no son mutuamente  excluyentes. 
Antes del suceso de Tunguska\index{Tunguska} se observaron fenómenos ópticos anómalos en diferentes partes del mundo. En particular, 7 horas antes de la explosión se Tunguska\index{Tunguska},  se observaron intensos resplandores (¿Auroras?) en el Polo Sur Antártico. Ello avala la hipótesis de que además de la coma o envoltura del fragmento de cometa que explotó en Tunguska\index{Tunguska} hubo otra importante nube de polvo que poco antes chocó contra la Tierra.
 
 En la fig. \ref{DosNubes1}, se  escenifica  la supuesta colisión de una  nube de polvo con la Tierra, ocurrida  7 horas antes de la explosión de Tunguska\index{Tunguska}. El ángulo que la Tierra rotó en el tiempo que duró la colisión fue seguramente pequeño y si el tamaño de la nube de polvo fue del orden o mayor que el tamaño de la Tierra, todo hemisferio terrestre diurno de ese momento acumuló el polvo que incidió sobre  su mesósfera. Probablemente la parte más densa de la nube interaccionó con la atmósfera cercana al polo Sur. Ello explicaría la gran cantidad de material cósmico encontrado en hielo Antártico  por el Dr. Ramachandran Ganapathy \cite{Ganapathy} y la intensa iluminación del cielo nocturno observada  7 horas antes de la explosión de Tunguska\index{Tunguska} por  expedicionarios que se encontraban en ese momento en la  Antártida\index{Antártida}. 
 
 En la fig. \ref{DosNubes2}, se escenifica  el resultado combinado de ambos encuentros, es decir el relacionado con el evento de Tunguska\index{Tunguska}  (fig. \ref{anillo}) y 
 el relacionado con el evento observado en  la Antártida (fig. \ref{DosNubes1}). 
Aquí como en la fig. \ref{DosNubes1} dibujamos un corte sobre el hemisferio norte paralelo al ecuador terrestre. Los puntos rojos representan a los granos de polvo que acompañaban al pequeño cometa que explotó  en Tunguska\index{Tunguska}. 
 Los granos de polvo   acumulados en la mesósfera debido al encuentro con la nube que chocó primero   son  representados   por puntos negros en la  fig. \ref{DosNubes2}, pero rotados con respecto a los mostrados en la fig. \ref{DosNubes1} en $105^{\circ}$, que es el ángulo que rotó la Tierra en 7 horas.

 \subsection{Conclusiones sobre el caso Tunguska} \index{Tunguska}
La explicación más probable del evento Tunguska\index{Tunguska} es que un cuerpo cósmico, tal como un cometa o asteroide, explotó en la atmósfera. Nosotros hemos argumentado  en favor de  la hipótesis de que un fragmento de núcleo cometario  impactó en Tunguska\index{Tunguska} y de que ese relativamente denso fragmento fue parte de una extensa y poco densa corriente de polvo que cruzó la Tierra por al menos un día. Por la fecha del evento, dicha corriente de polvo se corresponde con la corriente de meteoros ``las Beta Tauridas''\index{Tauridas Beta (corriente de meteoros)}.  La partículas de polvo de la corriente interceptadas por la atmósfera formaron  en la mesósfera una capa  de polvo que rodeó completamente a la Tierra y produjo efectos ópticos que se observaron aun antes de la explosión de  Tunguska\index{Tunguska}. La presencia de polvo a grandes alturas sobre la superficie puede reflejar la luz solar y producir noches luminosas. Sin embargo, la cantidad de polvo acumulada en la alta atmósfera por el paso de la corriente de las Beta Táuridas\index{Tauridas Beta (corriente de meteoros)} no es suficiente para explicar las singulares noches blancas que se observaron en toda Europa un día después de la explosión de Tunguska\index{Tunguska}. Por otra parte, la explosión de Tunguska\index{Tunguska} tuvo efectos solo locales y el polvo eyectado alcanzó distancias relativamente cortas.

Nuestra hipótesis es que el fragmento de cometa que impactó en Tunguska\index{Tunguska} estuvo rodeado por  una extensa envoltura  de polvo (coma) que cubrió en unos pocos minutos gran parte del hemisferio diurno de la Tierra. Los granos de polvo que chocaron en forma rasante con la alta atmósfera que se encontraba en el límite entre el hemisferio iluminado y el nocturno formaron un frente de gas y polvo que se trasladó a una velocidad de $\approx 100$ km h$^{-1}$ y parte del mismo cubrió al día siguiente el cielo nocturno de Europa, originando las noches blancas.

 La corriente de granos de polvo golpeó la parte de la atmósfera iluminada por el Sol y por lo tanto las lluvias de meteoros no se vieron a simple vista. En cambio, sobre una angosta franja de la atmósfera encima del plano que divide el hemisferio iluminado de la Tierra del hemisferio nocturno, los meteoros más luminosos debieron haberse visto. El gas de la alta atmósfera  excitado por los choques contra los granos de polvo se hace luminoso, como en una lámpara de neón.  Este resplandor del gas más  la reflexión de la luz solar por las partículas  de polvo en la alta atmósfera provocaron probablemente crepúsculos más luminosos de lo normal y enmascararon las lluvias de meteoros.

Los astrónomos australianos Steel \index{Steel D.} y Ferguson \index{Ferguson R.} revisaron informes de expedicionarios que en ese entonces se encontraban en la Antártida y encontraron que una aurora excepcional fue observada alrededor de 7 horas antes de la explosión de Tunguska\index{Tunguska}. Este hecho y otros fenómenos ópticos manifestados antes del suceso de Tunguska\index{Tunguska} refutan la idea de que el polvo que envolvió al planeta provino  del material eyectado por la explosión de Tunguska\index{Tunguska}.
 
 El Dr. Ramachandran Ganapathy\index{Ganapathy R.} \cite{Ganapathy} descubrió, en el hielo Antártico, una capa de elementos  abundantes en cometas y asteroides, depositada con la nieve  que cayó en 1908 y 1909.  y por lo tanto asociada con el evento Tunguska\index{Tunguska}. Nosotros proponemos que ese abundante material recolectado en la Antártida\index{Antártida} provino  de una nube de polvo que chocó 7 horas que la coma del fragmento del cometa que explotó en Tunguska\index{Tunguska}. Ello explicaría los intensos resplandores que vieron los expedicionarios en la Antártida y el abundante material cometario descubierto por Ganapathy.

 Tanto el Dr. Ganapathy, a partir del material cósmico precipitado  en la región Polar Sur, como nosotros a través de modelos que expliquen las noches blancas, concluimos que la masa total de granos de polvo que cubrió la Tierra fue de algunos millones de toneladas.

\section{Enjambres cósmicos y los cielos que vieron las civilizaciones antiguas}

El sistema solar\index{sistema solar} está inmerso en una extensa nube de cometas, remanentes de la nube primigenia que dio origen al sistema solar\index{sistema solar}. A dicha nube cometaría se le llama nube de Oort en honor al astrónomo holandés Jan Hendrik Oort (1900-1992)\index{Oort J.H.} que  fue el primero en proponer su existencia. Hoy sabemos casi con certeza que los cometas son fundamentalmente cuerpos del sistema solar\index{sistema solar} \footnote{Sin embargo, no podemos excluir la posibilidad de que  algunos cometas puedan provenir del medio interestelar. Un ejemplo de ello podrían ser los objetos llamados  21/Borisov\index{Borisov 21 (cometa)} y  Oumuamua\index{Oumuamua (cometa)}, los cuales están pasando por el sistema solar\index{sistema solar} con órbitas hiperbólicas.}. Cuando un cometa ingresa al interior del sistema solar\index{sistema solar}, los grandes planetas, esencialmente Júpiter y Saturno, pueden perturbar gravitacionalmente la órbita del cometa y convertirlo en un cometa de período corto. Parte del medio ambiente orbital de la Tierra, como el polvo que refleja la luz zodiacal y las lluvias de meteoros, puede explicarse por los escombros que dejan detrás los cometas de corto periodo en sus recurrentes pasos por las cercanías de la órbita terrestre.

El cometa con el periodo más corto es el débil Cometa Encke \footnote{Este cometa fue también estudiado desde Argentina\index{Argentina}, \cite{Paolantonio}} que completa una vuelta  alrededor del Sol en solo  3,3 años y es de hecho único entre los cometas conocidos. Fue detectado telescópicamente por primera vez en 1785 y desde entonces registrado varias veces sin saber que se trataba del mismo cometa. En 1818,  fue observado nuevamente por Jean Lois Pons\index{Pons J.L.} y su órbita fue calculada por Johann Encke\index{Encke J.}, un alumno del famoso matemático
Gauss\index{Gauss} que había desarrollado un método para determinar la órbita de un cuerpo a partir de  medidas de su posición en el cielo en al menos tres diferentes instantes de tiempo. Se puede trazar un paralelo histórico con el Cometa Halley\index{Halley (cometa)},  ambos cometas no fueron llamados con los nombres de sus descubridores, sino con los nombres de quienes calcularon sus órbitas.

La órbita  o trayectoria que sigue  un planeta es trazada por una línea imaginaria determinada por las ecuaciones de movimiento. En cambio, en un cometa periódico, la corriente de meteoros asociada con el cometa materializa su órbita. En efecto, el cometa va desprendiendo partículas  que como ``el hilo de Ariadna'' \index{Ariadna}  indican su camino.
La corriente de meteoros asociada con la órbita del cometa Encke\index{Encke (cometa)} es la corriente de la Tauridas\index{Tauridas (corriente de meteoros)}. La sección de esta corriente que se encuentra con la Tierra en junio/julio se la llama Beta Tauridas\index{Tauridas Beta (corriente de meteoros)}  e ingresa a la atmósfera desde  el lado diurno de la Tierra. Como hemos visto, el choque con un cuerpo de las Beta Tauridas\index{Tauridas Beta (corriente de meteoros)} podría ser la causa del evento de Tunguska\index{Tunguska} del 30 de junio de 1908.

Nuevos descubrimientos refuerzan la sospecha de que la corriente de la Tauridas\index{Tauridas (corriente de meteoros)} ha jugado un papel en el desarrollo de acontecimientos astronómicos que influyeron en la cultura de los pueblos antiguos. Se han encontrado 6 o 7 asteroides que orbitan dentro de la corriente de la Táuridas\index{Tauridas (corriente de meteoros)}, de los cuales el asteroide Hephaistos\index{Hephaistos (asteroide)} es el más grande con cerca de 10 km de diámetro. Otro de los asteroides de esta corriente es Oljato\index{Oljato (asteroide)} con un diámetro de 1.5 km y con una apariencia de  cometa difunto. Se estima que debe haber entre 100 y 200 asteroides de más de un 1 km de diámetro dentro de la corriente de la Tauridas\index{Tauridas (corriente de meteoros)}. Todo esto parece indicar que estamos presenciando los  escombros de un  enorme cometa que se fue desintegrando durante los últimos 20 mil años, según ciertas estimaciones.

Entonces, en estos últimos 20 mil años, la Tierra ha transitado por un camino muy peligroso y no es de extrañar que se haya encontrado con un fragmento del gran cometa. En efecto, hay evidencias geológicas de que un gran cuerpo cósmico chocó, alrededor de 13 mil años atrás, contra la Tierra desencadenando una catástrofe climática planetaria, conocida como ``Dryas Reciente''. Se descubrió  un delgado estrato sedimentario, con abundancia de materiales de origen cósmico y con una  amplia distribución sobre la Tierra, cuya datación se corresponde con el inicio del enfriamiento global ``Dryas Reciente''\index{Dryas Reciente} \cite{Dryas}. Esto abona la hipótesis de que un gran impacto cósmico contribuyó al enfriamento de la Tierra y a la extinción de la megafauna\index{megafauna}  del final del Pleistoceno\index{Pleistoceno} \cite{Firestone}.

Al comienzo, la presencia en el cielo de  tan extraños objetos luminosos debe haber infundido pánico en los hombres. El miedo a lo desconocido es una herramienta evolutiva de nuestro cerebro que nos previene sobre un peligro posible. Desde este punto de vista,  es natural que acontecimientos como la aparición de un gran cometa en el cielo despertaran temor en el hombre antiguo. Hoy que conocemos la potencialidad destructiva de la colisión con un cometa,  vemos que la intuición de aquellos hombres estaba acertada. Al establecerse el fenómeno, es probable que se lo haya divinizado.
En el proceso de  fragmentación del gran cometa, arriba mencionado, quizá dos conspicuos fragmentos del cometa resaltaban  en el cielo y pudieron haber inspirado en la civilización Babilónica  la idea de los dioses Anu\index{Anu y Enlil} y Enlil. Claro esto es una mera especulación, pues no disponemos  de evidencias arqueastronómicas para tal asociación.
Sin embargo, se descubrió en la tablillas cuneiformes que hubo,   entre el año 300 y año 0 antes de Cristo, un incremento de referencias a fenómenos astronómicos.  También, en la cultura China\index{China}, el registro de acontecimientos astronómicos, como la aparición de cometas, lluvias de meteoros y bólidos, era considerado de vital importancia por su interés astrológico. Los registros Chinos, suplementados por aquellos de Babilonia\index{Babilonia} y por escritos de autores clásicos, muestran un notable acrecentamiento de tales fenómenos astronómicos  hacia fines del primer milenio de nuestra era. Ĺas referencias bibliográficas sobre tales hechos históricos pueden encontrarse en los libros \cite{Bailey} y \cite{Clube1}.

\section*{Agradecimientos}
Agradezco al Dr. Andrés Colubri la lectura del manuscrito y sus comentarios.

\renewcommand{\refname}{bibliografía}

\addcontentsline{toc}{}{Índice alfabético de nombres}
\printindex 

\end{document}